\pdfoutput=1 
\documentclass[11pt]{article}
\usepackage{ragged2e}
\usepackage{fullpage}
\usepackage[pagebackref, colorlinks = true, linkcolor = blue, urlcolor  = blue, citecolor = purple, bookmarks, hypertexnames=false]{hyperref}
\usepackage[bookmarks]{hyperref}
\usepackage{amssymb}
\usepackage{amsmath}
\usepackage{amsthm}
\usepackage{graphicx,color,colordvi}
\usepackage{bbm}
\usepackage{varioref}
\usepackage[capitalise]{cleveref}
\usepackage{stmaryrd}
\usepackage[utf8]{inputenc}
\usepackage[blocks]{authblk}
\usepackage{dsfont}
\usepackage{mathtools}
\usepackage{authblk}
\usepackage{wrapfig}
\usepackage[english]{babel}
\usepackage{physics}
\usepackage{braket}
\usepackage[table]{xcolor}
\usepackage[center]{caption}
\usepackage{tikz}
\usetikzlibrary{shapes,arrows,patterns,positioning,shapes.geometric,arrows.meta,decorations.markings}
\usepackage{ifthen}
\usepackage{pgfplots}
\pgfplotsset{compat=1.9}
\usetikzlibrary{shapes,arrows.meta}
\usetikzlibrary{positioning}
\usetikzlibrary{shapes.geometric}
\RequirePackage[framemethod=default]{mdframed}
\usepackage[margin=1in]{geometry}
\usepackage{comment}
\usepackage{url}
\usepackage{ upgreek }
\usepackage{makecell}
\usepackage{fancyref} 
\usepackage{float}
\usepackage{enumitem}
\usepackage{autonum}
\usepackage{subcaption}
\newfloat{algorithm}{t}{lop}
\usepackage{array,rotating ,makecell, multirow, tabularx}
\usepackage{scrextend}
\usepackage{mathrsfs}
\usepackage{enumitem}
\usepackage{soul}

\usepackage{IEEEtrantools}
\usepackage[normalem]{ulem}
\usepackage{cancel}

\usepackage{multicol}

\DeclareMathOperator{\dif}{\textup{d}}

\DeclareMathOperator{\poly}{poly}

\DeclareMathOperator{\id}{id}

 \newcommand{\1}{\ensuremath{\mathbbm{1}}}

\usepackage{cancel}

\newtheoremstyle{newdefinition}{}{}{\normalfont}{}{\bfseries}{}
{ }
{\thmname{#1} \thmnumber{#2}\thmnote{ (#3)}}

\newtheoremstyle{newplain}{}{}{\itshape}{}{\bfseries}{}{1em}
{\thmname{#1} \thmnumber{#2}\thmnote{ (#3)}}

\newtheoremstyle{newremark}{}{}{\normalfont}{}{\bfseries}{}{1em}
{\thmname{#1}}

\theoremstyle{newdefinition}
\newtheorem{definition}{Definition}[section]

\newtheorem*{definition*}{Definition}

\theoremstyle{newplain}
\newtheorem{theorem}[definition]{Theorem}
\newtheorem{lemma}[definition]{Lemma}
\newtheorem{claim}[definition]{Claim}

\newtheorem{proposition}[definition]{Proposition}
\newtheorem{corollary}[definition]{Corollary}
\newtheorem{remark}[definition]{Remark}
\newtheorem{example}[definition]{Example}

\newtheoremstyle{myplain}{5pt}{5pt}{\itshape}{0pt}{\bfseries}{}{5pt plus 1pt minus 1pt}{}
\theoremstyle{myplain}
\newtheorem*{theorem*}{Theorem}
\newtheorem*{corollary*}{Corollary}

\DeclareMathOperator{\R}{\mathbb{R}} 
\DeclareMathOperator{\N}{\mathbb{N}}

\DeclareMathOperator{\C}{\mathbb{C}}

\DeclareMathOperator{\cH}{\mathcal{H}}
\DeclareMathOperator{\cS}{\mathcal{S}}
\DeclareMathOperator{\cK}{\mathcal{K}}

\DeclareMathOperator{\cB}{\mathcal{B}}

\DeclareMathOperator{\cX}{\mathcal{X}}
\DeclareMathOperator{\cY}{\mathcal{Y}}

\DeclareMathOperator{\cD}{\mathcal{D}}

\newcommand{\NP}{\mathsf{NP}}

\title{Complexity of mixed Schatten norms of quantum maps}

\date{\today}

\author[1]{Jan Kochanowski\thanks{Email: jan.kochanowski@inria.fr}}
\author[2]{Omar Fawzi\thanks{Email: omar.fawzi@ens-lyon.fr}}
\author[1]{Cambyse Rouzé\thanks{Email: cambyse.rouze@inria.fr}}

\affil[1]{Inria, Télécom Paris - LTCI, Institut Polytechnique de Paris, 91120 Palaiseau, France}

\affil[2]{Inria, ENS Lyon, UCBL, LIP, F-69342 Lyon Cedex 07, France}

\date{\today}

\begin{document}
\maketitle

\begin{abstract}
We study the complexity of computing the mixed Schatten $\|\Psi\|_{q\to p}$ norms of linear maps $\Psi$ between matrix spaces, as well as their completely bounded (cb) counterparts $\|\Psi\|_{cb,q\to p}$. 
For entanglement-breaking quantum channels (CPTP maps) $\Phi_1,\Phi_2$, we prove that computing $\| \Phi_1-\Phi_2 \|^+_{1 \to 1}$ is an $\NP$-complete problem. In addition we show that computing $\| \Phi \|_{1 \to p}$ for entanglement-breaking CPTP maps $\Phi$ is also an $\NP$-complete problem when $p>1$. In contrast, for the completely-bounded (cb) case, we describe a polynomial-time algorithm to compute $\|\Psi\|_{cb,1\to p}$ and $\|\Psi\|^+_{cb,1\to p}$ for any linear map $\Psi$ and $p\geq1$ extending a seminal result about the computability of the diamond norm ($p=1$) of hermiticity preserving maps [Watrous, 2009].
Finally, when $\Phi$ is completely positive, we show that $\| \Phi \|_{q \to p}$ can be computed efficiently when $q \geq p$, and give an explicit runtime. The regime $q \geq p$, known as the non-hypercontractive regime, is also known to be easy for the mixed vector norms $\ell_{q} \to \ell_{p}$ [Boyd, 1974].
\end{abstract}
\thispagestyle{empty}

\begin{multicols}{2}
\setcounter{tocdepth}{2}
\tableofcontents
\end{multicols}

\newpage

\setcounter{page}{1}

\section{Introduction}
Given two normed vector spaces $(\cX,\|\cdot\|_{\cX}), (\cY,\|\cdot\|_{\cY})$, the mixed norm (also called operator norm) of a linear map $\Phi:\cX\to\cY$ is defined as
\begin{align}
    \|\Phi\|_{\cX\to\cY}:=\sup_{X\in\cX}\frac{\|\Phi(X)\|_{\cY}}{\|X\|_{\cX}}.
\end{align}
The probably most well studied instance of this generic problem is the case where $(\cX,\|\cdot\|_{\cX})=(\C^n,\|\cdot\|_{\ell_q}), (\cY,\|\cdot\|_{\cY})=(\C^m,\|\cdot\|_{\ell_p})$, where the $\ell_p$ norm of a vector $v=(v_i)_{i\in[m]}$ is defined as
\begin{align}
    \|v\|_{\ell_p}:=\left(\sum_{i=1}^m|v_i|^p\right)^{\frac{1}{p}}
\end{align} and the linear map $\Phi$ is described by a matrix $A\in\C^{m\times n}$. The corresponding 
\begin{align}
    \|A\|_{\ell_q\to \ell_p}:=\sup_{v\in\C^{n}}\frac{\|Av\|_{\ell_p}}{\|v\|_{\ell_q}},
\end{align} is known as the \textit{mixed matrix norm} and has found many applications, such as to the study of convergence times of Markov chains via hypercontractivity \cite{Diaconis.1996},  robust optimization \cite{Daureen.2007}, machine learning \cite{Dan.2019, Flamary.2014, Saketha.2009}, robust control theory, oblivious routing \cite{Bhaskara.2011}, analysis of random walks \cite{Biswal.2011, Barak.2012}, and even certain quantum data processing problems, such as computing acceptance probabilities of quantum protocols with multiple unentangled provers \cite{Barak.2012}, to name just a few. 

This naturally raises the question of algorithms to compute or approximate such norms. The computational complexities of the mixed matrix norms have been extensively studied and reveal a very interesting structure \cite{Boyd.1974, Daureen.2007, Bhaskara.2011, Bhattiprolu.2019, Bhattiprolu.2023, Barak.2012}. On the one hand, algorithms have been designed for some values of $q,p$, specifically when $A$ has non-negative entries. On the other hand hardness results were established for other ranges of $q,p$. For a more thorough overview of known results, see \cref{sec:classicalComplexities} or \cref{fig:Intro.figure} column a). 

In quantum information theory, linear maps between matrix spaces—such as quantum channels or their differences—are of central importance. The natural extension of the classical (commutative) $\ell_p$ space is the non-commutative $p$-Schatten space $\cS_p(\C^n)=(\C^{n\times n},\|\cdot\|_p)$, where $\|X\|_p:=\Tr[|X|^p]^\frac{1}{p}$. 
These give a natural notion of \textit{mixed Schatten norms}
\begin{align}
    \|\Phi\|_{q\to p}:=\sup_{X\in\C^{n\times n}}\frac{\|\Phi(X)\|_p}{\|X\|_q}, \quad \|\Phi\|^+_{q\to p}:=\sup_{\underset{X\in\C^{n\times n}}{X\geq0}}\frac{\|\Phi(X)\|_p}{\|X\|_q}
\end{align} which have, with their \textit{completely bounded} counterparts,
\begin{align}
    \|\Phi\|_{cb,q\to p}:=\sup_{n\in\N}\|\id_n\otimes\Phi\|_{M_n(\cS_q)\to M_n(\cS_p)},
\end{align} 
already found numerous applications in the study of quantum systems and  quantum information processing. Here, the spaces $M_n(\cS_q)$ used to define the mixed cb norm of a map are special cases of the $2$-indexed Schatten norms and are introduced in Section \ref{sec:notation} below. 
Like in the mixed matrix setting, these norms find various applications, some of which are detailed below, however, unlike the mixed matrix norms, the computational complexity of the mixed Schatten norms remain largely unknown.

In this work, we initiate a systematic study of the computational complexities of computing the mixed Schatten norms and completely bounded mixed Schatten norms of linear maps between matrix spaces. Our main results are summarized in Figure \ref{fig:Intro.figure}.

\subsection{Our results}\label{sec:summary.of.results}

\newcommand{\ReferenceColor}{black!80}   

\begin{figure}[h]
\begin{minipage}{0.34\textwidth}
\centering
    \begin{tikzpicture}[scale=0.85, font=\small]

    \draw[->] (0,0) -- (4.3,0) node[right] {$q$};
    \draw[->] (0,0) -- (0,4.3) node[left] {$p$};
    \draw[-] (0,4) -- (4,4);
    \draw[-] (4,0) -- (4,4);
    
    \foreach \x in {0,2,4} {
        \draw[-] (\x,0) -- (\x,-0.1); 
    }
    \foreach \y in {0,2,4} {
        \draw[-] (0,\y) -- (-0.1,\y); 
    }
    \node[below] at (4,0) {$\infty$};
    \node[below] at (2,0) {$2$};
    \node[below] at (0,0) {$1$};
    \node[left]  at (0,4) {$\infty$};
    \node[left]  at (0,2) {$2$};
    \node[left]  at (0,0) {$1$};

    \fill[red!50] (0,0) rectangle (4,4);
    
    \draw[red!50!,ultra thick] (0,0) -- (4.03,0);
    \draw[red!50!,ultra thick] (4,-0.03) -- (4,4.03);
    \draw[green!55!black,ultra thick] (0,-0.03) -- (0,4.03);
    \draw[green!55!black,ultra thick] (-0.03,4) -- (4.03,4);
    
    \draw[dotted, thick] (0,0) -- (4,4);
    
    \fill[green!50!black] (2,2) circle (1.5pt);
    
    \node at (2,4.3) {$\ell_q  \to \ell_p$};
    \node[rotate=90] at (-1,2) {general $A: \C^n \to \C^m$};

\end{tikzpicture}
\begin{tikzpicture}[scale=0.85, font=\small]

    \draw[->] (0,0) -- (4.3,0) node[right] {$q$};
    \draw[->] (0,0) -- (0,4.3) node[left] {$p$};
    \draw[-] (0,4) -- (4,4);
    \draw[-] (4,0) -- (4,4);
    
    \foreach \x in {0,2,4} {
        \draw[-] (\x,0) -- (\x,-0.1); 
    }
    \foreach \y in {0,2,4} {
        \draw[-] (0,\y) -- (-0.1,\y); 
    }
    
    \node[below] at (4,0) {$\infty$};
    \node[below] at (2,0) {$2$};
    \node[below] at (0,0) {$1$};
    \node[left]  at (0,4) {$\infty$};
    \node[left]  at (0,2) {$2$};
    \node[left]  at (0,0) {$1$};

    \draw[green!55!black, thick] (0,0) -- (4,4);
    \draw[dotted, thick] (0,0) -- (4,4);

    \draw[green!55!black,ultra thick] (0,0) -- (0,4.03);
    \draw[green!55!black,ultra thick] (-0.03,4) -- (4,4);
    \draw[green!55!black,ultra thick] (-0.03,0) -- (4.03,0);
    \draw[green!55!black,ultra thick] (4,-0.03) -- (4,4.03);

    \fill[green!55!black] (0,0) -- (4,0) -- (4,4) -- cycle;

    \fill[green!55!black] (2,2) circle (1.5pt);
    
    \node at (1.2,2.8) {\Large\color{orange}?};
    
    \node at (2,4.3) {$\ell_q\to \ell_p$};
    \node[rotate=90] at (-1,2) {non-negative};
    \node[rotate=90] at (-0.5,2) {$A:\C^n\to \C^m$};
\end{tikzpicture}
(a)
\end{minipage}
%
%
\begin{minipage}{0.32\textwidth}
\centering
\begin{tikzpicture}[scale=0.85, font=\small]

    \draw[->] (0,0) -- (4.3,0) node[right] {$q$};
    \draw[->] (0,0) -- (0,4.3) node[left] {$p$};
    \draw[-] (0,4) -- (4,4);
    \draw[-] (4,0) -- (4,4);

    \foreach \x in {0,2,4} {
        \draw[-] (\x,0) -- (\x,-0.1); 
    }
    \foreach \y in {0,2,4} {
        \draw[-] (0,\y) -- (-0.1,\y); 
    }
    
    \node[below] at (4,0) {$\infty$};
    \node[below] at (2,0) {$2$};
    \node[below] at (0,0) {$1$};
    \node[left]  at (0,4) {$\infty$};
    \node[left]  at (0,2) {$2$};
    \node[left]  at (0,0) {$1$};
    
    
    \fill[red!50] (0,0) rectangle (4,4);

    \draw[red!50,ultra thick] (0,0) -- (4.02,0);
    \draw[red!50,ultra thick] (4,-0.02) -- (4,4);
    
    \draw[\ReferenceColor,line width = 2.5pt] (0,0) -- (0,4.04);
    \draw[\ReferenceColor,line width = 2.5pt] (-0.04,4) -- (4,4);
    \node[below right] at (0,4) {\textcolor{\ReferenceColor}{Theorem \ref{intro.thm.1top.hardness}}\footnotesize};
    \draw[red!50,ultra thick] (-0.02,4) -- (4,4);
    \draw[red!50,ultra thick] (0,0) -- (0,4.03);

    \draw[dotted, thick] (0,0) -- (4,4);
    
    \fill[\ReferenceColor] (0,0) circle (2pt);
     \fill[\ReferenceColor] (4,4) circle (2pt);
    \fill[orange] (0,0) circle (1.5pt);
    \node[below right] at (0,0) {\textcolor{\ReferenceColor}{Thm. \ref{intro.thm.1to1.hardness}}\footnotesize} ;
    \fill[orange] (4,4) circle (1.5pt);
    \fill[green!50!black] (2,2) circle (1.5pt);
    
    \node at (2,4.3) { $\mathcal{S}_q \to \mathcal{S}_p$};
    \node[rotate=90] at (-0.7,2) {$\Phi : \C^{n \times n} \to \C^{m \times m}$};

\end{tikzpicture}
\begin{tikzpicture}[scale=0.85, font=\small]

    \draw[->] (0,0) -- (4.3,0) node[right] {$q$};
    \draw[->] (0,0) -- (0,4.3) node[left] {$p$};
    \draw[-] (0,4) -- (4,4);
    \draw[-] (4,0) -- (4,4);

    \foreach \x in {0,2,4} {
        \draw[-] (\x,0) -- (\x,-0.1); 
    }
    \foreach \y in {0,2,4} {
        \draw[-] (0,\y) -- (-0.1,\y); 
    }
    
    \node[below] at (4,0) {$\infty$};
    \node[below] at (2,0) {$2$};
    \node[below] at (0,0) {$1$};
    \node[left]  at (0,4) {$\infty$};
    \node[left]  at (0,2) {$2$};
    \node[left]  at (0,0) {$1$};

    \draw[green!55!black, thick] (0,0) -- (4,4);
    \draw[dotted, thick] (0,0) -- (4,4);

    \fill[green!55!black] (2,2) circle (1.5pt);

    \draw[\ReferenceColor,line width = 2.5pt] (0,0) -- (0,4.04);
    \draw[\ReferenceColor,line width = 2.5pt] (-0.04,4) -- (4,4);
    \node[below right] at (0,4) {\textcolor{\ReferenceColor}{Theorem \ref{intro.thm.1top.hardness}}\footnotesize};

    \draw[red!60,ultra thick] (0,0) -- (0,4.03);
    \draw[red!60,ultra thick] (-0.02,4) -- (4,4);

 \begin{scope}
\clip (4,4.04) rectangle (0.04,0.04);
\fill[\ReferenceColor] (4,4) circle (2pt);
    \end{scope}

    \draw[green!55!black,ultra thick] (-0.03,0) -- (4.03,0);
    \draw[green!55!black,ultra thick] (4,-0.03) -- (4,4.03);
    
    
    \fill[green!55!black] (0,0) -- (4,0) -- (4,4) -- cycle;
    \fill[green!55!black] (2,0-0.03) -- (4.03,-0.03) -- (4.03,2) -- (2,2) -- cycle;
    \draw[\ReferenceColor, dotted, thick] (2,0-0.03) -- (4.03,-0.03) -- (4.03,2) -- (2,2) -- cycle ;
\node[rotate= 45, above left] at (2.625,2.625) {\textcolor{green!55!black}{\cref{intro.lem.q.top.heuristic}}};
\node[above left] at (4,0) {\textcolor{\ReferenceColor}{\cref{intro:q.Boyds}}};
 \begin{scope}
\clip (-0.04,0) rectangle (0.04,1);
\fill[\ReferenceColor] (0,0) circle (2pt);
    \end{scope}
\fill[green!55!black] (0,0) circle (1.5pt);

\fill[green!55!black] (4,4) circle (1.5pt);
    
    \node at (1.2,2.8) {\Large\color{orange}?};
    
    \node at (2,4.3) { $\mathcal{S}_q\to \mathcal{S}_p$};
    \node[rotate=90] at (-0.7,2) {CP $\Phi:\C^{n\times n}\to \C^{m\times m}$};

\end{tikzpicture}
(b)
\end{minipage}
%
%
\begin{minipage}{0.32\textwidth}
\centering
\begin{tikzpicture}[scale=0.85, font=\small]

    \draw[->] (0,0) -- (4.3,0) node[right] {$q$};
    \draw[->] (0,0) -- (0,4.3) node[left] {$p$};
    \draw[-] (0,4) -- (4,4);
    \draw[-] (4,0) -- (4,4);

    \foreach \x in {0,2,4} {
        \draw[-] (\x,0) -- (\x,-0.1); 
    }
    \foreach \y in {0,2,4} {
        \draw[-] (0,\y) -- (-0.1,\y); 
    }
    
    \node[below] at (4,0) {$\infty$};
    \node[below] at (2,0) {$2$};
    \node[below] at (0,0) {$1$};
    \node[left]  at (0,4) {$\infty$};
    \node[left]  at (0,2) {$2$};
    \node[left]  at (0,0) {$1$};
    
    \draw[\ReferenceColor,line width = 2.5pt] (0,0) -- (0,4.04);
    \draw[\ReferenceColor,line width = 2.5pt] (-0.04,4) -- (4,4);
    \node[below right] at (0,4) {\textcolor{\ReferenceColor}{ \cref{intro.thm.efficient.cb.1top}}\footnotesize};
    
    \draw[green!55!black,ultra thick] (0,0) -- (0,4.03);
    \draw[green!55!black,ultra thick] (-0.03,4) -- (4,4);

    \draw[dotted, thick] (0,0) -- (4,4);

\node at (1.2,2.8) {\Large\color{orange}?};
\node at (2.8,1.2) {\Large\color{orange}?};
    
    \fill[green!55!black] (2,2) circle (1.5pt);
    \begin{scope}
\clip (-0.04,0) rectangle (0.04,1);
\fill[\ReferenceColor] (0,0) circle (2pt);
    \end{scope}

 \begin{scope}
\clip (3.96,4.04) rectangle (2,3);
\fill[\ReferenceColor] (4,4) circle (2pt);
    \end{scope}

\fill[green!55!black] (0,0) circle (1.5pt);
\fill[green!55!black] (4,4) circle (1.5pt);
    
    \node at (2,4.3) {\footnotesize{CB  $\cS_q \to \cS_p$ }};
    \node[rotate=90] at (-0.7,2) {$\Phi : \C^{n \times n} \to \C^{m \times m}$};

\end{tikzpicture}
\begin{tikzpicture}[scale=0.85, font=\small]
    \draw[->] (0,0) -- (4.3,0) node[right] {$q$};
    \draw[->] (0,0) -- (0,4.3) node[left] {$p$};
    \draw[-] (0,4) -- (4,4);
    \draw[-] (4,0) -- (4,4);

    \foreach \x in {0,2,4} {
        \draw[-] (\x,0) -- (\x,-0.1); 
    }
    \foreach \y in {0,2,4} {
        \draw[-] (0,\y) -- (-0.1,\y); 
    }   
    \node[below] at (4,0) {$\infty$};
    \node[below] at (2,0) {$2$};
    \node[below] at (0,0) {$1$};
    \node[left]  at (0,4) {$\infty$};
    \node[left]  at (0,2) {$2$};
    \node[left]  at (0,0) {$1$};
    \draw[green!55!black, thick] (0,0) -- (4,4);
    \draw[dotted, thick] (0,0) -- (4,4);
    \draw[\ReferenceColor,line width = 2.5pt] (0,0) -- (0,4.04);
    \draw[\ReferenceColor,line width = 2.5pt] (-0.04,4) -- (4,4);
    \node[below right] at (0,4) {\textcolor{\ReferenceColor}{Theorem \ref{intro.thm.efficient.cb.1top}}\footnotesize};
    \draw[green!55!black,ultra thick] (0,0) -- (0,4.02);
    \draw[green!55!black,ultra thick] (-0.03,4) -- (4,4);
    \draw[green!55!black,ultra thick] (-0.03,0) -- (4.03,0);
    \draw[green!55!black,ultra thick] (4,-0.03) -- (4,4.03);  

 \begin{scope}
\clip (4,4.04) rectangle (0.04,0.04);
\fill[\ReferenceColor] (4,4) circle (2pt);
    \end{scope}
    
    \fill[green!55!black] (0,0) -- (4,0) -- (4,4) -- cycle;
    \fill[green!55!black] (2,0-0.03) -- (4.03,-0.03) -- (4.03,2) -- (2,2) -- cycle;
    \draw[\ReferenceColor, dotted, thick] (2,0-0.03) -- (4.03,-0.03) -- (4.03,2) -- (2,2) -- cycle ;
\node[rotate= 45, above left] at (2.625,2.625) {\textcolor{green!55!black}{\cref{intro.lem.q.top.heuristic}}};
\node[above left] at (4,0) {\textcolor{\ReferenceColor}{\cref{intro:q.Boyds}}};
 \begin{scope}
\clip (-0.04,0) rectangle (0.04,1);
\fill[\ReferenceColor] (0,0) circle (2pt);
    \end{scope}
\fill[green!55!black] (0,0) circle (1.5pt);
\fill[green!55!black] (4,4) circle (1.5pt);    
    \node at (1.2,2.8) {\Large\color{orange}?};   
     \node at (2,4.3) {\footnotesize{CB  $\cS_q\to\cS_p$}};
    \node[rotate=90] at (-0.7,2) {CP $\Phi:\C^{n\times n}\to \C^{m\times m}$};
\end{tikzpicture}
(c)
\end{minipage}
    \caption{\footnotesize\raggedright(a) known complexity results on approximating mixed $q\to p$ matrix norms of any matrix on top and of a positivity preserving one on bottom, see \cref{sec:classicalComplexities} for details \cite{Bhattiprolu.2019,Bhattiprolu.2023, Bhaskara.2011, Barak.2012, Daureen.2007, Brandao.2015}. In (b) and (c) our new complexity results for computing, respectively, the mixed Schatten (\cref{sec:mixed.Schatten}), and mixed completely bounded Schatten norms (\cref{sec:mixed.cb.Schatten}) of a linear map $\Phi:\C^{n\times n}\to \C^{m\times m}$ on top and on bottom when the map is promised to be completely positive (CP). Included the references to our results in \cref{sec:summary.of.results}
    \textcolor{red!70}{Red} hereby signifies $\NP$-hardness and \textcolor{green!50!black}{green} the existence of an efficient algorithm.
    The white areas represent unknown complexities. The {\color{orange} orange} dot represents hardness for the positive version of the problem, i.e. the $\|\cdot\|^+_{1\to 1}$.} 
    \label{fig:Intro.figure}
\end{figure}

Of particular importance to quantum information processing are $1\to p$-norms for $p\in[1,\infty]$. 
Specifically, for $p=1$, the variant $\| \cdot \|^+_{1\to 1}$ of the $1\to 1$-norm where the optimization is restricted to positive semidefinite inputs has a clear operational interpretation due to the seminal Holevo-Helstrom theorem. For a difference of two quantum channels $\Phi,\Psi$, it governs the optimal success probability for  distinguishing between these two channels when no auxiliary system is allowed, via
\begin{align}
    \|\Phi-\Psi\|^+_{1\to 1}:=\sup_{\rho\geq 0}\frac{\|(\Phi-\Psi)(\rho)\|_1}{\|\rho\|_1}.\label{extabs:def1to1}
\end{align} 
The completely bounded version of this norm, denoted $cb,1\to1$, but better known as the diamond norm, similarly governs the distinguishability with access to an arbitrary auxiliary system and entanglement, via
\begin{align}
    \|\Phi-\Psi\|_\diamond=\|\Phi-\Psi\|_{cb,1\to 1}:=\sup_{n\in\mathbb{N}}\|\id_n\otimes(\Phi-\Psi)\|^+_{1\to 1}.
\end{align}

Interestingly, while it is a well known result of \cite{Watrous.2012} that the diamond norm of a difference of channels is efficiently computable, we show a hardness result for its non-cb analogue:
\begin{theorem}[$\NP$-completeness of $\|\cdot\|^+_{1\to 1}$, informal, \cref{thm:thm1to1}]\label{intro.thm.1to1.hardness}
Computing $\|\Phi-\Psi\|_{1\to 1}^+$ is $\NP$-complete for quantum channels $\Phi, \Psi$. In addition the channels may be assumed to be entanglement-breaking, see \eqref{equ:def.eb}.
\end{theorem} 
We are not aware of this result having being established, despite that fact that the clear operational significance of this quantity was already established in the last century.
Note that this hardness is in contrast to the mixed matrix setting, where $\|A\|^{(+)}_{\ell_1\to\ell_1}$ is efficiently computable for any matrix $A$, implying the analogs for classical-classical channels.

In quantum cryptography and information processing, $1\to p$-norms are related to minimal output entropies of quantum channels \cite{Beigi.2008}. Likewise, the $cb,1\to p$-norms correspond to the entanglement assisted versions \cite{Devetak.2006} of the latter, and the $1\to2$- and $1\to\infty$-norms have been studied in the context of product state testing and the analysis of certain quantum protocols among others \cite{Harrow.2013, Brandao.2015}. Concerning these we show
\begin{theorem}[$\NP$-completeness of CP $\|\cdot\|_{1\to p}$, $p>1$, informal, \cref{thm:1top.hardness}]\label{intro.thm.1top.hardness}
Computing $\|\Phi\|_{1\to p}$ for a completely positive map $\Phi$ is $\NP$-complete.
In addition $\Phi$ may be assumed entanglement-breaking, see \eqref{equ:def.eb}.    
\end{theorem} 
This is a strengthening of a result of~\cite{Harrow.2013} that showed $\NP$-hardness for general CPTP maps. Note that~\cite{Harrow.2013} also showed that computing $\|\Phi\|_{1\to p}$ within a constant factor is hard under the exponential time hypothesis.

We prove these results by reducing a decision problem formulation (\cref{def:decicion problem}) of the norm computations to the $\NP$-complete 2-Out-of-4-SAT problem. In some more details, the proof idea of  \cref{thm:1top.hardness} is to construct a quantum channel (CPTP map) that it outputs a pure state iff a given instance of the 2-Out-of-4-SAT problem is satisfiable. The construction of such a  channel is inspired by \cite{Beigi.2008} and contains a partial trace channel. A modified analysis of a suitable mixing of this partial trace component with the replacer channel with a maximally mixed state then leads to the entanglement-breaking result. A slight modification of this channel is also core to the proof of \cref{intro.thm.1to1.hardness}.
Hence effectively we are showing that given some CPTP map, deciding the existence of a pure output is an $\NP$-complete problem. For a more detailed sketch of the proof see \cref{sec:Proof.Sketches}.

General mixed Schatten norms, for $1\leq q,p\leq \infty$, also find  applications, for example in the study of open systems dynamics and thermalization of quantum systems via the concept of hypercontractivity \cite{Bardet.2022, Beigi.2015}. 
The mixed norms under consideration are those where $q\leq p$. Showing that a given quantum Markov semigroup $\{\Phi_t\}_{t\geq0}$ satisfies $\|\Phi_t\|_{q\to p}\leq1$, or $\|\Phi_t\|_{cb,q\to p}\leq1$ for some time $t>0$ and certain $q<p$ allows bounds on so-called (complete) modified logarithmic Sobolev constants, which constitute a powerful tool in the study of their mixing times \cite{Bardet.2024, Bardet.2022, Beigi.2015, Capel.2021}. Concerning these we show the following.

\begin{proposition}[$\NP$-hardness of mixed Schatten norms, informal, \cref{cor:generic.Schatten.hardness}]
Computing $\|\Psi\|_{q\to p}$ for any $1\leq q,p\leq\infty, (q,p)\neq (2,2)$, is $\NP$-hard.
\end{proposition}
Notably the case $1\to p$ here follows from the previous \cref{intro.thm.1top.hardness}, while the rest follows from a reduction to the already established classical setting. 

Next, we study an algorithm in the non-hypercontractive case.
\begin{lemma}[Efficient algorithm for CP $q\geq p$, informal, \cref{thm:ellipsoid.result}, \cref{lem:cb.equals.non.cb}]\label{intro.lem.q.top.heuristic}
For any CP map $\Phi$,
$\|\Phi\|_{cb,q\to p} = \|\Phi\|_{q\to p}$ can be computed efficiently.
\end{lemma}
We do this by showing that mixed norm computation for these parameters $q,p$ is a convex optimization problem and give an explicit algorithm with runtime through use of the ellipsoid method.

In addition, for the special case $1\leq p\leq 2\leq q\leq \infty$, we analyze a quantum version of Boyd's iterative power method \cite{Boyd.1974}. In particular, we revisit the pre-print \cite{Shahverdikondori.2022} in depth and give an explicit runtime, inspired by an analysis of \cite{Bhaskara.2011}. Concerning the result of \cite{Shahverdikondori.2022} we find a gap in their derivation which shows that their original claim is incorrect for any $p\leq q$ other than $1\leq p\leq 2\leq q\leq \infty$. We give a counterexample in \cref{lem:contraction.counterexample}. 
\begin{theorem}[Quantum Boyd's algorithm, informal, \cref{thm:Quantum.Boyds} and \cref{thm:strict.positive.q.Boyds}]\label{intro:q.Boyds}
For any CP map $\Phi:\mathbb{C}^{d\times d}\to \mathbb{C}^{d\times d}$ and $1\leq p\leq 2\leq q\leq\infty$ we can compute $\|\Phi\|_{q\to p}$ to relative error $1+\epsilon$ in polynomial time $\poly(d,\frac{1}{\epsilon})$.  
\end{theorem}

Explicit runtimes can be found in \cref{thm:Quantum.Boyds} and \cref{thm:strict.positive.q.Boyds}, while the proof is sketched in \cref{sec:q.Boyds}, and details are deferred to \cref{app:Boyds.analysis}. 
Perhaps surprisingly, and in contrast with  \cref{intro.thm.1top.hardness}, we show that the corresponding completely bounded versions can also be efficiently computed.
\begin{theorem}[Efficient algorithm for ${cb,1\to p}$, informal, \cref{thm:efficient.cb.1top}]\label{intro.thm.efficient.cb.1top} For any linear map $\Psi$ and any $p \in [1,\infty]$,
$\|\Psi\|_{cb,1\to p}$ can be efficiently computed.    
\end{theorem}
Recall that for $p=1$ and $\Psi=\Phi_1-\Phi_2$ a difference of quantum channels this is a seminal result of Watrous' \cite{Watrous.2012}, hence this theorem should be regarded as a generalization of it, both in terms of parameter regions and types of maps.
An explicit runtime can be found in \cref{thm:efficient.cb.1top}.
Note that from the definition of the cb-norm, computability is not obvious since the norm is defined as a supremum over reference systems of arbitrarily large dimension. However, for the case of $1 \to p$ norms, we generalize a connection between the above mixed cb norm and a particular Pisier (2-indexed Schatten) norm of its Choi operator to arbitrary linear maps (see \cite[Theorem 10]{Devetak.2006} for CP maps), and then establish concavity of a suitable function which leads to the above completely bounded mixed norms being given after non-trivial rewriting as convex optimization problems, which we deal with with the ellipsoid method, see \cref{sec:cb.1top.computability}. 

Our results are summarized in Fig.~\ref{fig:Intro.figure}.

\subsection{Sketch of the proof of \cref{thm:1top.hardness} and \cref{thm:thm1to1}}\label{sec:Proof.Sketches}

The hardness results in these theorems come from reduction to the 2-Out-of-4-SAT problem \cite{Khanna.2001} formulated as in \cite{Beigi.2008}.
\begin{definition*}[2-Out-of-4-SAT]\label{def:2OutOf4SAT}
    Given a $d-$dimensional Hilbert space $\mathcal{K}$ with ONB $\{|i\rangle\}_{i=1}^d$ and $m=\poly(d)$ vectors of the form \begin{align}
    |A_k\rangle=\sum_{i=1}^da_i^k|i\rangle \in \mathcal{K},
\end{align} where for each $k\in[m]$ we have $a_i^k\in\{0,\pm\frac{1}{2}\}$ s.t. exactly four $a_i^k$ are non-zero, decide whether there exists a vector of the form
\begin{align}
    |\tilde{\psi}\rangle = \frac{1}{\sqrt{d}}\sum_{i=1}^dx_i|i\rangle \in \mathcal{K},
    \label{def:VectorSpecialForm}
\end{align} with $x_i\in\{\pm 1\}$ for all $i\in[d]$, that is orthogonal to all $A_k$, i.e., $\langle A_k|\tilde{\psi}\rangle=0$ $\forall k\in[m]$.
\end{definition*}
A vector of this from \eqref{def:VectorSpecialForm} is sometimes also referred to as a \textit{proper state} in this context.
This decision problem is known to be $\NP$-complete \cite{Khanna.2001}.

In the following we will sketch the proof of \cref{thm:1top.hardness}. The proof of \cref{thm:thm1to1} is very analogous up to slight modification of the channels involved, see \cref{sec:proof.1to1.hardness}.
Fix some $p\in(1,\infty]$.
We will, in \cref{sec:2-out-of-4.setup}, show that deciding whether $\|\Phi\|_{1\to p} = c$, for some specific CPTP map $\Phi$ and some value $c$ reduces to deciding whether an instance of the 2-Out-of-4 SAT problem is satisfiable or not. Given such an an instance we will define four channels $\Phi_\text{trace}$ \eqref{equ:Phi.trace}, $\Phi_{\text{swap}}$\eqref{equ:Phi.swap}, $\Phi_{\text{cube}}$\eqref{equ:Phi.cube}, and $\Phi_{H}$ \eqref{equ:def:Phi.H}, where $H$ is a function of said instance and the last three of them entanglement-breaking \eqref{equ:def.eb}. They will be constructed in such a manner as to effectively measure the existence of a satisfying proper state $|\tilde{\psi}\rangle$. Specifically we will analyze $\Psi_H:=\Phi_\text{trace}\otimes\Phi_{\text{swap}}\otimes\Phi_{\text{cube}}\otimes\Phi_{H}: \mathcal{B}((\mathcal{K}\otimes\mathcal{K})^{\otimes 4})\to \mathcal{B}(\mathcal{K}\otimes(\mathbb{C}^2)^{\otimes3})$ and the problem of deciding $\|\Psi_H\circ\Lambda_4\|_{1\to p}$, where $\Lambda_4:\mathcal{B}((\mathcal{K}\otimes\mathcal{K})^{\otimes 4})\to \mathcal{B}((\mathcal{K}\otimes\mathcal{K})^{\otimes 4})$, \eqref{equ:def.Lambda} is a channel s.t. the 4 marginals on each copy of $(\mathcal{K}\otimes\mathcal{K})$ is equal to every other.  
First of all, if the 2-Out-of-4 SAT problem is satisfiable, then there exists a pure proper state $|\tilde{\psi}\rangle\langle\tilde{\psi}|\in\mathcal{B}(\mathcal{K})$ of form \eqref{def:VectorSpecialForm}, such that $\|\Psi_H\circ \Lambda_4((\tilde{\psi}\otimes\tilde{\psi})^{\otimes 4})\|_p=\|\Psi_H(\tilde{\psi}\otimes\tilde{\psi})^{\otimes 4})\|_p=1$ is optimal and hence $\|\Psi_H\circ\Lambda_4\|_{1\to p}=1$.
On the other hand, if $\|\Psi_H\circ\Lambda_4\|_{1\to p}=1$, then there exists a state $\tau$, such that $\|\Psi_H\circ\Lambda_4(\tau)\|_{p}=1$, i.e. $\Psi_H\circ\Lambda_4(\tau)$ is pure.
In Lemma \ref{lemma:MainCorrect} it will follow that if $\Psi_H$ outputs a pure state, i.e. one maximizing the $p$-norm, then we may w.l.o.g. assume the input state to be a product state, due to its entanglement-breaking nature. Thus it will follows that a maximal $p$-norm, whcih is equivalent to a pure state output, implies
\begin{align}
    \Psi_H(\Lambda_4(\tau))= \Psi_H(\tilde{\tau}^{\otimes 4})=  \Phi_\text{trace}(\tilde{\tau})\otimes\Phi_{\text{swap}}(\tilde{\tau})\otimes\Phi_{\text{cube}}(\tilde{\tau})\otimes\Phi_{H}(\tilde{\tau}),
\end{align} where we denote with $\tilde{\tau}$ the equal local marginals of $\Lambda_4(\tau)$.
So we found a state $\tilde{\tau}\in\mathcal{D}(\mathcal{H})$ that is optimal w.r.t all $\Phi_\text{trace},\Phi_\text{swap},\Phi_\text{cube},\Phi_H$, which by their construction implies that it has to be of the form $\tilde{\psi}\otimes\tilde{\psi}$, where the pure state $\tilde{\psi}$ is a proper state of the form \eqref{def:VectorSpecialForm} and satisfies the 2-Out-of-4 SAT problem. The idea is that the optimizers of $\Phi_\text{trace}$ are product states with one pure marginal, the ones of $\Phi_\text{swap}$ further have to be pure product states of the from $\psi\otimes\psi$, and by $\Phi_\text{cube}$ these optimal pure states have to be of form \eqref{def:VectorSpecialForm} and if $\Phi_H$ achieves its optimum on such states, then we have found that there exists a vector $\tilde{\psi}$ of form \eqref{def:VectorSpecialForm} that is orthogonal to all $|A_k\rangle$, i.e. satisfies this SAT problem.
Hence we will have shown that $\|\Psi_H\circ\Lambda_4\|_{1\to p}= 1$ is equivalent to being in the ``Yes'' instance of the 2-Out-of-4-SAT problem. If the 2-Out-of-4-SAT instance is not satisfiable, then the idea is that there is at least one constraint that is not satisfied, which will imply that the corresponding $\|\Psi_H\circ\Lambda_4\|_{1\to p} < 1-\frac{1}{\poly(d)}$. 
A slight modification generalized this proof to the entanglement breaking and difference of quantum channels. 
The omitted details and the full proof may be found in \cref{sec:2-out-of-4.setup}.

\subsection{Literature on the complexity of mixed matrix norms}\label{sec:classicalComplexities}
The computational complexity of mixed matrix norms has received a great deal of attention in the past 20 years. Here, we adumbrate the known complexities of computing mixed matrix norms. By this we mean the problem of computing $\|A\|_{\ell_q\to \ell_p}$ for some $A\in\C^{m\times n}$ up to at least inverse polynomial relative factor, i.e., factor $(1+\epsilon)$, where $\epsilon=\frac{1}{\poly(m,n)}$.\footnote{Though all presented complexities also hold, when requiring arbitrary small constant factor, i.e. up to $(1+\epsilon)$ for any prior fixed $\epsilon>0$.} 

More detailed results about hardness of constant-, or sublinear factor and easiness of constant-, or polynomial/rational-factor approximations do exist under reasonable complexity assumptions. We refer to \cite{Bhattiprolu.2019,Bhattiprolu.2023, Bhaskara.2011, Barak.2012, Daureen.2007, Brandao.2015, Hendrickx.2010} for more details.\footnote{Specifically see
\cite[Chapter 2]{Daureen.2007}, \cite[Theorems 1.1, 1.2, 1.3]{Bhattiprolu.2019}, \cite[Theorems 2.4, 2.5]{Brandao.2015}, see also \cite[Theorem 1.3 and Section 4.2]{Bhattiprolu.2023}, \cite{Barak.2012}, and \cite[Theorems 6.6, 6.8 and Proposition 6.1]{Bhaskara.2011}.}

There exist efficient algorithms (in the dimensions $n,m$) that for any matrix $A$ compute $\|A\|_{\ell_1\to \ell_p}$ for any $1\leq p\leq\infty$, and $\|A\|_{\ell_q\to \ell_\infty}$ for any $1\leq q\leq\infty$\footnote{Note that these are just maximizations over, respectively, column sums achieved on standard basis vectors or row sums achieved on standard basis colums vectors.}, see e.g., \cite{Daureen.2007}, as well as $\|A\|_{\ell_2\to \ell_2}$.~\footnote{Note that $\|A\|_{\ell_2\to \ell_2}=\|A\|_\infty$ is equal to the largest singular values of $A$, which is efficiently computable.} 
On the other hand, approximating the $q\to p$ norm for any other values of $q,p$ is known to be $\NP$-hard. 
Computing hypercontractive norms, i.e. $\|A\|_{\ell_q\to \ell_p}$ for $1<q<p<\infty$ to any constant factor precision is hard under reasonable complexity assumptions, such as the exponential time hypothesis (ETH)\footnote{The assumption of ETH is only required to show hardness of the hypercontractive norms with $2\in[q,p]$
} 
\cite{Barak.2012, Bhattiprolu.2019, Bhattiprolu.2023, Brandao.2015, Daureen.2007, Bhaskara.2011}. These hypercontractive mixed matrix norms are of particular interest due to connections with the analysis of random walks, set expanders \cite{Biswal.2011,Barak.2012}, gadget reduction \cite{Bhattiprolu.2019, Biswal.2011}, the study of Markov processes through log-Sobolev inequalities \cite{Gautier.2020} and even analysis of certain quantum protocols \cite{Barak.2012,Brandao.2015}.  Computing the non-hypercontractive norms of a matrix to arbitrary relative precision are also $\NP$-hard problems \cite{Daureen.2007, Hendrickx.2010, Bhaskara.2011}.

If the matrix under consideration has only non-negative entries the problems of computing the mixed matrix norms are in principle simplified. The non-hypercontractive setting now becomes efficiently computable \cite{Bhaskara.2011, Boyd.1974}, via e.g. Boyd's power iterations method. Surprisingly, in the hypercontractive setting it is unknown if this is also the case. 
See \cref{fig:Intro.figure} column $a)$ for a simplified graphical representation of these results.

In contrast to these, the computability of their "quantum"-extensions had received little systematic attention in the literature. Some individual results had been known, the ones mentioned above in \cref{sec:summary.of.results}. 

\subsection{Structure of the paper}

In the following we first set the notations adopted in this paper and define the considered norms, in particular the Schatten and completely bounded norms. 
In \cref{sec:mixed.Schatten}, we present our results on the computational complexities of mixed Schatten norms. We first state and prove our $\NP$-completeness results before introducing an ellipsoid method- and a quantum Boyd's algorithm and proving its efficiency in the non-hypercontractive regime. The derivation of the latter and detailed analysis can be found in \cref{app:Boyds.analysis}.
In \cref{sec:mixed.cb.Schatten}, we present our results on mixed completely bounded Schatten norms.
Lastly, in \cref{sec:outlook}, we present some remaining open questions. For a visual summary of our results, see \cref{fig:Intro.figure}.


\subsection{Notation}\label{sec:notation}
We denote with $[n]$ the indexing set $\{1,...,n\}\subset\N$.
Hilbert spaces, assumed finite dimensional in this work, are denoted either with $\cH,\cK$ or $\C^n$ for some $n\in\N$. The algebra of bounded operators $X:\cH\to\cK$ 
with the operator norm $\|\cdot\|_\infty$ is denoted as $\cB(\cH,\cK)$. We simplify $\cB(\cH,\cH)\equiv \cB(\cH)$. We will use the terms matrices and operators interchangeably and typically denote them with upper case Latin letters, e.g., $A,B,X,Y$. We denote the identity operator in $\cB(\cH)$ with $\1$. The canonical trace is denoted as $\Tr[\cdot]$. It induces the Schatten-$p$ norms, see below, and the so-called Hilbert-Schmidt inner product $\langle A,B\rangle:=\Tr[A^*B]$, where $A^*$ is the adjoint of the matrix/operator $A$ w.r.t this inner product.
We denote the partial trace $\tr_2[\cdot]:\cB(\cH_1\otimes\cH_2)\to \cB(\cH_1)$. The reduced state $\tr_2[\rho]$ is at times denoted by $\rho_1$.
An element $X \in\cB(\cH)$ is (spectrally) \textit{positive (semi-definite)} 
 if and only if (iff) it satisfies $X=Y^*Y$ for some $Y\in\cB(\cH)$. 
We write $X \geq 0$ and $X\in\mathcal{B}^+(\cH)$ in this case. 
A \textit{(quantum) state} on $\cH$ is a positive operator with trace 1, i.e., $\rho\geq 0$, such that $\Tr[\rho]=1$. The set of quantum states over $\cH$ is denoted with $\cD(\cH):=\{\rho\in\cB(\cH)|\rho\geq 0, \Tr[\rho]=1\}$. When a state is rank-1, we call it a \textit{pure} quantum state and it can be written as $\rho=|\psi\rangle\langle\psi|\equiv \psi$, where $\ket{\psi} \in\cH$. Pure quantum states will in abuse of notation sometimes be denoted as $\psi$ or $|\psi\rangle$. Quantum states are usually denoted with lower case Greek letters $\rho,\sigma,\omega,\tau$.
The von Neumann entropy of a state $\rho$ is denoted by $S(\rho):=-\Tr(\rho\log\rho)$.

We denote the identity map $\cB(\cH)\to \cB(\cH)$ with $\id$, or sometimes with $\id_d$, when $\cH\simeq\C^d$.
A linear map $\Phi:\cB(\cH) \to \cB(\cK)$ is called \textit{positive} (or positivity preserving) if $\Phi(X)\geq0$ for all $X\geq 0$. They are further called \textit{completely positive (CP)} if $\id_{n}\otimes\Phi \in \cB(\C^n\otimes\cH,\C^n\otimes\cK)$ is a positive map $\forall n\in\N$. The dual of a linear map $\Phi$ with respect to the Hilbert-Schmidt inner product $\langle .,.\rangle$ is denoted by $\Phi^*$. A linear map $\Phi$ is called \textit{trace preserving (TP)} if $\Tr[\Phi(\rho)]=\Tr[\rho]$ holds $\forall \rho$ and \textit{unital} if $\Phi(\1)=\1$.
A \textit{quantum channel} is a linear CPTP map. It is further called \textit{entanglement-breaking} or \textit{measure and prepare} if it is of the form
\begin{equation}
    \label{equ:def.eb} \Phi(\rho):=\sum_{i}\Tr[M_i\rho]\,\sigma_i
\end{equation} where $\{M_i\}_i$ form a positive operator valued measure (POVM), i.e., $M_i\geq 0\  \forall i$ and $\sum_iM_i=\1$, and $\{\sigma_i\}_i$ a set of quantum states. Equivalently entanglement-breaking channels are those who's outputs are separable w.r.t some environment with which the input was possibly entangled with.
If the states $\sigma_i$ are pure states forming some distinguished basis and the measurements $M_j$ are orthogonal rank-1 projections $|j\rangle\langle j|$, then we call such a map further \textit{classical-classical} (c-c) map. If only the sates are, then it is a \textit{quantum-classical} (q-c) map and if only the measurements are a \textit{classical-quantum} (c-q)  map. 

Given a linear map $\Phi:\cB(\cH)\to \cB(\cK)$, its \textit{Choi operator} is given by
\begin{align}
    J_\Phi:=\sum_{ij}|i\rangle\langle j|\otimes\Phi(|i\rangle\langle j|) \in \cB(\cH \otimes \cK),
\end{align} which is a spectrally positive operator iff $\Phi$ is CP. Here $\{|i\rangle\}_i$ is an orthonormal basis (ONB) of $\cH$. 

\medskip 
In this work we consider mixed norms of linear maps between finite dimensional matrix spaces. These spaces will generically be $\cB(\C^n)\equiv\C^{n\times n}$ for some $n\in\N$, though the results equally hold with $\R$ instead of $\C$, unless explicitly specified otherwise.

Given an $n$-dimensional vector $v\in\C^n$, its $\ell_p$ norm, for $1\leq p\le \infty$, is defined as
\begin{align}
    \|v\|_p:=\bigg(\sum_{i=1}^n|v_n|^p\bigg)^\frac{1}{p}, \quad \|v\|_\infty:=\max_i|v_i|.
\end{align}
This norm naturally induces a mixed norm on linear maps $A:\C^n\to\C^m$, namely
\begin{align}
    \|A\|_{q\to p}:=\max_{v\in\C^n}\frac{\|Av\|_p}{\|v\|_q}.
\end{align}
These norms are called \textit{mixed matrix norms}. 
We will consider the problem of approximating this value to inverse polynomial (in the dimension) relative error on arbitrary matrices $A\in\C^{n\times m}$ and call it the problem of ``approximating'' or ``computing'' the mixed matrix norm. 

For matrices $X\in\C^{n\times n}$, the natural analog of the $\ell_p$ norm is the so-called \textit{Schatten-p} norm and the Banach space $\C^{n\times n}$ endowed with this norm is denoted with $\cS_p\equiv\cS_p(\C^n)$. The norm is defined as
\begin{align}
    \|X\|_p:=\Tr\Big[|X|^p\Big]^\frac{1}{p},
\end{align} where $|X|:=\sqrt{X^*X}$ and $\Tr[\cdot]$ is the canonical trace on $\C^{n\times n}$. Likewise, it induces a notion of mixed norm for linear maps $\Phi:\C^{n\times n}\to\C^{m\times m}$ between matrix spaces.
\begin{align}
    \|\Phi\|_{q\to p}:=\max_{X\in\C^{n\times n}}\frac{\|\Phi(X)\|_p}{\|X\|_q}, \quad \|\Phi\|^+_{q\to p}:=\max_{X\ge 0}\frac{\|\Phi(X)\|_p}{\|X\|_q},
\end{align} 
where the second optimization is restricted to only (spectrally) positive matrices $X\geq 0$. 
We note here that the notions of \textit{(spectral) positivity} and \textit{positivity preserving}, i.e. positivity of entries, are very different and independent properties of matrices.
We will equally consider the complexities of computing these mixed Schatten norms up to inverse polynomial precision, either for any linear map $\Phi$ or under the promise that the map is completely positive.

The last class of mixed norms we will consider is that of \textit{completely bounded} (cb) norms. 
In order to define these, we need to introduce the notion of a \textit{2-indexed Schatten norm}, also called Pisier norm \cite{Book.Pisier.1998, Devetak.2006, Beigi.2023}.
Given an element $X\in \cB(\cH\otimes\cK)$ we take the following as the definition of its 2-indexed $(q,p)-$norm:
\begin{align}
\label{equ:def-multi-index}
\|X\|_{(q,p)}
:=
    \begin{cases}
        \inf_{\underset{X=(A\otimes\1_{\cK})Y(B\otimes\1_{\cK})}{A,B\in \mathcal{B}(\mathcal{H}),Y\in \mathcal{B}(\mathcal{H}\otimes\cK)}} \|A\|_{2r}\|B\|_{2r}\|Y\|_{p}\,, & \quad \textup{for } q< p \\
        \|X\|_p\,, & \quad \textup{for } q=p \\
        \sup_{A,B\in\mathcal{B}(\mathcal{H})} \|A\|^{-1}_{2r}\|B\|^{-1}_{2r}\|(A\otimes\1_{\cK})X(B\otimes\1_{\cK})\|_p, & \quad \textup{for } q> p
    \end{cases}
\end{align}
where the infimum and supremum are over $A,B\in \mathcal{B}(\mathcal{H})$ acting only on the first Hilbert space and $Y\in \mathcal{B}(\mathcal{H}\otimes\cK)$ with $\frac{1}{r}:=\left|\frac{1}{q}-\frac{1}{p}\right|$. 
\newline 
Given these the \textit{completely bounded} $q\to p$ norm of the linear map $\Phi:\cB(\cH)\to\cB(\cK)$ is defined as
\begin{align}
    \|\Phi\|_{cb,q\to p}:=\sup_{n\in\mathbb{N}}\|\id_n\otimes\Phi\|_{(\infty,q)\to (\infty,p)}&=\sup_{n\in\mathbb{N}}\|\id_n\otimes\Phi\|_{(t,q)\to (t,p)}\\
    &= \sup_{n\in\N}\sup_{X\in\cB(\C^n\otimes\cH)}\frac{\|(\id_n\otimes\Phi)(X)\|_{(t,p)}}{\|X\|_{(t,q)}},
\end{align} for any $1\leq t\leq \infty$.

We remark here that there exists a similar, yet  different notion of ``entanglement assisted'' mixed norms, namely the \textit{stabilized} $q\to p$-norms \cite{Watrous.2004}, defined as
\begin{align}\label{def:stabilized.mixed.norm}
    \|\Phi\|_{st,q\to p}:=\sup_{n}\|\id_n\otimes\Phi\|_{q\to p}= \sup_{n}\|\id_n\otimes\Phi\|_{(q,q)\to (p,p)}.
\end{align} They are in general, e.g. when $q\neq p$, different to their completely bounded counterparts.

\subsection{Visualization}
A useful way to visualize these mixed $q\to p$ norms is in a 2-axis diagram, that has the input index $q\in[1,\infty]$ as the horizontal axis and the output norm index $p\in[1,\infty]$ as the vertical axis, see \cref{fig:Nomenclature}. 
The region $1\leq p\leq q\leq \infty$, including the diagonal, is sometimes called the \textit{non-hypercontractive region} whereas the complement region $1<q<p<\infty$ is called the \textit{hypercontractive region}. 
\begin{figure}[h]
\begin{minipage}{0.45\textwidth}
    \centering
    \begin{tikzpicture}[scale=1, font=\small]

    \draw[->] (0,0) -- (4.3,0) node[right] {$q$};
    \draw[->] (0,0) -- (0,4.3) node[left] {$p$};
    \draw[-] (0,4) -- (4,4);
    \draw[-] (4,0) -- (4,4);

    \foreach \x in {0,2,4} {
        \draw[-] (\x,0) -- (\x,-0.1); 
    }
    \foreach \y in {0,2,4} {
        \draw[-] (0,\y) -- (-0.1,\y); 
    }
    
    \node[below] at (4,0) {$\infty$};
    \node[below] at (2,0) {$2$};
    \node[below] at (0,0) {$1$};
    \node[left]  at (0,4) {$\infty$};
    \node[left]  at (0,2) {$2$};
    \node[left]  at (0,0) {$1$};
    
    \fill[purple, opacity=0.55] (0,0) -- (0,4) -- (4,4) -- cycle;
    \fill[orange, opacity=0.55] (0,0) -- (4,0) -- (4,4) -- cycle;

    \draw[orange!55, thick] (0,0) -- (4,4);
    \draw[orange,  opacity=0.55] (0,0) -- (4,0);
    \draw[orange,  opacity=0.55] (4,0) -- (4,4);
    \draw[black, dotted, thick] (0,0) -- (4,4);

    \node at (2,4.7) {\color{purple} hypercontractive};
    \node at (2,4.2) {\color{purple} region};
    \node at (5.3,2.5) {\color{orange}non-};
    \node at (5.3,2) {\color{orange}hypercontractive};
    \node at (5.3,1.5) {\color{orange}region};

    \draw[black, dotted, thick] (0,4) -- (4,0);
    
    \fill[black] (2,2) circle (1.5pt);
    \node[left] at (2,2) {\footnotesize{$(2\to 2)$}};
    \fill[black] (2,3) circle (1.5pt);
    \node[left] at (2,3) {\footnotesize{$(2\to 4)$}};
    \fill[black] (4,4) circle (1.5pt);
    \node[below left] at (4,4) {\footnotesize{$(\infty\to \infty)$}};
    \fill[black] (1.5,0.5) circle (1.5pt);
    \node[left] at (1.5,0.5) {\footnotesize{$(q\to p)$}};
    \fill[black] (3.5,2.5) circle (1.5pt);
    \node[left] at (3.5,2.5) {\footnotesize{$(p^\prime\to q^\prime)$}};
    \draw[<->, thick] (1.6,0.5) .. controls (3.2,0.5) and (3.5,0.8) .. (3.5,2.4);
    \node at (2.5,1.4) {Hölder};
    \node at (2.5,1) {duality};

\end{tikzpicture}
\end{minipage}
\begin{minipage}{0.54\textwidth}
    \caption{A diagram visualizing mixed $q\to p$ norms, with $1\leq q\leq\infty$ on the horizontal and $1\leq p\leq \infty$ on the vertical axese. The region $\infty>p>q>1$, (the upper left triangle without borders) is called the \textcolor{purple}{\textit{hypercontractive} region}. Its complement is the \textcolor{orange}{\textit{non-hypercontractive} region}, i.e., the lower right triangle with borders. By Hölder duality computing the $q\to p$ norm of some linear map is equivalent to computing the $q^\prime \to p^\prime$ of its adjoint map. 
    }
     \label{fig:Nomenclature}
\end{minipage}
\end{figure}

The next result is a direct consequence of Hölder's duality.
\begin{lemma}[Hölder duality simplification]
If $1\leq q,q^\prime$ and $1\leq p,p^\prime$ are each dual indices, i.e. $q^{-1}+q^{\prime-1}=p^{-1}+p^{\prime-1}=1$, then the following equalities hold:
\begin{align}
    \|A\|_{\ell_q\to \ell_p}&=\|A^*\|_{\ell_{p^\prime}\to \ell_{q^\prime}}, \\
    \|\Phi\|_{q\to p} &= \|\Phi^*\|_{{p^\prime}\to {q^\prime}}, \\
    \|\Phi\|_{cb, q\to p} &= \|\Phi^*\|_{cb,{p^\prime}\to {q^\prime}},
\end{align} where $A^*$ is the adjoint matrix of $A$ w.r.t. the canonical inner product on $\C^n$ and $\Phi^*$ the adjoint map of $\Phi$, w.r.t. the Hilbert-Schmidt inner product. 
\end{lemma}
\begin{proof}
This follows directly from Hölder duality of the underlying norms. For the 2-indexed Schatten norms see e.g. \cite{Book.Pisier.1998, Beigi.2023}. 
\end{proof}

Further, if $A$ is positivity preserving (i.e., has only non-negative entries), then so is $A^*$ and hence the computational complexity of $\| \cdot \|_{\ell_q\to \ell_p}$ and its Hölder related $\|\cdot \|_{\ell_{p^\prime}\to \ell_{p^\prime}}$ are equivalent. 
The same applies to $\Phi$, i.e. $\Phi$ is CP iff $\Phi^*$ is CP.
Note that this change of indices leaves the hypercontractive and non-hypercontractive regions invariant, as can be seen in \cref{fig:Nomenclature}.

\section{Mixed Schatten norms}\label{sec:mixed.Schatten}
In the following, we present our main results on the computational complexities of mixed norms of linear operators between matrix spaces, i.e. $\Phi:\cS_q(\C^n)\to \cS_p(\C^m)$. 
We consider the problem of computing or approximating this quantity, for some fixed $q,p$ in the following sense: we either find an algorithm that outputs $\|\Phi\|_{q\to p}$ up to relative error $\epsilon$ in polynomial (in $n,m,1/\epsilon$) time, or prove that it is $\NP$-complete to decide whether it is above some $c>0$ or below some $s>0$ where $c-s=\frac{1}{\poly(n,m)}>0$. 
The latter would imply that approximating it up to additive inverse polynomial precision in polynomial time is $\NP$-hard. Since we show this for (normalized) quantum channels, this will directly also  $\NP$-hardness of approximation to arbitrary small relative precision.
To simplify notations, we hence model the computability problem as the following decision problem.
\begin{definition}\label{def:decicion problem} $(q \to p)$-\textsc{Schatten norm}: \hfill \\
    \textbf{Input: } The input and output dimensions $n,m$, the linear map $\Phi:\C^{n\times n}\to \C^{m\times m}$ given by polynomially many bits, and two numbers $c>s>0$ such that $c-s=\frac{1}{\poly(n,m)}$ for some fixed polynomial. \\
    \textbf{Promise: } $\|\Phi\|_{q\to p}\geq c$ or $\|\Phi\|_{q\to p}<s$. Optionally whether $\Phi$ is CP.\\
    \textbf{Output: } Yes iff $\|\Phi\|_{q\to p}\geq c$.
\end{definition}
We will also write $(q^+ \to p)$-\textsc{Schatten norm}, when referring to the above problem with $\|\Phi\|^+_{q\to p}$ instead of $\|\Phi\|_{q\to p}$.

In particular, we first prove that mixed Schatten norms are at least as hard to compute as mixed matrix norms. This follows via the simple observation that when restricting to diagonal matrices the mixed Schatten norm is just a mixed matrix norm. 

\begin{proposition}\label{prop:classical.quantum.embedding}
Computing or approximating quantum $q\to p$-Schatten norms $\|\Phi\|_{q\to p}$ is at least as hard as computing mixed norms $\|A\|_{\ell_q\to \ell_p}$. This holds equally under the restriction of positivity preservation. 
\end{proposition}
This directly lifts all $\NP$-hardness results form the matrix case to the Schatten case, we have
\begin{corollary}[generic $\NP$-hardness of mixed Schatten norms]\label{cor:generic.Schatten.hardness}
Given $1<q\leq \infty, 1\leq p<\infty, (p,q)\neq (2,2)$ then it is $\NP$-hard to decide $(q\to p)-$\textsc{Schatten Norm} for arbitrary maps $\Phi$. \\
In other words, computing $\|\Phi\|_{q\to p}$ up to (sufficiently small) constant factor precision is $\NP$-hard for all $1<q\leq \infty, 1\leq p<\infty, (p,q)\neq (2,2)$.
This directly implies that it is $\NP$-hard to approximate them to additive inverse polynomial precision for these values of $q,p$. 
\end{corollary}

\begin{remark}
    For CP maps this result does not give us anything new, since for positivity preserving matrices no hardness results are known. In this work we will prove hardness results even for CP maps complementing the above ones.
\end{remark}

The proof of \cref{prop:classical.quantum.embedding} consists in embedding the classical $\ell_q\to \ell_p$ problem into the mixed Schatten norm $\mathcal{S}_q\to \mathcal{S}_p$ problem and checking that non-diagonal inputs do not contribute. It may be found in \cref{app:classicalembedding}.

The second simple observation is the well-known fact that computing the $2\to 2$ norm is always efficiently computable for any linear map $\Phi$, since $\cS_2(\C^n)$ is a Hilbert space, with the Hilbert-Schmidt inner product. Via the Hilbert-Schmidt isomorphism $\mathcal{I}:\cS_2(\cH) \to \cH\otimes\cH, |\varphi\rangle\langle\psi|\mapsto |\varphi\rangle\otimes|\psi\rangle$ this becomes explicit and we have
\begin{align}
    \|\Phi\|_{2\to 2}= \|\mathcal{I}\circ \Phi\circ\mathcal{I}^{-1}\|_{\ell_2\to \ell_2} = \|\mathcal{I}\circ \Phi\circ\mathcal{I}^{-1}\|_\infty = \|J_\Phi^\Gamma\|_\infty,
\end{align} where $\mathcal{I}\circ \Phi\circ\mathcal{I}^{-1} = J_\Phi^\Gamma$ is the on the second subsystem partially transposed Choi operator of $\Phi$. 

\subsection{$\NP$-completeness of $1\to p$-Schatten norms, $p>1$}

We now consider the $1 \to p$ Schatten norms and show hardness, even for entanglement-breaking CPTP maps.
\begin{theorem}[$\NP$ completeness of CP $1\to p$]
\label{thm:1top.hardness} Fix $1<p\leq\infty$. The problem $(1\to p)-$\textsc{Schatten Norm} is $\NP$-hard for entanglement-breaking CPTP maps.
In addition, deciding whether $\|\Phi\|_{1\to p}=1$ or $\leq 1-\frac{1}{\poly(n,m)}$ is an $\NP$-complete problem for CPTP $\Phi:\mathbb{C}^{n\times n}\to\mathbb{C}^{m\times m}$.
\end{theorem}
The complexity of $1 \to p$ norm of CP maps was previously considered in \cite{Harrow.2013,Brandao.2015}. In particular, the authors of \cite{Harrow.2013} showed that computing $\|\Phi\|_{1\to p}$ for $p \in \{2, \infty\}$ is equivalent to the problem of linear optimization over the set of separable states, and is hence $\NP$-hard to compute within an additive error of $\frac{1}{\poly(n,m)}$. Our result shows that hardness remains for entanglement-breaking channels. Note that for entanglement-breaking channels, \cite{Brandao.2015} designed an algorithm to compute $\| \Phi \|_{1 \to p}$ within an additive error $\epsilon$ in time polynomial in the dimension but exponential in $1/\epsilon$. Our result shows that, unless $\mathsf{P} = \mathsf{NP}$, this exponential dependence is necessary. The necessity of this exponential dependence on $\epsilon$ was actually already known from the results of \cite{Barak.2012, Brandao.2015}, but under the stronger exponential hypothesis (ETH). In quantum information theory, the quantity $\|\Phi\|_{1\to p}$ for $p>1$ is interpreted (up to a logarithm) as the minimum output $p$-R\'enyi entropy of the channel $\Phi$ which plays an important role in quantum Shannon theory \cite{shor2004equivalence}.

Our proof is inspired by the reduction of \cite{Beigi.2008}: we will map an instance of the 2-Out-of-4-SAT problem to an entanglement-breaking channel $\Phi$ such that if the formula is satisfiable $\|\Phi\|_{1\to p}>c$ and otherwise $\|\Phi\|_{1\to p}<c-\frac{1}{\poly(n)}$. 

\subsubsection{Proof of \cref{thm:1top.hardness}}\label{sec:2-out-of-4.setup}
We may restrict ourselves to optimization over quantum states, since for completely positive (CP) maps $\Lambda$, it is known that for any $1\leq p,q\leq\infty$, $\|\Lambda\|_{q\to p}=\sup_{X, \|X\|_q=1}\|\Lambda(X)\|_p=\sup_{X\geq 0, \|X\|_q=1}\|\Lambda(X)\|_p =\|\Lambda\|^+_{q\to p}$, see e.g., \cite[Theorem 1]{Watrous.2004} or \cite[Corollary 6]{Devetak.2006}. \\
We also recall, that given a quantum state $\rho$ and a $p>1$, it holds that $\|\rho\|_p\leq \|\rho\|_1=1$ with equality iff $\rho$ is pure.

We will reduce to the so-called 2-Out-of-4-SAT problem \cite{Khanna.2001} formulated as in \cite{Beigi.2008}, see \cref{def:2OutOf4SAT}, repeated here for convenience. 
\begin{definition*}[2-Out-of-4-SAT]
    Given a $d-$dimensional Hilbert space $\mathcal{K}$ with ONB $\{|i\rangle\}_{i=1}^d$ and $m=\poly(d)$ vectors of the form \begin{align}
    |A_k\rangle=\sum_{i=1}^da_i^k|i\rangle \in \mathcal{K},
\end{align} where for each $k\in[m]$ we have $a_i^k\in\{0,\pm\frac{1}{2}\}$ s.t. exactly four $a_i^k$ are non-zero, decide whether there exists a vector of the form
\begin{align}
    |\tilde{\psi}\rangle = \frac{1}{\sqrt{d}}\sum_{i=1}^dx_i|i\rangle \in \mathcal{K},
\end{align} with $x_i\in\{\pm 1\}$ for all $i\in[d]$, that is orthogonal to all $A_k$, i.e., $\langle A_k|\tilde{\psi}\rangle=0$ $\forall k\in[m]$.
\end{definition*}
This decision problem is known to be $\NP$-complete \cite{Khanna.2001}. A vector of this from \eqref{def:VectorSpecialForm} is sometimes also referred to as a \textit{proper state} in this context.

As in \cite{Beigi.2008} given an instance of this problem, that is the tuple $(d,m,\{|A_k\rangle\}_{k=1}^m)$ we define the associated ``Hamiltonian'' as
\begin{align}\label{equ:Hamiltonian}
    H:=\frac{1}{m}\sum_{i=1}^m |A_k\rangle\langle A_k|\otimes |A_k\rangle\langle A_k| \in \mathcal{B}(\mathcal{K}\otimes \mathcal{K})\equiv \mathcal{B}(\mathcal{H}),
\end{align} 
where we set $\mathcal{H}:=\mathcal{K}_1\otimes\mathcal{K}_2$, where $\mathcal{K}_1,\mathcal{K}_2$ are copies of $\mathcal{K}$.
Now we define four entanglement-breaking channels, as in \cite{Beigi.2008}, that the proof hinges on. They will all be mapping $\mathcal{B}(\mathcal{H})=\mathcal{B}(\mathcal{K}\otimes\mathcal{K})$ into $\mathcal{B}(\mathcal{K})$.
\begin{remark}
The idea behind the following is that each of the maps achieves its $1\to p$ norm on a certain subset of states, such that the intersection of these subsets singles out exactly states of the form $\tilde{\psi}\otimes\tilde{\psi}$, where $\tilde{\psi}=|\tilde{\psi}\rangle\langle\tilde{\psi}|$ is a pure state of from \eqref{def:VectorSpecialForm}. Thus, we can check whether we are in a ``Yes'', or ``No'' instance of the problem.
\end{remark}

\begin{definition}
    Given the Hamiltonian associated to some instance of the 2-Out-of-4-SAT problem, we define the entanglement-breaking CPTP map $\Phi_H:\mathcal{B}(\mathcal{H})\to \mathcal{B}(\mathbb{C}^2)$ as
    \begin{align}\label{equ:def:Phi.H}
    \Phi_H(\rho):=\frac{1}{2}\Tr[H\rho]|0\rangle\langle0|+\Tr\left[\left(\1_\mathcal{H}-\frac{1}{2}H\right)\rho\right]|1\rangle\langle1|\,,
\end{align}

\end{definition}
By construction $\|\frac{1}{2}H\|_\infty\leq \frac{1}{2}$ 
, hence $|1\rangle\langle1|$ is always in the support of $\Phi_H(\rho)$ and the output is just $|1\rangle\langle1|$ if and only if $\rho$ is a ground state of $H$ with eigenvalue $0$. 
We also make use of the following entanglement-breaking version of the partial trace \cite[Lemma 4.5]{Beigi.2008}:

\begin{definition}
For $0\leq\eta\leq d^2$, we define the channel $\Phi_{\textup{trace}}^{(\eta)}:\cB(\cH)\to \cB(\cK_1)$ to be
\begin{align}\label{equ:Phi.trace}
    \Phi^{(\eta)}_{\textup{trace}}(\rho):=\tr_2\left[\left(1-\frac{\eta}{d^2}\right)\frac{\1_\mathcal{H}}{d^2}+\frac{\eta}{d^2}\rho\right]=\left(1-\frac{\eta}{d^2}\right)\frac{\1_1}{d}+\frac{\eta}{d^2}\rho_1\,.\label{def:EntanglementBreakingPartialTrace}
\end{align} 
And for $0<\eta<1$ is entanglement-breaking by \cite[Lemma 4.5]{Beigi.2008}.
\end{definition}

It is easy to check that for $1<p<\infty$ and $0<\eta\leq d^2$ on states it satisfies

\begin{align}
    &\|\Phi^{(\eta)}_\textup{trace}(\rho)\|_p^p \leq (d-1)\left(\frac{1-\eta d^{-2}}{d}\right)^p+\left(\frac{1-\eta d^{-2}}{d}+\eta d^{-2}\right)^p =:f(\eta,d,p)^p, \\
    &\|\Phi^{(\eta)}_\textup{trace}(\rho)\|_\infty \leq \frac{1-\eta d^{-2}}{d}+\eta d^{-2} = \frac{1}{d}-\frac{\eta}{d^3}+\frac{\eta}{d^2} =:f(\eta,d,\infty),
\end{align} with equality iff $\rho_1$ is pure, i.e. whenever $S(\rho_1)=0$. 
We note here that for $\eta=d^2$ we have $f(d^2,d,p)=1$ for all $p\in[1,\infty]$.

We also define the partial trace channel as
\begin{align}
    \Phi_{\textup{trace}}(\rho):=\tr_2[\rho].
\end{align}

Next we recall the measure-and-prepare channel that implements the SWAP-test.

\begin{definition}
The swap channel $\Phi_{\text{swap}}:\cB(\cH)\to\cB(\mathbb{C}^2)$ is defined on pure product states $\rho=\psi_1\otimes\psi_2$ as
\begin{align}\label{equ:Phi.swap}
    \Phi_\text{swap}(\psi_1\otimes\psi_2)=\frac{1}{2}(1+|\langle\psi_1|\psi_2\rangle|^2)|0\rangle\langle 0|+\frac{1}{2}(1-|\langle\psi_1|\psi_2\rangle|^2)|1\rangle\langle 1|\,.
\end{align}
It extends to all of $\mathcal{B}(\mathcal{H})$ by linearity. 
\end{definition}
It is easy to check that $|0\rangle\langle 0|$ is always in the support of $\Phi_\text{swap}(\rho)$ and that the output on product states is $|0\rangle\langle0|$ if and only if the input was of the form $\psi\otimes\psi$, where $\psi$ is a pure state.

\medskip 

The last channel we need is the following:

\begin{definition}
   The cube channel $\Phi_{\text{cube}}:\cB(\cH)\to \cB(\mathbb{C}^2)$ is defined as
    \begin{align}\label{equ:Phi.cube}
   \Phi_\text{cube}(\rho) := \frac{1}{d(d-1)}\sum_{i\neq j}^d\Tr[(\Pi_{ij}\otimes\Pi_{ij}^\prime) \rho]|0\rangle\langle 0| + \Tr[M\rho]|1\rangle\langle 1|,    
    \end{align} where $\Pi_{ij}:=\frac{1}{2}(|i\rangle+|j\rangle)(\langle i|+\langle j|), \Pi^\prime_{ij}:=\frac{1}{2}(|i\rangle-|j\rangle)(\langle i|-\langle j|),$ and $M:=\1_\mathcal{H}-\frac{1}{d(d-1)}\sum_{ij}\Pi_{ij}\otimes\Pi_{ij}^\prime$.
\end{definition}
By construction, see \cite{Beigi.2008}, $M>0$ is positive definite, and hence $|1\rangle\langle1|$ is always in the support of $\Phi_\text{cube}(\rho)$, and the output on pure i.i.d. state inputs is $|1\rangle\langle 1|$ if and only if the input takes the form $\tilde{\psi}\otimes\tilde{\psi}$, where $\tilde{\psi}$ is a proper state of the form \eqref{def:VectorSpecialForm}.


The following lemma is key to establishing \cref{thm:1top.hardness}.
\begin{lemma}\label{lemma:MainCorrect}
 Let $1<p\leq\infty$, $0<\eta\leq d^2$ and consider the map $\Psi_H:=\Phi_\textup{trace}^{(\eta)}\otimes\Phi_\textup{swap}\otimes\Phi_{\textup{cube}}\otimes\Phi_H:\cB(\cH\otimes \cH\otimes \cH\otimes \cH)\to\cB(\cK_1\otimes \mathbb{C}^2\otimes\mathbb{C}^2\otimes\mathbb{C}^2)$. 
  Then $\|\Psi_{H}\|_{1\to p} \leq f(\eta,d,p)$. Moreover, if there exists a state $\tau$ such that $\| \Psi_{H}(\tau) \|_{p} = f(\eta,d,p)$, then
\begin{align}
\Psi_H(\tau)&=\Phi^{(\eta)}_\textup{trace}(\tau_1)\otimes\Phi_\textup{swap}(\tau_2)\otimes\Phi_{\textup{cube}}(\tau_3)\otimes\Phi_H(\tau_4)
    \end{align}where $\tau_i\in\mathcal{D}(\mathcal{H}_i)$ are the marginals of $\tau$.
\end{lemma}

\begin{remark} 
Note that for $\eta=d^2$ $\Phi^{(\eta)}_{\textup{trace}}(\cdot)=\Phi_{\textup{trace}}(\cdot)=\tr_{1_2}[\cdot]$ is just the partial trace and the optimal value becomes $f(d^2,d,p)=1 \quad \forall p\in[1,\infty]$ and for $0<\eta<1$ the channel $\Psi_H$ is entanglement-breaking.
Note also that the above result does not imply that optimal input states are all in tensor product form, but rather that their outputs cannot be distinguished from those on the tensor product of their marginals.
\end{remark}
In order to prove this we first collect some observations.
\begin{claim}\label{prop:Claim1}
    Let $1<p\leq\infty$.
    For a collection of mutually orthogonal pure quantum states $\{\sigma_i\}_i\in\mathcal{D}(\mathcal{H})$ and a probability distribution $\{p_i\}_i$ it holds that if we have equality in  $\|\sum_ip_i\sigma_i\|_p\leq\sum_ip_i\|\sigma_i\|_p=1$, 
    then $p_i=\delta_{i,n}$ for some $n$, i.e. the convex combination is singular. 
\end{claim}
\begin{proof}
    Since all $\sigma_i$ are pure, we have $\|\sigma_i\|_p=1$ and since the $\sigma_i$ are in addition orthogonal we have 
    \begin{align}
        \bigg\|\sum_i p_i\sigma_i\bigg\|_p = \Tr\bigg[\sum_ip_i^p\sigma_i\bigg]^\frac{1}{p} = \bigg(\sum_ip_i^p\bigg)^\frac{1}{p} = \|\vec{p}\|_p \overset{!}{=} 1 \\
        \bigg\|\sum_i p_i\sigma_i\bigg\|_\infty = \max_i p_i\|\sigma_i\|_\infty = \max_i p_i = \|\vec{p}\|_\infty \overset{!}{=} 1.
    \end{align} Both of these imply the claim, since $1<p$.
\end{proof}
We further require the following well known result.
\begin{claim}\label{prop:Claim2}
    Given a bipartite quantum state $\rho_{12}\in\mathcal{D}(\mathcal{H}_1\otimes\mathcal{H}_2)$, then if one of its marginals, say $\rho_1:=\tr_2\rho_{12}$ is pure, then $\rho_{12}=\rho_1\otimes\rho_2$ is a product state.
\end{claim}
\begin{proof}  By purity $S(\rho_1)=0$ and by the triangle inequality and subadditivity of the von Neumann entropy it further follows that
    \begin{align}
        S(\rho_2)=|S(\rho_1)-S(\rho_2)|\leq S(\rho_{12})\leq S(\rho_1)+S(\rho_2) \Rightarrow S(\rho_{12})=S(\rho_1)+S(\rho_2).
    \end{align} This is equivalent to the mutual information between the systems $1$ and $2$ being zero, which implies the claim.
\end{proof}
Having established these two we may now prove Lemma \ref{lemma:MainCorrect}.
\begin{proof}[Proof of Lemma \ref{lemma:MainCorrect}]
We start by proving the claim with $\eta=d^2$: for a given state $\tau=\tau_{1234}$ on $\cH^{\otimes 4}$, and denoting $\tr_{1_2}$ the partial trace over the second subsystem of the first system $\cH$, we write 
\begin{align}
\Psi_H(\tau)&:=(\Phi_\text{trace}\otimes\Phi_\text{swap}\otimes\Phi_{\text{cube}}\otimes\Phi_H)(\tau) \\ &=\sum_{j_2j_3j_4}\tr_{1_2}\tr_{234}[\tau_{1234}(M^{(2)}_{j_2}\otimes M^{(3)}_{j_3}\otimes M^{(4)}_{j_4})]\otimes\sigma^{(2)}_{j_2}\otimes\sigma^{(3)}_{j_3}\otimes\sigma^{(4)}_{j_4} \\ &\equiv \sum_{j_2j_3j_4}p_{j_2j_3j_4}\tau_{1_1|j_2j_3j_4}\otimes\sigma^{(2)}_{j_2}\otimes\sigma^{(3)}_{j_3}\otimes\sigma^{(4)}_{j_4},
\end{align} where $p_{j_2j_3j_4}:=\Tr_{1234}[\tau_{1234}(M^{(2)}_{j_2}\otimes M^{(3)}_{j_3}\otimes M^{(4)}_{j_4})] = \Tr_{234}[\tau_{234}(M^{(2)}_{j_2}\otimes M^{(3)}_{j_3}\otimes M^{(4)}_{j_4})]$ is a probability distribution and $\tau_{1_1|j_2j_3j_4}$ a quantum state on $\cK_1$. Above, we used that the channels $\Phi_{\text{swap}}$, $\Phi_{\text{cube}}$ and $\Phi_H$ are measure-and-prepare, and denoted by $M^{(i)}_{j_i}$ and $\sigma^{(i)}_{j_i}$ their POVM elements and corresponding output states. Also recall that each $\sigma^{(i)}_{j_i}$ is a pure state by the construction of $\Phi_\text{swap},\Phi_\text{cube},\Phi_H$. Now we study the following inequality
\begin{align}
\|\Psi_H(\tau)\|_p&=\bigg\|\sum_{j_2j_3j_4}p_{j_2j_3j_4}\tau_{1_1|234}\otimes\sigma^{(2)}_{j_2}\otimes\sigma^{(3)}_{j_3}\otimes\sigma^{(4)}_{j_4}\bigg\|_p \\ &\overset{\text{}}{\leq} \sum_{j_2j_3j_4}p_{j_2j_3j_4}\big\|\tau_{1_1|j_2j_3j_4}\otimes\sigma^{(2)}_{j_2}\otimes\sigma^{(3)}_{j_3}\otimes\sigma^{(4)}_{j_4}\big\|_p \\ &=\sum_{j_2j_3j_4}p_{j_2j_3j_4}\|\tau_{1_1|j_2j_3j_4}\|_p\\
&\leq \sum_{j_2j_3j_4}p_{j_2j_3j_4}\\
&=1\,,
\end{align} 
where the first inequality is an application of the triangle inequality and the second follows by monotonicity of Schatten norms. Note that $1$ is the optimal value of $\|\Psi_H(\tau)\|_p$, since it is a quantum channel. Equality in the last inequality implies that all $\tau_{1_1|j_2j_3j_4}$ must be pure, hence for the optimal states \cref{prop:Claim1} applies and $p_{j_2j_3j_4}=\delta_{j_2,k_2}\delta_{j_3,k_3}\delta_{j_4,k_4}$ is a Dirac distribution. Then, since positive operators of zero trace are identically zero,  
\begin{align}
\tr_{234}\big(\tau_{1_1234}(M^{(2)}_{j_2}\otimes M^{(3)}_{j_3}\otimes M^{(4)}_{j_4})\big)=0\qquad \forall (j_{2},j_3,j_4)\ne (k_2,k_3,k_4)\,.
\end{align}
Therefore we showed that 
\begin{align}
\tau_{1_1}&=\tr_{234}(\tau_{1_1234})\\
&=\sum_{j_2j_3j_4}\tr_{234}\big(\tau_{1_1234}(M^{(2)}_{j_2}\otimes M^{(3)}_{j_3}\otimes M^{(4)}_{j_4})\big)\\
&=\tr_{234}\big(\tau_{1_1234}(M^{(2)}_{k_2}\otimes M^{(3)}_{k_3}\otimes M^{(4)}_{k_4})\big)\\
&=\frac{1}{p_{k_2k_3k_4}}\tr_{234}\big(\tau_{1_1234}(M^{(2)}_{k_2}\otimes M^{(3)}_{k_3}\otimes M^{(4)}_{k_4})\big)\\
&=\tau_{1_1|k_2k_3k_4}
\end{align}
Thus, $\tau_{1_1|k_2k_3k_4}=\tau_{1_1}$ is pure and hence by \cref{prop:Claim2} it follows that 
\begin{align}
    \tau_{1234}= \tau_{1_1}\otimes \tau_{1_2234}\,.
\end{align}
Due to the singularity of the distribution $p_{j_2j_3j_4}=\delta_{j_2j_3j_4,k_2k_3k_4}=\delta_{j_2,k_2}\delta_{j_3,k_3}\delta_{j_4,k_4}=p_{j_2}p_{j_3}p_{j_4}$  we further have for any $j_2,j_3,j_4$ that  
\begin{align}
    0&=p_{j_2j_3j_4}-p_{j_2}p_{j_3}p_{j_4}\\&=\Tr[\tau_{1234}(M^{(2)}_{j_2}\otimes M^{(3)}_{j_3}\otimes M^{(4)}_{j_4})]-\Tr[\tau_2M^{(2)}_{j_2}]\Tr[\tau_3M^{(3)}_{j_3}]\Tr[\tau_4M^{(4)}_{j_4}] \\ &= \Tr[(\tau_{234}-\tau_2\otimes\tau_3\otimes\tau_4)M^{(2)}_{j_2}\otimes M^{(3)}_{j_3}\otimes M^{(4)}_{j_4}].
\end{align} 
Next, we define $\delta\tau_{234}:=(\tau_{234}-\tau_2\otimes\tau_3\otimes\tau_4)\in\mathcal{M}^\perp_{234}:=\left(\text{span}_{j_2j_3j_4}\{M^{(2)}_{j_2}\otimes M^{(3)}_{j_3}\otimes M^{(4)}_{j_4}\}\right)^\perp$.
Then, if $\|\Psi_H(\tau)\|_p=1$ is optimal, 
\begin{align}
    \Psi_H(\tau) &= \sum_{j_2j_3j_4}\tr_{1_2}\tr_{234}[\tau_{1234}(M^{(2)}_{j_2}\otimes M^{(3)}_{j_3}\otimes M^{(4)}_{j_4})]\otimes\sigma^{(2)}_{j_2}\otimes\sigma^{(3)}_{j_3}\otimes\sigma^{(4)}_{j_4} \\ 
    &\overset{(1)}{=} \tau_{1_1}\otimes \Tr_{234}[\tau_{234}(M^{(2)}_{k_2}\otimes M^{(3)}_{k_3}\otimes M^{(4)}_{k_4})]\otimes\sigma^{(2)}_{k_2}\otimes\sigma^{(3)}_{k_3}\otimes\sigma^{(4)}_{k_4} \\ 
    &= \tau_{1_1}\otimes\Tr_{234}[(\tau_2\otimes\tau_3\otimes\tau_4+\delta\tau_{234})(M^{(2)}_{k_2}\otimes M^{(3)}_{k_3}\otimes M^{(4)}_{k_4})]\otimes\sigma^{(2)}_{k_2}\otimes\sigma^{(3)}_{k_3}\otimes\sigma^{(4)}_{k_4} \\
    & \overset{(2)}{=} \tau_{1_1}\otimes\Tr_{234}[\tau_2\otimes\tau_3\otimes\tau_4(M^{(2)}_{k_2}\otimes M^{(3)}_{k_3}\otimes M^{(4)}_{k_4})]\otimes\sigma^{(2)}_{k_2}\otimes\sigma^{(3)}_{k_3}\otimes\sigma^{(4)}_{k_4} \\ &= \tau_{1_1}\otimes\Tr[\tau_2M^{(2)}_{k_2}]\sigma^{(2)}_{k_2}\otimes\Tr[\tau_3M^{(3)}_{k_3}]\sigma^{(3)}_{k_3}\otimes\Tr[\tau_4M^{(4)}_{k_4}]\sigma^{(4)}_{k_4} \\ &= \Phi_\text{trace}(\tau_1)\otimes\Phi_\text{swap}(\tau_2)\otimes\Phi_{\text{cube}}(\tau_3)\otimes\Phi_H(\tau_4),
\end{align}
which is the claim of the Lemma. Here in $(1)$ we used the product state nature of $\tau_{1234}$ along the cut $1_1,1_2234$, which follows from \cref{prop:Claim2} and the Dirac nature of $p_{j_2j_3j_4}=\Tr_{1234}[\tau_{1234}(M^{(2)}_{j_2}\otimes M^{(3)}_{j_3}\otimes M^{(4)}_{j_4})]=\delta_{j_2n_2}\delta_{j_3n_3}\delta_{j_4n_4}$; in equality $(2)$ we used that $\delta\tau_{234}\in\mathcal{M}_{234}^\perp$.

The above given proof still goes through for arbitrary $\eta\leq d^2$ and in particular $\eta<1$. Doing this we effectively replace $\tau_{1_1|j_2j_3j_4}$
by $\tau^{(\eta)}_{1_1|j_2j_3j_4}
= \left(1-\frac{\eta}{d^2}\right)\frac{\1}{d}+\frac{\eta}{d^2}\tau_{1_1|j_2j_3j_4}$ 
and for this and any $1<p\leq\infty$ it holds that 
\begin{align}
\big\|\tau^{(\eta)}_{1_1|j_2j_3j_4}\big\|_p=\bigg\|\left(1-\frac{\eta}{d^2}\right)\frac{\1}{d}+\frac{\eta}{d^2}\tau_{1_1|j_2j_3j_4}\bigg\|_p\leq f(\eta,d,p),
\end{align} with equality iff $\tau_{1_1|j_2j_3j_4}$ is pure. The rest thus follows exactly as above.
\end{proof}

We need one last ingredient to prove the main theorem of this section, because we need to ensure that all these channels receive the same input state.
\begin{definition}\label{def:Lambda}
    Define the following unital CPTP mixing channel $\Lambda_k:\mathcal{D}(\otimes_{i=1}^k\mathcal{H}_i) \to \mathcal{D}(\otimes_{i=1}^k\mathcal{H}_i)$ via
    \begin{align}\label{equ:def.Lambda}
        \Lambda_k(\rho_{1...k}):=\frac{1}{k!}\sum_{s\in \Omega_k}\rho_s,
    \end{align} where the sum goes over $\Omega_k$, the set of  all permutations of $k$ elements.
\end{definition}
It is easy to check that all the local marginals of the output are equal, in particular $\tilde{\tau}:=\tr_{1^c}\Lambda_k(\tau)=\tr_{2^c}\Lambda_k(\tau)=...=\tr_{k^c}\Lambda_k(\tau)$, where $\tr_{i^c}$ traces out all subsystems except the $i$-th one.
\begin{example}
For $k=2$, we have $\Lambda_2:\rho_{12}\mapsto \frac{1}{2}(\rho_{12}+\rho_{21})$, i.e. it is a uniform mixture of the identity channel and the swap channel. 
Here it is straightforward to check that $\tr_2\Lambda_2(\tau)=\frac{1}{2}\tr_2[\tau_{12}+\tau_{21}]= \frac{1}{2}(\tau_1+\tau_2)= \tr_1\Lambda_2(\tau)$.
\end{example}
We are now ready to prove Theorem \ref{thm:1top.hardness}.
\begin{proof}[Proof of Theorem \ref{thm:1top.hardness}]
We will show that deciding whether $\|\Phi\|_{1\to p} = c$, for some specific CPTP map $\Phi$, is equal to some given value $c$ reduces to deciding whether an instance of the 2-Out-of-4 SAT problem is satisfiable or not. Fix some $p\in(1,\infty]$. Given an instance of 2-Out-of-4-SAT, take $\Phi_\text{trace},\Phi_{\text{swap}},\Phi_{\text{cube}}, $ and $\Phi_{H}$ as defined above. Now we set $\Psi_H:=\Phi_\text{trace}\otimes\Phi_{\text{swap}}\otimes\Phi_{\text{cube}}\otimes\Phi_{H}$ which is by linearity defined on all of $\mathcal{B}(\mathcal{M})$, where $\mathcal{M}:=\otimes_{i=1}^4\mathcal{H}_i$ where all $\mathcal{H}_i$ are copies of $\mathcal{H}=\mathcal{K}\otimes\mathcal{K}$.
This so defined $\Psi_H$ is the CPTP map considered in  Lemma \ref{lemma:MainCorrect}. We consider the problem of computing the norm of $\Psi_H\circ\Lambda_4$.
First of all, if the 2-Out-of-4 SAT problem is satisfiable, then there exists a pure state $\tilde{\psi}\in\mathcal{D}(\mathcal{K})$ of form \eqref{def:VectorSpecialForm}, such that $\|\Psi_H\circ \Lambda_4((\tilde{\psi}\otimes\tilde{\psi})^{\otimes 4})\|_p=\|\Psi_H(\tilde{\psi}\otimes\tilde{\psi})^{\otimes 4})\|_p=1$ is optimal and hence $\|\Psi_H\circ\Lambda_4\|_{1\to p}=1$.
On the other hand, if $\|\Psi_H\circ\Lambda_4\|_{1\to p}=1$, then there exists a state $\tau$, such that $\|\Psi_H\circ\Lambda_4(\tau)\|_{p}=1$. 
By Lemma \ref{lemma:MainCorrect}, it follows that
\begin{align}
    \Psi_H(\Lambda_4(\tau))= \Psi_H(\tilde{\tau}^{\otimes 4})=  \Phi_\text{trace}(\tilde{\tau})\otimes\Phi_{\text{swap}}(\tilde{\tau})\otimes\Phi_{\text{cube}}(\tilde{\tau})\otimes\Phi_{H}(\tilde{\tau}),
\end{align} where we denote with $\tilde{\tau}$ the local marginals of $\Lambda_4(\tau)$.
So we found a state $\tilde{\tau}\in\mathcal{D}(\mathcal{H})$ that is optimal w.r.t all $\Phi_\text{trace},\Phi_\text{swap},\Phi_\text{cube},\Phi_H$, which by the construction of these implies that it has to be of the form $\tilde{\psi}\otimes\tilde{\psi}$, where the pure state $\tilde{\psi}$ is of the form \eqref{def:VectorSpecialForm} and satisfies the 2-Out-of-4 SAT problem. This is because by the initial remarks, the optimizers of $\Phi_\text{trace}$ are product states with one pure marginal, by $\Phi_\text{swap}$ these further have to be pure product states of the from $\psi\otimes\psi$, and by $\Phi_\text{cube}$ these pure states have to be of form \eqref{def:VectorSpecialForm} and if $\Phi_H$ achieves its optimum on such states, then we have found that there exists a vector $\tilde{\psi}$ of form \eqref{def:VectorSpecialForm} that is orthogonal to all $|A_k\rangle$, i.e. satisfies this SAT problem.
Hence we have shown that $\|\Psi_H\circ\Lambda_4\|_{1\to p}= 1$ is equivalent to being in the ``Yes'' instance of the 2-Out-of-4-SAT problem. Now, if the 2-Out-of-4-SAT instance is not satisfiable, then there is at least one constraint that is not satisfied, which implies that the corresponding $\|\Psi_H\circ\Lambda_4\|_{1\to p} < 1$.  In fact, we show below that it implies $\|\Psi_H\circ\Lambda_4\|_{1\to p} < 1 - \frac{1}{\poly(d)}$, i.e. completeness. (cf. $m=\operatorname{poly}(d)$). 

Replacing in this argument $\Phi_\textup{trace}$ by $\Phi^{(\eta)}_\textup{trace}$ for some $0<\eta<1$ we get that the so constructed channel $\Psi^{(\eta)}_H\circ\Lambda$ of which we compute the $1\to p$-norm is entanglement-breaking and hence we get in that case that the ``Yes'' instance of the 2-Out-of-4 SAT problem is equivalent to $\|\Psi_H^{(\eta)}\circ\Lambda_4\|_{1\to p}=f(\eta,d,p)$ and the ``No'' instance to $\|\Psi_H^{(\eta)}\circ\Lambda_4\|_{1\to p}<f(\eta,d,p)$. By the completeness argument we will give below we will have that the ``No'' instance will correspond to $\|\Psi_H^{(\eta)}\circ\Lambda_4\|_{1\to p}<f(\eta,d,p)-\epsilon$ for some $\epsilon\geq \frac{1}{\poly(d)}$. 

We show the claimed completeness by showing that the above argument (for $\eta=d^2$) we gave is robust using the gentle measurement Lemma and related standard tools. Explicitly we show that there exists a suitable polynomial and $\epsilon\geq \frac{1}{\poly(d)}$ s.t. if $\|\Psi_H\circ\Lambda_4\|_{1\to p}\geq 1-\epsilon$ we can still infer satisfiability. The case $\eta<d^2$ works analogously. Assume there is an input state $\tau\in\cD(\cH^{\otimes4})$ s.t. $\|\Psi_H\circ\Lambda_4(\tau)\|_p\geq 1-\epsilon$ holds for a later to be determined $\epsilon$. We will first show that this state must be approximately a product state along the cut $1_1,1_2234$.
We first require the following simple result.
\begin{claim}\label{claimholder}
Let $q$ be a $d$-dimensional probability distribution that satisfies $\|q\|_{p}\geq 1-\epsilon$ for some $p>1$ and $\epsilon\geq 0$. Then it holds that the largest entry satisfies $\|q\|_\infty\geq (1-\epsilon)^{\frac{p}{p-1}}=1-\frac{p}{p-1}\epsilon+\mathcal{O}(\epsilon^2)$.   
The analogous statement also holds for quantum states and their Schatten norms.
\end{claim}
\begin{proof}
By Hölder's inequality
\begin{align}
  \|q\|_p=\|q^{1-\frac{1}{p}}q^{\frac{1}{p}}\|_p\leq \|q\|_\infty^{1-\frac{1}{p}}\|q^{\frac{1}{p}}\|_{p}=\|q\|_\infty^{1-\frac{1}{p}}\|q\|_{1}^{\frac{1}{p}}\,.
\end{align} Since by assumption $\|q\|_1=1$ and $(1-\epsilon) \leq  \|q\|_p$ rearranging yields the claim.
\end{proof}
Now by assumption on $\tau$ we have
\begin{align}
1-\epsilon \leq \|\Psi_H(\Lambda_4(\tau))\|_p &=\Big\|\sum_{j_2j_3j_4}p_{j_2j_3j_4}(\Lambda_4(\tau))_{1_1|j_2j_3j_4}\otimes \sigma^{(2)}_{j_2}\otimes \sigma^{(3)}_{j_3}\otimes \sigma^{(4)}_{j_4}\Big\|_p \\ &= \|(p_{j_2j_3j_4}\|(\Lambda_4(\tau))_{1_1|j_2j_3j_4}\|_p)_{j_2j_3j_4}\|_{\ell_p},
\end{align} 
which implies, by the Claim \ref{claimholder} above, that there must exist some index $k_2k_3k_4$ such that
\begin{align}\label{equ:polyproof0}
    1-\delta &\leq p_{k_2k_3k_4}\|\Lambda_4(\tau)_{1_1|k_2k_3k_4}\|_p
\end{align} 
for $\delta=\mathcal{O}(\epsilon)$. Now since $0\leq \|\Lambda_4(\tau)_{1_1|k_2k_3k_4}\|_p\leq 1$ we have
\begin{align}\label{equ:polyproof1}
    1-\delta \leq p_{k_2k_3k_4}
\end{align} which implies via the gentle measurement lemma, see e.g. \cite[Corollary 3.15]{book.Watrous.2018} with \cite[Theorem 3.33]{book.Watrous.2018}, that 
\begin{align}
    \|\Lambda_4(\tau)_{1_1|k_2k_3k_4}-\Lambda_4(\tau)_{1_1}\|_1\leq 2\sqrt{\delta}=\mathcal{O}(\sqrt{\epsilon}).
\end{align}
Also since $0\le p_{k_2k_3k_4} \le 1$, the above implies that 
\begin{align}\label{equ:polyproof2}
 1-\delta\leq \|\Lambda_4(\tau)_{1_1|k_2k_3k_4}\|_p  \Rightarrow (\Lambda_4({\tau}))_{1_1|k_2k_3k_4}=(1-\delta^\prime)\psi+\delta^\prime\sigma,
\end{align}
by another use of Claim \ref{claimholder}, where $\sigma,\psi\in\cD(\cH)$, with $\psi$ pure, and $\delta^\prime=\mathcal{O}(\epsilon)$. Hence by triangle inequality
\begin{align}
    \|(\Lambda_4(\tau))_{1_1}-\psi\|_1\leq \|(\Lambda_4(\tau))_{1_1}-(\Lambda_4(\tau))_{1_1|k_2k_3k_4}\|_1+\|(\Lambda_4(\tau))_{1_1|k_2k_3k_4}-\psi\|_1 = \mathcal{O}(\sqrt{\epsilon}).
\end{align}
Next, we show the existence of some state $\rho_{1_2}$ s.t.
\begin{align}
   \|(\Lambda_4(\tau))_{1}-\psi\otimes \rho_{1_2}\|_1= \mathcal{O}(\epsilon^{\frac{1}{4}}). 
\end{align}
For simplicity we denote in the following $\tilde{\tau}:=(\Lambda_4(\tau))_1$ the local marginal.
Indeed, By Uhlmann's theorem \cite[Theorem 3.22]{book.Watrous.2018} and data processing for the fidelity we have, for some purification $\Phi_{R1}$ of $\tilde{\tau}$, i.e. also of $(\Lambda_4(\tau))_{1_1}$, that
\begin{align}
    F(\tilde{\tau}_{1_1},\psi) = \max_{\phi_{R1_2}}F(\Phi_{R1},\psi\otimes\phi_{R1_2}) \leq \max_{\sigma_{1_2}}F(\tilde{\tau},\psi\otimes\sigma_{1_2}) = F(\tilde{\tau},\psi\otimes\rho_{1_2}),
\end{align} where we denoted the maximizer with $\rho_{1_2}$. Now by Fuchs van de Graaf inequality we have $1-\mathcal{O}(\sqrt{\epsilon})= F(\tilde{\tau}_{1_1},\psi)$ and hence 
\begin{align}
 \|\tilde{\tau}_{1}-\psi\otimes \rho_{1_2}\|_1\leq 2\sqrt{1-F^2(\tilde{\tau},\psi\otimes\rho_{1_2})} = 2\sqrt{\mathcal{O}(\sqrt{\epsilon})} = \mathcal{O}(\epsilon^{\frac{1}{4}}).   
\end{align}
Next we show that this state $\psi\otimes \rho_{1_2}$ and hence $\tilde{\tau}$ is close to a tensor product product of symmetric pure states. 
Another consequence of \eqref{equ:polyproof1} is that
\begin{align}
   \Tr[M_{k_2}^{(2)}\tilde{\tau}]= p_{k_2}(\tilde{\tau})\equiv p_{k_2}=\sum_{j_3j_4}p_{k_2j_3j_4}\geq 1-\delta= 1-\mathcal{O}(\epsilon),
\end{align}
where by the closeness in trace distance and $\|M^{(2)}_{k_2}\|_\infty\leq 1$ we have
\begin{align}
    \Tr[M^{(2)}_{k_2}(\psi\otimes\rho_{1_2})] = \underbrace{\Tr[M_{k_2}^{(2)}\tilde{\tau}]}_{p_{k_2}\geq 1-\delta} - \mathcal{O}(\epsilon^\frac{1}{4}) =1-\mathcal{O}(\epsilon)-\mathcal{O}(\epsilon^\frac{1}{4}) = 1-\mathcal{O}(\epsilon^\frac{1}{4}).
\end{align}
By construction, since $\Tr[M^{(2)}_{k_2}(\psi\otimes\rho_{1_2})]>1/2$ for small enough $\epsilon$, we must have
\begin{align}
1-\mathcal{O}(\epsilon^\frac{1}{4}) = \Tr[M^{(2)}_{k_2}(\psi\otimes\rho_{1_2})]=\frac{1}{2}\left(1+F^2(\psi,\rho_{1_2})\right)\,.
\end{align} 
This, again by Fuchs van de Graaf, implies 
\begin{align}
    \|\psi\otimes\psi-\psi\otimes\rho_{1_2}\|_1 = \|\psi-\rho_{1_2}\|_1\leq 2\sqrt{1-F^2(\psi,\tilde{\tau}_{1_2})} = 2\sqrt{2\mathcal{O}(\epsilon^\frac{1}{4})} = \mathcal{O}(\epsilon^\frac{1}{8}),
\end{align} and hence with the above
\begin{align}\label{equ:polyproof3}
    \|\tilde{\tau}-\psi\otimes\psi\|_1= \mathcal{O}(\epsilon^{\frac{1}{8}}).
\end{align}
Analogously we will now show that this product state $\psi$ that is close must be approximately proper $\tilde{\psi}$, i.e. of form \eqref{def:VectorSpecialForm}.
For the probability of the measurement outcome of the $\Phi_{\text{cube}}$ we have the equivalent bound. Let $\delta<\frac{1}{2}$ then via \eqref{equ:polyproof3} and $\|\Pi_{ij}\otimes\Pi_{ij}^\prime\|_\infty\leq 1$ we have
\begin{align}
    \delta d(d+1)\geq \Tr[(\Pi_{ij}\otimes \Pi^\prime_{ij})\tilde{\tau}] =\Tr[(\Pi_{ij}\otimes\Pi^\prime_{ij})(\psi\otimes\psi)]-\mathcal{O}(\epsilon^\frac{1}{8})\,.
\end{align} 
It follows that 
\begin{align}\label{equ:approx0}
   \mathcal{O}(\epsilon^\frac{1}{8}\poly(d)) = \Tr[\Pi_{ij}\otimes\Pi_{ij}^\prime (\psi\otimes\psi)] = \frac{1}{4}|\psi_i+\psi_j|^2|\psi_i-\psi_j|^2 \quad \forall i,j\in[d],
\end{align} where $|\psi\rangle=\sum_i\psi_i|i\rangle$.
This means that there exists a proper state $\tilde{\psi}=\sum_i\tilde{\psi}_i|i\rangle$, s.t. $|\psi_i-\tilde{\psi}_i|\leq \mathcal{O}(\epsilon^{\frac{1}{16}}\poly(d))$ for all $i$ which implies that
\begin{align}\label{equ:approx1}
    \|\psi\otimes\psi-\tilde{\psi}\otimes\tilde{\psi}\|_1\leq 2\|\psi-\tilde{\psi}\|_1\leq 4\||\psi\rangle-|\tilde{\psi}\rangle\|_{l_2} \leq 4\sqrt{d}\||\psi\rangle-|\tilde{\psi}\rangle\|_{l_\infty}
    =\mathcal{O}(\epsilon^\frac{1}{16}\poly(d)).
\end{align}
Hence by triangle inequality again this gives 
\begin{align}\label{equ:approx2}
    \|\tilde{\tau}-\tilde{\psi}\otimes\tilde{\psi}\|_1 = \mathcal{O}(\epsilon^{\frac{1}{16}}\poly(d)).
\end{align}
Repeating this argument now for the last measurement we get, completely analogously,
\begin{align}
    \frac{1}{2}\Tr[H(\tilde{\psi}\otimes\tilde{\psi})] = \underbrace{\frac{1}{2}\Tr[H\tilde{\tau}]}_{=1-p_{k_4} \leq \delta=\mathcal{O}(\epsilon)<\frac{1}{2}} +  \mathcal{O}(\epsilon^\frac{1}{16}\poly(d))  \equiv \eta = \mathcal{O}(\epsilon^\frac{1}{16}\poly(d)).
\end{align}
Considering finally the integrality of the problem, which implies that either $\frac{1}{2}\Tr[H(\tilde{\psi}\otimes\tilde{\psi})]=0$ or $\geq \frac{1}{m}=\frac{1}{\poly(d)}$, for some prior chosen $m=\poly(d)$ depending on whether the SAT problem is satisfiable or not. Thus if we choose $\epsilon$ s.t. the last error term $\eta=\mathcal{O}(\epsilon^\frac{1}{16}\poly(d))\overset{!}{=}\frac{1}{2m}<\frac{1}{m}$ we have that existence of a state $\tau$ s.t. $\|\Psi_H(\Lambda_4(\tau))\|_p\geq 1-\epsilon$ implies that the given instance of the 2-Out-of-4-SAT problem must be satisfiable. Clearly this works for $\epsilon=\frac{1}{\poly(d)}$ as $m=\poly(d)$, which completes the proof.
\end{proof}

\subsection{$\NP$-completeness of $1^+\to 1$-norm}
While the hardness established for $1 \to p$ is clearly not true for the $1\to 1$ norm of a CP map, since that is given by a semidefinite program, the problem becomes $\NP$-complete again when dropping the CP assumption. 
\begin{theorem}[Computing the $1\to 1$ norm over states is $\NP$-complete]
\label{thm:thm1to1}
Deciding $(1^+\to 1)-$\textsc{Schatten Norm} for differences of entanglement-breaking quantum channels is $\NP$-complete.
\end{theorem}
In other words given a two entanglement-breaking quantum channels $\Phi,\Psi$ and a real number $c>0$, then deciding whether $\|\Phi-\Psi\|^+_{1\to1}>c$ is $\NP$-complete.

\begin{remark}
We note that the quantity $\|\Phi_1-\Phi_2\|^+_{1\to 1}$ is an operationally significant one since it is directly related to the optimal probability of successfully distinguishing between $\Phi_1$ and $\Phi_2$, without having access to an auxiliary system. For the diamond norm, i.e. the $cb,1\to 1$ norm, it is known that $\|\Phi_1-\Phi_2\|_{cb,1\to 1}=\|\Phi_1-\Phi_2\|^+_{cb,1\to 1}$, when $\Phi_1,\Phi_2$ are quantum channels \cite[Theorem 3.51]{book.Watrous.2018}. This is not expected to hold for the non-cb norm considered here. In fact, in general it holds that for a difference of quantum channels $\|\Phi_1-\Phi_2\|^+_{1\to 1}<\|\Phi_1-\Phi_2\|_{1\to 1}$, see \cite[Proposition 3]{Watrous.2004}. Whether the latter is also $\NP$-complete for a difference of quantum channels remains open.
\end{remark}

\subsubsection{Proof of \cref{thm:thm1to1}}\label{sec:proof.1to1.hardness}
To prove this we will use similar ingredients as in the proof of Theorem \ref{thm:1top.hardness}, but first we recall some simple facts. 
\begin{claim}\label{claimm3.19}
Given two quantum states $\rho,\sigma$ it holds that $\|\rho-\sigma\|_1\leq 2$ with equality iff they are orthogonal w.r.t the Hilbert Schmidt inner product. Furthermore it holds that $\|\rho-\frac{\1}{d}\|_1\leq 2-\frac{2}{d}$, with equality iff $\rho$ is pure. Here $\rho$ and $\1$ are acting on a $d$-dimensional Hilbert space.    
\end{claim}

Like for \cref{thm:1top.hardness} we reduce the $2$-Out-of-$4$ SAT problem to our problem. 
Specifically we will analyze, for some fixed $0<\eta\leq d^2$, the map 
\begin{align}
  \tilde{\Psi}^{(\eta)}_H&:=(\Phi^{(\eta)}_\text{trace}-\frac{\1_{1_1}}{d}\Tr(\cdot ))\otimes(\Phi_\text{swap}-|1\rangle\langle 1|\Tr(\cdot ))\otimes(\Phi_\text{cube}-|0\rangle\langle 0|\Tr(\cdot ))\otimes(\Phi_H-|0\rangle\langle 0|\Tr(\cdot )) \\ &\equiv \tilde{\Psi}^{+(\eta)}_H-\tilde{\Psi}^{-(\eta)}_H\,,
\end{align}
where the maps $\Phi^{(\eta)}_{\text{trace}},\,\Phi_\text{swap},\,\Phi_\text{cube},\,\Phi_H$ are the ones defined in \cref{sec:2-out-of-4.setup}. 
\begin{remark}
   It is easy to check that $\frac{1}{8}\tilde{\Psi}^{+(\eta)}_H$ and $\frac{1}{8}\tilde{\Psi}^{-(\eta)}_H$ are entanglement-breaking quantum channels and hence $\frac{1}{8}\tilde{\Psi}^{(\eta)}_H$ is a difference of entanglement-breaking CPTP maps. Since $\|\frac{1}{8}\tilde{\Psi}^{(\eta)}_H\|^+_{1\to 1}=\frac{1}{8}\|\tilde{\Psi}^{(\eta)}_H\|^+_{1\to1}$ is only constant rescaling it suffices to prove that deciding $\|\tilde{\Psi}^{(\eta)}_H\|^+_{1\to1}>c$ is NP-complete.
\end{remark}

To do so we require, analogous to \cref{lemma:MainCorrect} the following Lemma.
\begin{lemma}\label{lemma:1to1correct}
    Let $0<\eta\leq d^2$. For $\tilde{\Psi}^{(\eta)}_H$ it holds that $\|\tilde{\Psi}^{(\eta)}_H\|^+_{1\to 1}\leq2^3(2-\frac{2}{d})\frac{\eta}{d^2}$ and that when $\|\tilde{\Psi}^{(\eta)}_H(\tau)\|_1$ achieves equality, then
    \begin{align}
\tilde{\Psi}^{(\eta)}_H(\tau)=\tilde{\Psi}^{(\eta)}_H(\tau_1\otimes\tau_2\otimes\tau_3\otimes\tau_4)\end{align}where $\tau_i\in\mathcal{D}(\mathcal{H}_i)$ are the marginals of $\tau$.
    Hence if there exists a state $\tau\in\mathcal{D}(\mathcal{H}_1\otimes\mathcal{H}_2\otimes\mathcal{H}_3\otimes\mathcal{H}_4)$ that is optimal, then its marginals are optimal w.r.t. the individual tensor factor channels. 
\end{lemma}
Again, notice that for $0<\eta<1$ this constructed map can be written as a difference of entanglement-breaking channels, up to a normalization.
\begin{proof}[Proof of Lemma \ref{lemma:1to1correct}]
Notice that
\begin{align}
 \Phi_H(\rho)-|0\rangle\langle0| &= \left(\frac{1}{2}\Tr[H\rho]|0\rangle\langle0|+\Tr[(\1-\frac{1}{2}H)\rho]|1\rangle\langle1|\right)-|0\rangle\langle0| \\&= \Tr[(\1-\frac{1}{2}H)\rho](|1\rangle\langle1|-|0\rangle\langle0|)\,\\
 &\equiv\Tr(M^{(4)}\rho)(|1\rangle\langle 1|-|0\rangle\langle 0|),
\end{align} 
is a measure and prepare map with one measurement and one (non-positive) output operator. Moreover, if the output trace norm is $2$, which is maximal, then the input lies in the kernel of $H$. The same holds for $\Phi_\text{swap}-|1\rangle\langle1|\Tr(\cdot )$ and $\Phi_\text{cube}-|0\rangle\langle0|\Tr(\cdot )$, where an optimal output trace norm implies the same properties on the input as were discussed in their respective definitions:
\begin{align}
&\Phi_{\text{swap}}(\rho)-|1\rangle\langle 1|= \Tr\bigg[\frac{1}{2}\Big(\1+\sum_{ij}|i\rangle\langle j|\otimes|j\rangle\langle i|\Big)\rho\bigg](|0\rangle\langle 0|-|1\rangle\langle 1|) \equiv \Tr(M^{(2)}\rho)(|0\rangle\langle 0|-|1\rangle\langle 1|)\\
&\Phi_{\text{cube}}(\rho)-|0\rangle\langle0|=\Tr(M\rho)(|1\rangle\langle 1|-|0\rangle\langle 0|)\equiv \Tr(M^{(3)}\rho)(|1\rangle\langle 1|-|0\rangle\langle 0|)\\
&\Phi_{\text{trace}}^{(\eta)}(\rho)-\frac{\1_{1_1}}{d}=\frac{\eta}{d^2}\Big(\rho_{1_1}-\frac{{\1}_{1_1}}{d}\Big)\,.
\end{align}
Therefore, we can write
\begin{align}
   - \tilde{\Psi}^{(\eta)}_H(\tau) &= p_{234}\frac{\eta}{d^2}\left(\tau_{1_1|234}-\frac{\1_{1_1}}{d}\right)\otimes(|1\rangle\langle 1|-|0\rangle\langle 0|)^{\otimes 3},
\end{align} 
with $p_{234}=\Tr_{234}[\tau_{234}(M_{}^{(2)}\otimes M_{}^{(3)}\otimes M_{}^{(4)})]\leq1$ and $\tau_{1_1|234}=\tr_{1_2234}[\tau_{1234}(\1_1\otimes M^{(2)}\otimes M^{(3)}\otimes M^{(4)})]$ the associated post-measurement state. 
By Claim \ref{claimm3.19}, we have
\begin{align}
\|\tilde{\Psi}^{(\eta)}_H(\tau)\|_1= p_{234}\frac{\eta}{d^2} \left\|\tau_{1_1|234}-\frac{\1_{1_1}}{d}\right\|_1\||1\rangle\langle 1|-|0\rangle\langle 0|\|_1^3 \leq 2^3\left(2-\frac{2}{d}\right)\frac{\eta}{d^2},
\end{align}
where we have equality in the last inequality iff $\tau_{1_1|234}= \tau_{1_1}$ is pure and iff $p_{234}=1$. This, analogous to in the proof of \cref{lemma:MainCorrect}, implies that  
\begin{align}
   \tilde{\Psi}^{(\eta)}_H(\tau) 
   &= \ \tilde{\Psi}^{(\eta)}_H(\tau_1\otimes\tau_2\otimes\tau_3\otimes\tau_4)\,.
\end{align} 
\end{proof}

With this we can now, similarly to the previous section, prove the $\NP$-hardess of computing the $1^+\to1$ norm for differences of channels.

\begin{proof}[Proof of Theorem \ref{thm:thm1to1}]
    We will show that deciding whether the $\|\Phi_1-\Phi_2\|^+_{1\to 1}$, for some specific differences of CPTP maps $\Phi_1,\Phi_2$, is larger than some constant is equivalent to deciding whether an instance of the 2-Out-of-4 SAT problem is satisfiable. Given an instance of 2-Out-of-4-SAT, fix some $0<\eta\leq d^2$ and take $\Phi^{(\eta)}_\text{trace}, \Phi_{\text{swap}}, \Phi_{\text{cube}}$ and $\Phi_{H}$ as defined above. Now we consider, as in Lemma \ref{lemma:1to1correct}, $\tilde{\Psi}_H:=(\Phi^{(\eta)}_\text{trace}-\frac{\1_{1_1}}{d}\Tr(\cdot ))\otimes(\Phi_\text{swap}-|1\rangle\langle 1|\Tr(\cdot ))\otimes(\Phi_\text{cube}-|0\rangle\langle 0|\Tr(\cdot ))\otimes(\Phi_H-|0\rangle\langle 0|\Tr(\cdot))$ which is by linearity defined on all of $\mathcal{B}(\mathcal{M})$, where $\mathcal{M}:=\otimes_{i=1}^4\mathcal{H}_i$ where all $\mathcal{H}_i$ are copies of $\mathcal{H}=\mathcal{K}\otimes\mathcal{K}$.
We consider the problem of computing the norm of $\tilde{\Psi}_H\circ\Lambda_4$.
First of all, if the 2-Out-of-4 SAT problem is satisfiable, then there exists a pure state $\tilde{\psi}\in\mathcal{D}(\mathcal{K})$ of form \eqref{def:VectorSpecialForm}, such that $\|\tilde{\Psi}_H\circ \Lambda_4((\tilde{\psi}\otimes\tilde{\psi})^{\otimes 4})\|_1=\|\tilde{\Psi}_H(\tilde{\psi}\otimes\tilde{\psi})^{\otimes 4})\|_1=2^3(2-\frac{2}{d})\frac{\eta}{d^2}$ is optimal and hence $\|\Psi_H\circ\Lambda_4\|^+_{1\to 1}=2^3(2-\frac{2}{d})\frac{\eta}{d^2}$.
On the other hand if $\|\Psi_H\circ\Lambda_4\|^+_{1\to 1}=2^3(2-\frac{2}{d})\frac{\eta}{d^2}$, then there exists a state $\tau$, such that $\|\tilde{\Psi}_H(\Lambda_4(\tau))\|_1=2^3(2-\frac{2}{d})\frac{\eta}{d^2}$. 
By Lemma \ref{lemma:1to1correct} it follows that
\begin{align}
    \tilde{\Psi}_H(\Lambda_4(\tau))&= \tilde{\Psi}_H(\tilde{\tau}^{\otimes 4})  \\&=(\Phi^{(\eta)}_\text{trace}(\tilde{\tau})-\frac{\1_{1}}{d})\otimes(\Phi_{\text{swap}}(\tilde{\tau})-|1\rangle\langle1|)\otimes(\Phi_{\text{cube}}(\tilde{\tau})-|0\rangle\langle0|)\otimes(\Phi_{H}(\tilde{\tau})-|0\rangle\langle0|),
\end{align} 
where we denote with $\tilde{\tau}$ the identical local marginals of $\Lambda_4(\tau)$.
So, exactly as in the proof of Theorem \ref{thm:1top.hardness}, we find a state $\tilde{\tau}\in\mathcal{D}(\mathcal{H})$ that is optimal w.r.t all $\Phi^{(\eta)}_\text{trace}-\frac{\1_1}{d},\Phi_\text{swap}-|1\rangle\langle1|,\Phi_\text{cube}-|0\rangle\langle0|,\Phi_H-|0\rangle\langle0|$ in trace norm, which by the construction of these implies that it has to be of the form $\tilde{\psi}\otimes\tilde{\psi}$, where the pure state $\tilde{\psi}$ is of the form \eqref{def:VectorSpecialForm} and satisfies the 2-Out-of-4 SAT problem. 
Hence we have shown that $\|\tilde{\Psi}_H\circ\Lambda_4\|_{1\to 1}= 2^3(2-\frac{2}{d})\frac{\eta}{d^2}$ is equivalent to being in the ``Yes'' instance of the 2-Out-of-4-SAT problem. Now by integrality of the problem, analogously as in the proof of \cref{thm:1top.hardness}, there exists a $\epsilon>\frac{1}{\poly(d)}$, s.t. we are in a ``Yes'' instance when $\|\tilde{\Psi}_H\circ\Lambda_4\|_{1\to 1}> 2^3(2-\frac{2}{d})\frac{\eta}{d^2}-\epsilon$ and else in the ``No'' instance when $\|\Psi_H\circ\Lambda_4\|_{1\to 1}\leq 2^3(2-\frac{2}{d})\frac{\eta}{d^2}-\epsilon$.
Since this works for any $0<\eta\leq d^2$ and for $0<\eta<1 \ \tilde{\Psi}_H$ is a difference of entanglement-breaking CPTP maps, the claim in the entanglement-breaking setting follows.
\end{proof}

\subsection{Efficient algorithm in the non-hypercontractive region}\label{sec:q.Boyds}

We now consider the non-hypercontractive case with $1 \leq p \leq q \leq \infty$. For the classical case (i.e., $\ell_q \to \ell_p$), Boyd's algorithm \cite{Boyd.1974} efficiently computes these norms. We show that analogous results also hold for the matrix case (i.e., $\cS_q \to \cS_p$).

\begin{proposition}\label{thm:non.hypercontractive.concavity}
For a linear CP map $\Phi:\mathcal{B}(\mathcal{H})\to \mathcal{B}(\mathcal{K})$, 
$\|\Phi\|_{q\to p}$  is given as a convex optimization problem over quantum states when $1\leq p\leq q \leq \infty$. 
    \label{thm:HeuristicQBoyds}
\end{proposition}
\begin{proof}
    We show that the problem amounts to maximizing a concave function over a convex set. We first remark that by complete positivity, $\|\Phi\|_{q\to p}=\|\Phi\|_{q\to p}^+$, see \cite{Watrous.2004}. Now we may rewrite
    \begin{align}
        \|\Phi\|^p_{q\to p}=\max_{\omega\geq 0, \|\omega\|_q\leq 1} \|\Phi(\omega)\|^p_p &= \max\{\Tr[\Phi(\omega)^p]|\Tr[\omega^q]\leq 1, \omega\geq 0\} \\&\leq \max\{\Tr[\Phi(\tau^{\frac{1}{q}})^p]|\Tr[\tau]\leq 1,\tau\geq 0\} \\ &=\max_{\tau\geq 0, \|\tau\|_1\leq1}\Tr[\Phi(\tau^{\frac{1}{q}})^p].
    \end{align}
    The concavity of the map $\tau\mapsto \Tr[\Phi(\tau^{\frac{1}{q}})^p]$ for $p\leq q$ follows from \cite[Theorem 4.1 (1)]{Hiai.2013}.
\end{proof}

The above formulation can be used to get polynomial-time algorithms. Next, we present two different algorithms for computing such mixed-Schatten norms. The first one goes via the ellipsoid method \cite{book.Grotschel.1993,book.Bubeck.2015}.

\begin{theorem}\label{thm:ellipsoid.result}
Let $\Phi:\C^{d\times d}\to \C^{d\times d}$ be a CP map such that $1\leq p\leq q\leq\infty$, then there exists an explicit polynomial algorithm which computes $\|\Phi\|^p_{q\to p}$ to multiplicative error $(1+\epsilon)$ in time $\mathcal{O}(d^7\log\frac{\sqrt{d}}{\epsilon})$.  
\end{theorem}

Note that this yields a $1+\frac{\epsilon}{p}+\mathcal{O}(\epsilon^2)$ multiplicative approximation of $\|\Phi\|_{q\to p}$ in the above quoted time.

\begin{proof}
We prove this by showing that the convex optimization problem in \cref{thm:HeuristicQBoyds} can be efficiently solved using the ellipsoid method. 
In the following denote $F(X):=-\Tr[\Phi(X^{\frac{1}{q}})^p]$, which by \cref{thm:HeuristicQBoyds} is a convex function on $\mathcal{D}(\C^d):=\{X\in\C^{d\times d}| X\geq 0, \|X\|_1\leq 1\}$, the positive trace-normalized $d\times d$-matrices. It satisfies $F(\cD(\C^d))\subset [-\|\Phi\|^p_{q\to p},0]$. 
Following the black box approach from \cite[Section 2.2]{book.Bubeck.2015} the ellipsoid algorithm outputs a $1+\epsilon$ multiplicative approximation of $\|\Phi\|^p_{q\to p}$ using $t=2d^4\log\frac{4\sqrt{d}}{\epsilon}$ steps, with each step requiring access to a separation oracle and a sub-gradient oracle. The $\sqrt{d}$ factor comes from the fact that $\|\cdot\|_2\leq \|\cdot\|_1\leq \sqrt{d}\|\cdot\|_2$. We note here that although we optimize over $\C^{d\times d}$ matrices instead of $\R^d$ vectors in e.g. \cite{book.Bubeck.2015}, the proof goes through, up to a squaring of the dimension $d\mapsto d^2$.
Hence the proof is complete after showing that both these oracles can be implemented efficiently. In \cref{app:ellipsoid.method.details} we construct both these oracles as algorithms with runtimes bounded by $\mathcal{O}(d^3)$, making the overall complexity of this algorithm $\mathcal{O}(d^7\log\frac{\sqrt{d}}{\epsilon})$.
\end{proof}

We now consider a simpler algorithm for this problem based on Boyd's non-linear power iteration algorithm \cite{Boyd.1974}, see also~\cite{Gautier.2020, Bhaskara.2011}. In the preprint \cite{Shahverdikondori.2022}, Boyd's algorithm was extended to norms $\cS_q \to \cS_p$, i.e., the ones we analyze here, where the authors prove that for any CP map $\Phi$ such that $\Phi(X)>0 \quad \forall X\geq0, X\neq 0$, and all $1\leq p\leq 2\leq q\leq \infty$, $\|\Phi\|_{q\to p}$ is  efficiently computable.
The algorithm of \cite{Shahverdikondori.2022} is based on a fixed point iteration, for which we provide a concrete runtime, both in the case when the CP map is non-trivially bounded from below, i.e. $\Phi(\cdot)\geq \epsilon\1\Tr[\cdot]$ for some $\epsilon>0$, and when it is not. The latter had not been explicitly considered before, as far as we are aware. 
We note that in \cite{Shahverdikondori.2022} the authors claim convergence for $1\leq p\leq q\leq\infty$, however, their proof turns out to only work for $p\le 2\le q$. In \cref{lem:contraction.counterexample} below, we provide a counter example showing this restriction cannot be removed in general, although it does not preclude convergence of the iteration algorithm outside the region $1\leq p\leq 2\leq q\leq\infty$ for a possibly restricted class of maps. We note that in the classical mixed matrix norm case, the region $1\leq p\leq 2\leq q\leq\infty$ coincides with the region where constant factor approximations are known to exist for any matrix \cite{Daureen.2007, Bhattiprolu.2019}, which is not the case for other $p\leq q$.

\begin{theorem}[Extended quantum Boyd's algorithm]\label{thm:Quantum.Boyds}
    Let $\Lambda:\C^{d\times d}\to \C^{d\times d}$ be a positive map s.t. $c\leq\|\Lambda\|_{\infty\to\infty}\leq 1$. Then, for $1\leq p\leq 2\leq q\leq \infty$ there exists an efficient classical algorithm with a runtime 
    $T=\mathcal{O}\left(d^{6}\frac{c^3}{\epsilon^3}\log(\poly\left(d,\frac{c}{\epsilon}\right))\right)$ that approximates $\|\Lambda\|^+_{q\to p}$ up to multiplicative error $1+\epsilon$.
    If further $\Lambda$ is TP, then the runtime reduces to $T=\mathcal{O}\left(d^{5.5}\frac{1}{\epsilon^2}\log\left(\poly\left(d,\frac{1}{\epsilon}\right)\right)\right)$.
\end{theorem}
\begin{remark}

We note that this algorithm computes $\|\Lambda\|^+_{q\to p}$, which for CP maps is equal to $\|\Lambda\|_{q\to p}$. For positive maps this equality is surprisingly still an open question, i.e. whether $\|\Lambda\|_{q\to p}= \|\Lambda\|^+_{q\to p}$. We also note that we do not require the positivity improving assumption present in \cite{Shahverdikondori.2022}. Last but not least,  classically, the dependence on $\epsilon$ scales as $\epsilon^{-1}$, see \cite[Theorem 3.7]{Bhaskara.2011}. It would hence be interesting to see whether the same dependency can be obtained in the matrix setting. 
\end{remark}

\cref{thm:Quantum.Boyds} will follow from the following more canonical formulation of quantum Boyds algorithm \cref{thm:strict.positive.q.Boyds}  via \cref{lem:Schatten.mixed.continuity.bound} which is essentially continuity bounds for Schatten norms. The former, \cref{thm:strict.positive.q.Boyds} follows from an explicit analysis of the algorithm for positivity improving maps from \cite{Shahverdikondori.2022}, inspired by \cite{Gautier.2020, Bhaskara.2011}, which we defer to \cref{Hilbertmetric}.
 
\begin{lemma}[Quantum Boyd's algorithm]\label{thm:strict.positive.q.Boyds}
     Let $\Lambda:\C^{d\times d}\to \C^{d\times d}$ be a positivity improving map that satisfies, for some $N>0$, $\Lambda(\omega)\geq \frac{\1}{Nd}\Tr[\omega]$ $\forall \omega\geq 0$ and $\|\Lambda\|_{\infty\to\infty}\leq 1$. Then  for $1\leq p\leq 2\leq q\leq \infty$ there exists an efficient classical algorithm with a runtime 
    $T=\mathcal{O}((Nd)^{1+\frac{p}{q-1}}\log(N,d,\frac{1}{\epsilon}))=\tilde{\mathcal{O}}(N^3d^3)$, that approximates $\|\Lambda\|_{q\to p}$ up to multiplicative error $1-\epsilon$. The algorithm also outputs an approximate optimizer $\omega_T$, s.t. $\|\Lambda\|_{q\to p}(1-\epsilon)\leq \|\Lambda(\omega_T)\|_{p}\leq \|\Lambda\|_{q\to p}$.
Further if the map $\Lambda$ is assumed to be trace preserving,  the runtime reduces to \\ $T=\mathcal{O}\left(N^{1+\frac{p-1}{q-1}}d^{\frac{p-1}{q-1}+\frac{1}{q}}\sqrt{d}\log(N,d,\frac{1}{\epsilon})\right)=\tilde{\mathcal{O}}(N^2d^2\sqrt{d}).$
\end{lemma}

The proof of \cref{thm:Quantum.Boyds} follows with the following Lemma.

\begin{lemma}\label{lem:Schatten.mixed.continuity.bound}
Let $\Lambda\neq 0$ be a positive map s.t. $c\leq \|\Lambda\|_{\infty\to\infty}\leq 1$, $\epsilon>0$. Set 
\begin{align}
    \Lambda_\delta(\cdot):=(1-\delta)\Lambda(\cdot)+\delta\frac{\1}{d}\Tr[\cdot],
\end{align}
then $\|\Lambda\|^+_{q\to p}$ can be approximated to multiplicative error $(1+\epsilon+\mathcal{O}(\epsilon^2))$ if for $\delta=\frac{\epsilon}{4dc}$, 
$\|\Lambda_\delta\|_{q\to p}$ can be approximated to multiplicative error $(1+\frac{\epsilon}{2})$. If $\Lambda$ is trace-preserving, then $\delta=\frac{\epsilon}{4d}$ suffices.
\end{lemma}
The proof of \cref{lem:Schatten.mixed.continuity.bound} can be found in \cref{lemcontinuity.bound}. It can be easily checked that if $\Lambda$ satisfies the assumption in \cref{thm:Quantum.Boyds} then $\Lambda_\delta$ satisfies the ones of \cref{thm:strict.positive.q.Boyds} with $N\geq\frac{1}{\delta}$. Further if $\Lambda$ is trace-preserving, so is $\Lambda_\delta$.
Hence this directly implies the runtimes claimed in \cref{thm:Quantum.Boyds}, when replacing $N$ by $\frac{1}{\delta}=\mathcal{O}\left(\frac{d}{\epsilon}\right)$ and $\epsilon$ by $\frac{\epsilon}{2}$. Next section is devoted to the description of the algorithm claimed in \cref{thm:strict.positive.q.Boyds}. The proof of the latter is explained in \cref{Hilbertmetric}.

\subsubsection{Quantum Boyd's algorithm}

Let a linear map $\Lambda:\mathcal{B}(\mathcal{H})\to \mathcal{B}(\mathcal{\cH})$ between finite-dimensional Hilbert spaces as in the statement of Lemma \ref{thm:strict.positive.q.Boyds} be given. Without loss of generality, we may assume the input and output dimensions to be equal. We seek to find the maximum of the positive homogeneous function $f:\cB^+(\mathcal{H})\to \mathbb{R}_+$ on the set $\cB^+(\cH)$ of positive matrices on $\cH$:
\begin{align}
    f(\omega):=\frac{\|\Lambda(\omega)\|_p}{\|\omega\|_q}.
\end{align} 
This is done by rewriting this optimization problem as the one of finding a fixed point to a map which is shown to be contractive in projective Hilbert metric. 
In what follows, we fix $\dim\mathcal{H}=d$.

\begin{definition}\label{def:S}
Given $1\leq p\leq  2\leq  q\leq\infty$, we define the non-linear map:
\begin{align}
    S : \cB^+(\mathcal{H})\to \cB^+(\mathcal{H}),\qquad \omega\mapsto \Lambda^*\left(\Lambda(\omega)^{p-1}\right)^{\frac{1}{q-1}}  \,.
\end{align}    
\end{definition}
Given this map we may now describe the algorithm as consisting of the following steps.
\begin{enumerate}
    \item Initialize $\omega_0:=\frac{\1}{\|\1\|_q}$;
    \item Apply the non-linear map $S(\cdot)$;
    \item re-normalize and set $\omega_{i+1}:=\frac{S(\omega_i)}{\|S(\omega_i)\|_q}$;
    \item iterate (2) and (3) $T$ times and output $\omega_T$\,.
\end{enumerate}

The derivation and analysis of the convergence behavior may be found in \cref{app:Boyds.analysis}. It is rather lengthy, but effectively based on showing contractivity of $\omega\mapsto\frac{S(\omega)}{\|S(\omega)\|_q}$ in projective Hilbert metric, by appropriate applications of operator monotonicity of the map $t\mapsto t^r$ for $r\in[0,1]$ \cite{book:Bhatia.1997} and applications of the positivity improving property, see also \cite{Shahverdikondori.2022, Bhaskara.2011}.
We present here a simplified version.

\subsubsection{Proof of \cref{thm:strict.positive.q.Boyds}}\label{Hilbertmetric}

The proof of \cref{thm:strict.positive.q.Boyds} makes use of the projective Hilbert metric, which we introduce here. For more on its properties and uses in quantum information theory, see e.g. \cite{Reeb.2011}.
\begin{definition}[Projective Hilbert metric]
The projective Hilbert metric between two positive matrices $A,B\in \C^{n\times n}$ is defined as
    \begin{align}
d_H(A,B):=\begin{cases}
    \log\|A^{-1/2}BA^{-1/2}\|_\infty+\log\|B^{-1/2}AB^{-1/2}\|_\infty \quad &\text{ if } A\sim B, \\
    \infty \quad &\text{ else },
\end{cases}
    \end{align}
    where $A \sim B$ iff $\exists c,C$ s.t. $0<c\leq C <\infty$ and $cB\leq A \leq C B$, i.e. their supports coincide. The inverses here are the generalized Moore-Penrose inverses. 
\end{definition}

\begin{definition} Given a (not necessarily linear) map $\Phi:\C^{d\times d}\to \C^{m\times m}$, its projective Hilbert metric contraction coefficient is defined as
\begin{align}\label{equ:def.contraction.coeff}
    \kappa(\Phi):=\inf\left\{C > 0 :\, d_H(\Phi(A),\Phi(B)) \leq C\,d_H(A,B)\quad \forall A,B\ge 0\right\}.
\end{align}
\end{definition}
For linear positive maps, the contraction coefficient $\kappa(\Phi)$ can be re-expressed as follows:
\begin{lemma}[Birkhoff-Hopf \cite{Eveson.Nussbaum.1995}]
Let $\Lambda$ be a positive map, then
\begin{align}
    \kappa(\Lambda)=  \tanh\left(\frac{1}{4}\Delta(\Phi)\right) \leq 1,
\end{align} where $\Delta(\Phi):=\sup\{d_H(\Phi(A),\Phi(B))|A,B>0\}$ is the projective diameter of $\Phi$.
\end{lemma}
For the power maps appearing in $S$ we show that $\kappa(A\mapsto A^{p-1})\kappa(A\mapsto A^{\frac{1}{q-1}})=\frac{p-1}{q-1}$ iff $1\leq p\leq 2\leq q\leq\infty$, which follows from
\begin{lemma}\label{lem:contraction.counterexample}
$\kappa(A\mapsto A^r)=r$ if  $0\leq r\leq1$ and $\kappa(A\mapsto A^r)>r$ for $r > 1$.
\end{lemma}

\begin{proof}[Proof of \cref{lem:contraction.counterexample}]
Let $0\leq r\leq1$, then it follows either via interpolation and the observation that for $r=0,1$ it clearly holds, or via operator monotonicity of $t\mapsto t^r$, see also \cite[Proof of Theorem IX.2.10]{book:Bhatia.1997}. 
For $0\leq r\leq 1$, by monotonicity $B\leq \lambda^\frac{1}{r}A\Rightarrow B^r\leq\lambda A^r$ and hence
\begin{align}
    \|A^\frac{-1}{2}BA^\frac{-1}{2}\|_\infty^r=\inf\{\lambda\geq 0| B\leq \lambda^\frac{1}{r}A\}\geq \inf\{\lambda\geq 0| B^r\leq \lambda A^r\} = \|A^{\frac{-r}{2}}B^rA^{\frac{-r}{2}}\|_\infty.
\end{align} 
Hence $\kappa(A\mapsto A^r)\le r$. Achievability follows from the classical achievability for diagonal matrices.
For $1\leq r$, we first notice that the opposite bound holds, by the above argument for $0\leq \frac{1}{r}\leq 1$.
\begin{align}
    \|A^{-r/2}B^rA^{-r/2}\|_\infty = \inf\{\lambda\geq 0|B^r\leq \lambda A^r  \} \geq \inf\{\lambda\geq 0| B\leq \lambda^\frac{1}{r} A\} = \|A^{-1/2}BA^{-1/2}\|_\infty^r.
\end{align} 
To see that there exists a strict inequality take two non-commuting operators $A,B$ e.g., $d=2, A=\text{diag}(1,2)$ and $B=HAH$, where $H$ is the Hadamard matrix. Then one can check that
\begin{align}
    \|A^\frac{-1}{2}BA^\frac{-1}{2}\|_\infty^r &=  \|B^\frac{-1}{2}AB^\frac{-1}{2}\|_\infty^r = 2^{-4r}(15+\sqrt{97})^r, \\
    \|A^\frac{-r}{2}B^rA^\frac{-r}{2}\|_\infty &=  \|B^\frac{-r}{2}A^rB^\frac{-r}{2}\|_\infty = \\ &= 2^{-2-2r}(1+2^r+2^{2r}+2^{3r}+(2^r-1)\sqrt{1+2^{4r}+2^{2+r}+5\cdot2^{1+2r}+2^{2+3r}}),
\end{align} where the latter is strictly larger than the former for any $r>1$.
\end{proof}
\begin{remark}\label{rem:kappa.greater.1}
In the classical case, the equality in \cref{lem:contraction.counterexample} is true for all values of $r$ since the map $t\mapsto t^r$ is monotone for any $0\leq r$, however, it is only operator monotone for $0\leq r\leq 1$. In the matrix case, the equality fails to hold for $r>1$.
To the best of our knowledge, the exact value of $\kappa(A\mapsto A^r)$ remains unknown for $1<r$. 
\end{remark}
Now it is clear that $\kappa(S)\leq \kappa(\Lambda^*)\kappa(\Lambda)\kappa(A\mapsto A^{p-1})\kappa(A\mapsto A^{\frac{1}{q-1}})=\kappa(\Lambda^*)\kappa(\Lambda)\frac{p-1}{q-1}$, where the last equality holds when $p\leq 2\leq q$. As observed in \cite{Shahverdikondori.2022} having a map $\Lambda$ be positivity improving implies that $\kappa(\Lambda)<1$. It is easy to show that if $\Lambda$ is positivity improving, then so is its adjoint, see \cref{rmk:positivity improving}. Getting a quantitative bound on the contractivity remains however more challenging.

\medskip

In \cref{app:Boyds.analysis} we first show that the invariant states of $\omega\mapsto S(\omega)/\|S(\omega)\|_q$ correspond to the optimizers of $f$ (see also \cite[Lemma 3.1]{Shahverdikondori.2022}):

\begin{lemma}
\label{lem:1}
    The stationary points of $f$ on $\cB^+(\mathcal{H})$ are the fixed points of $\omega\mapsto S(\omega)/\|S(\omega)\|_q$.
\end{lemma}

Next, in \cref{cor:unique.optimizer} we show that the positivity improving property implies that the optimizer is unique, i.e. $\exists! \tilde{\omega}$, s.t. $\|\Lambda(\tilde{\omega})\|_p=\|\Lambda\|_{q\to p}$.
Then, to show that the algorithm converges we identify a Cauchy sequence in \cref{lem:BoydsMonotonicity} w.r.t the projective Hilbert metric. Then in \cref{lem:5} and \cref{lem:eigenvalue.lower.bound} we give a concrete bound on the contraction coefficient, leading to the following Lemma, whose proof can be found in \cref{app:Boyds.analysis}: 
\begin{lemma}\label{lemcontractivitykappa}
Let $\Lambda$ be a positive map s.t. $\|\Lambda\|_{\infty\to\infty}\leq1$ and that is positivity improving i.e. $\Lambda(\rho)\geq \frac{\1}{Nd}\Tr[\rho]$ holds for all $\rho\geq0$. Then,
\begin{align}
    \kappa(S)\leq \begin{cases}
        &1-\frac{1}{N^2d^2\sqrt{d}} \quad \text{if } \Lambda \text{ is trace-preserving}, \\  &1-\frac{1}{N^3d^3} \quad \text{ else}.
    \end{cases}
\end{align}
\end{lemma}
 To get a convergence time out of the contractivity result of \cref{lemcontractivitykappa}, we require a bound on the initial distance $d_H(\omega_0,\omega_1)$, which we give in \cref{cor:initial.distance}. Specifically we show
\begin{lemma}\label{lem:initial.distance.boyds}
Let $\Lambda$ be as in \cref{thm:Quantum.Boyds}, i.e. $\Lambda(\rho)\geq \frac{\1}{Nd}\Tr[\rho]$ is positivity improving and s.t. $\|\Lambda\|_{\infty\to\infty}\leq 1$, then it holds that for $p\leq q$
\begin{align}
    d_H(\omega_0,\omega_1)=d_H(\1,S(\1))\leq \begin{cases}
       & \log N \quad \text{ if } \Lambda \text{ is TP,} \\ &\log(dN^2) \quad \text{ else.}
    \end{cases} 
\end{align}
\end{lemma}
We also require a way to relate closeness in (exponential) projective Hilbert metric to fidelity of the approximation of $\|\Lambda\|_{q\to p}$. 
To do this we use the following property of the projective Hilbert metric.
\begin{claim}\label{claimdHrhosig}
Let $\rho, \sigma\in\cB^+(\cH)$ be two invertible operators (or operators with the same support) over a finite dimensional Hilbert space $\cH$. If $\|\rho\|_q=\|\sigma\|_q\neq 0$ for some $q\in[1,\infty]$, then
\begin{align}
    \rho \leq \exp(d_H(\rho,\sigma))\sigma, \text{ and } \sigma \leq \exp(d_H(\rho,\sigma))\rho
\end{align}
\end{claim}
\begin{proof}
Notice that $\rho\leq\lambda\sigma$ holds, in particular, for $\lambda:=\|\sigma^{-1/2}\rho\sigma^{-1/2}\|_\infty$. Since $\rho,\sigma$ have the same Schatten $q$ norm it follows that $\lambda\geq 1$, because $\|\cdot\|_q$ respects the matrix order, i.e. $\rho\leq\lambda\sigma \Rightarrow \|\rho\|_q\leq \|\lambda\sigma\|_q$. 
The same obviously holds when exchanging $\rho$ and $\sigma$ and so $\mu:=\|\rho^{-1/2}\sigma\rho^{-1/2}\|_\infty\geq 1$. Thus we have
\begin{align}
    \rho \leq \lambda \sigma \leq \lambda\mu\sigma = \exp(d_H(\rho,\sigma))\sigma.
\end{align} The second statement follows when exchanging $\rho$ for $\sigma$.
\end{proof}

To prove \cref{thm:strict.positive.q.Boyds}, we combine the above results. 
\begin{proof}[Proof sketch of \cref{thm:strict.positive.q.Boyds}]
Denote the state after $T$ iterations of the algorithm as $\omega_T$.
On the one hand obviously $\|\Lambda(\omega_T)\|_p\leq \|\Lambda\|_{q\to p}$ holds. On the other we have by \cref{claimdHrhosig} that
$\tilde{\omega} \leq \exp(d_H(\omega_K,\tilde{\omega})) \omega_K$, where $\tilde{\omega}$ is the unique positive optimizer (\cref{cor:unique.optimizer}). Then, by \cref{lem:1}, we get
\begin{align}
    d_H(\omega_T,\tilde{\omega})\leq \sum_{j=T}^\infty d_H(\omega_j,\omega_{j+1})\leq d_H(\omega_0,\omega_1)\sum_{j=T}^\infty \kappa(S)^j \leq d_H(\omega_0,\omega_1)\,\frac{\kappa(S)^T}{1-\kappa(S)}\overset{!}{\leq} \epsilon\,,
\end{align} which holds for $T\geq \frac{1}{\log(\kappa(S))}\log(\frac{(1-\kappa(S))\epsilon}{d_H(\omega_0,\omega_1)})= \mathcal{O}\left(\frac{1}{(1-\kappa(S))}\log\frac{d_H(\omega_0,\omega_1)}{(1-\kappa(S)) \epsilon}\right)_{\kappa(S)\to 1}$.
The above inequality now directly implies that
\begin{align}
    \|\Lambda(\omega_T)\|_p \geq e^{-\epsilon}\|\Lambda(\tilde{\omega})\|_p= e^{-\epsilon}\|\Lambda\|^+_{q\to p}.
\end{align} 
Since $e^{-\epsilon}=1-\epsilon+\mathcal{O}(\epsilon^2)$ the claim in \cref{thm:strict.positive.q.Boyds} follows when plugging in the bounds claimed in \cref{lemcontractivitykappa} and \cref{lem:initial.distance.boyds}.
\end{proof}


\section{Mixed completely bounded norms}\label{sec:mixed.cb.Schatten}

In this section, we study the completely bounded version of $q \to p$ norms. For $q=p=1$, \cref{thm:thm1to1} shows that computing the optimal one shot indistinguishability of two quantum channels is an $\NP$-hard problem. In contrast, allowing arbitrary large auxiliary systems, as governed by the diamond norm
\begin{align}
    \|\Phi-\Psi\|_\diamond:=\sup_n\|\id_n\otimes(\Phi-\Psi)\|_{1\to 1},
\end{align} 
leads to efficient algorithms based on a semidefinite programming, as shown in \cite{Watrous.2012}.
Hence there is a clear separation between the $1\to 1$ Schatten norm problem and its completely bounded norm variant. In this section we show that this extends to the $cb, 1\to p$ case.

\subsection{Efficient computability of $cb,1\to p$ norms for linear maps}\label{sec:cb.1top.computability}

In contrast to the non-cb norms here we will show that the cb ones are given as convex optimization problems over quantum states. 
The following theorem is one of the main results of this section.

\begin{theorem}[Concavity for $cb,1\to p$-norm of linear maps]\label{thm:efficient.cb.1top}
Given a (not necessarily CP) linear map $\Psi:\C^{n\times n}\to\C^{m\times m}$, $\|\Psi\|_{cb,1\to p}$ and $\|\Psi\|^+_{cb,1\to p}$ can be rewritten as a convex optimization problems, in particular as the maximization of a concave function over, respectively the set of quantum states on $\C^{2n}$, or $\C^{n}$ in the positive case.
In both cases the convex optimization problems are efficiently computable.
\end{theorem}

\begin{remark}
For the case $p=1$, as previously mentioned, this is a well-known result by Watrous \cite{Watrous.2012}. Note that in that case, for hermiticity preserving maps, e.g. differences of channels, it is known that the optimization over all states and the optimization over positive states are equivalent \cite[Theorem 3.51]{book.Watrous.2018}. 

Notably our result is that it is not just easy to compute the $cb,1\to p$ norm of a quantum channel, but of any possibly non CP linear map (like a difference of quantum channels), over both, quantum states or arbitrary trace normalized operators. This is in stark contrast to the non cb norm as we showed in \cref{thm:1top.hardness} and in fact also the stabilized $1\to p$-norm (see \cref{def:stabilized.mixed.norm}) for $p>2$, since there it coincides with the regular mixed one \cite{Watrous.2004}. 
This gives the notion of completely bounded norms a computational advantage beyond its operational interpretation. 

\end{remark}

We prove this by relating the $cb,1\to p$ norm to the $(\infty,p)$ 2-indexed Schatten norm of its Choi operator, generalizing \cite[Theorem 10]{Devetak.2006} where the map was assumed CP. We then show that this norm is efficiently computable as a convex optimization problem using the ellipsoid method. This works both in the case where the initial optimization is over positive ($\|\Phi\|^+_{cb,1\to p}$) or general trace-normalized operators ($\|\Phi\|_{cb,1\to p}$). That one can find such a suitable rewriting of these mixed norms as convex optimization problems is a priori not clear, since the initial cb norm problem is an optimization over auxiliary systems of possibly unbounded dimension and convex in the input states over which one maximizes.
\begin{lemma}\label{lem:cb.norm.Choi.norm}
Let $\Psi:\cS_1(\cH)\to \cS_p(\cK)$ be a linear map. Then it holds that
\begin{align}
    \|\Psi\|_{cb,1\to p}=\|J_\Psi\|_{(\infty,p)},
\end{align}
where the norm $(\infty,p)$ is defined in \eqref{equ:def-multi-index}.
\end{lemma}

\begin{proof}
Let $d:=\dim\cH$, then via a standard Schmidt decomposition argument, see e.g. \cite[Lemma A.1]{Fawzi.2025}, the cb norm under consideration can be rewritten as
\begin{align}
    \|\Psi\|_{cb,1\to p} = \sup_{|\phi\rangle,|\varphi\rangle, \||\phi\rangle\langle\varphi|\|_1=1}\|\id_d\otimes\Psi(|\phi\rangle\langle\varphi|)\|_{\cS_1[\C^d,\cS_p(\cK)]},
\end{align} where $|\psi\rangle,|\varphi\rangle\in \C^d\otimes\cH$ are normalized states, i.e. $\langle\psi|\psi\rangle=\langle\varphi|\varphi\rangle=1$.
Any such vector can be, up to applying a local unitary on the first subsystem, written as a purification of some quantum state $\rho\in\cD(\C^d)$. Since the norm is unaffected by such local unitaries, we may assume $|\phi\rangle=|\psi_{\sqrt{\rho}}\rangle$, where $|\psi_{\sqrt{\rho}}\rangle:=(\sqrt{\rho}\otimes\1)\sum_{i=1}^d|i\rangle\otimes|i\rangle$ is a canonical purification of $\rho$. Arguing similarly for $|\varphi\rangle=|\psi_{\sqrt{\sigma}}\rangle$, we directly get that $1=\langle\psi_{\sqrt{\rho}}|\psi_{\sqrt{\rho}}\rangle=\|\rho\|_1$ and thus $\||\psi_{\sqrt{\rho}}\rangle\langle\psi_{\sqrt{\sigma}}|\|_1=\sqrt{\langle\psi_{\sqrt{\rho}}|\psi_{\sqrt{\rho}}\rangle\langle\psi_{\sqrt{\sigma}}|\psi_{\sqrt{\sigma}}\rangle}=\|\sqrt{\rho}\|_2\|\sqrt{\sigma}\|_2$.
Putting these together we get
\begin{align}
  \|\Psi\|_{cb,1\to p} &= \sup_{\underset{\rho,\sigma\geq 0, \|\rho\|_1\|\sigma\|_1=1}{\rho,\sigma\in\cS_1(\C^d)}}\|(\id_d\otimes\Psi)(|\psi_{\sqrt{\rho}}\rangle\langle\psi_{\sqrt{\sigma}}|)\|_{(1,p)} \\ 
  &= \sup_{\underset{\|\rho\|_1\|\sigma\|_1=1}{\rho,\sigma\in\cS_1(\C^d)}} \|(\sqrt{\rho}\otimes\1)J_\Psi(\sqrt{\sigma}\otimes\1)\|_{(1,p)} \\
  &= \sup_{A,B\geq 0}\|A\|_2^{-1}\|B\|_2^{-1}\|(A\otimes\1)J_\Psi(B\otimes\1)\|_{(1,p)} \\
  &= \|J_\Psi\|_{(\infty,p)},
\end{align}
where in the second to last line we replaced $\sqrt{\rho}, \sqrt{\sigma}$, by $A$, respectively $B$. The last line follows from \cite[Lemma 3.1]{Fawzi.2025}, where we remark that it suffices to restrict to positive $A,B$ by an argument as in \cite[Corollary 2.4]{Fawzi.2025}, see also \cite{Devetak.2006, Beigi.2023}. We also used that 
\begin{align}
    (\id_d\otimes\Psi)(|\psi_A\rangle\langle\psi_B|)=\sum_{ij}A|i\rangle\langle j|B\otimes\Psi(|i\rangle\langle j|)= (A\otimes\1)J_\Psi(B\otimes\1).
\end{align}
In the case of $\|\Psi\|^+_{cb,1\to p}$, one ends up with the same optimization problem in the end, just with the simplification $A=B$. We note here that when $J_\Psi\geq 0$, i.e. $\Psi$ is CP, then this recovers $\|J_\Psi\|_{(\infty,p)}$.
\end{proof}

\begin{remark}
We note here that the proof of \cref{lem:cb.norm.Choi.norm} extends when replacing the output space $\cS_p(\cK)$ by any operator space $\cX$, i.e. then proving that $ \|\Psi\|_{cb,1\to \cX}=\|J_\Psi\|_{\cS_\infty[\cH^\prime,\cX]}$ for some $\mathcal{H}\simeq\mathcal{H}^\prime$. 
\end{remark}

\cref{thm:efficient.cb.1top} will follow directly from next Lemma.
\begin{lemma}\label{lem:computing.two.indexed.Schatten.norns}
The two-indexed Schatten norm $\|X\|_{(q,p)}$ of an operator $X$ is, for any $1\leq q,p\leq\infty$, given as a convex optimization problem over quantum states.
\end{lemma}

\begin{proof} We start with the well known variational formulas for the 2-indexed Schatten norms from \cite{Devetak.2006, book:Pisier.2003}, note though that through a standard polar decomposition argument as in \cite[Corollary 2.4]{Fawzi.2025} we may restrict the supremum and infimum to positive semi-definite operators, i.e. we have for $q\leq p$, and $q\geq p$, respectively,
\begin{align}
    \|X\|_{(q,p)} &= \inf_{A,B\geq 0}\|A\|_{2r}\|B\|_{2r}\|A^{-1}XB^{-1}\|_p = \inf_{A,B\geq 0, \|A\|_1\leq 1,\|B\|_1\leq 1}\|A^{-\frac{1}{2r}}XB^{-\frac{1}{2r}}\|_p, \\
    \|X\|_{(q,p)} &= \sup_{A,B\geq 0}\|A\|^{-1}_{2r}\|B\|^{-1}_{2r}\|AXB\|_p = \sup_{A,B\geq 0, \|A\|_1\leq 1,\|B\|_1\leq 1}\|A^{\frac{1}{2r}}XB^{\frac{1}{2r}}\|_p,
\end{align} where $\frac{1}{r}=\big|\frac{1}{q}-\frac{1}{p}\big|$. Here $A^{-1}$ denotes the generalized Moore-Penrose inverse of $A$.
Rewriting
\begin{align}
    \|A^\xi XB^\xi\|_p^p=\Tr[|A^\xi XB^\xi|^p]= \Tr[(B^\xi X^*A^{2\xi}XB^\xi)^\frac{p}{2}]
\end{align} we have in the case where $q\leq p$, that
\begin{align}
    \|X\|^p_{(q,p)}= \inf_{A,B\geq 0, \|A\|_1\leq 1,\|B\|_1\leq 1}\Tr[(B^{-\frac{1}{2r}}X^* A^{-\frac{1}{r}}XB^{-\frac{1}{2r}})^\frac{p}{2}]. 
    \end{align} The map over which the infimum is taken is by \cite[Theorem 1.1]{Zhang.2020} jointly convex in $A,B$ since $\frac{p}{2}>0$ and $-1\leq-\frac{1}{r}\leq 0$. The latter follows since $0\leq \frac{1}{r}=\big|\frac{1}{q}-\frac{1}{p}\big|\leq 1$, as both $1\leq q,p$. The set over which the supremum is taken is clearly convex and compact, hence $\|X\|_{(q,p)}^p$ is given as a convex optimization problem.
Similarly in the case where $q\geq p$ we have that
\begin{align}
    \|X\|^p_{(q,p)}=\sup_{A,B\geq 0, \|A\|_1\leq 1,\|B\|_1\leq 1}\Tr[(B^\frac{1}{2r}X^*A^\frac{1}{r}XB^\frac{1}{2r})^\frac{p}{2}],
\end{align} which is a maximization over a convex compact set of a function, that due to \cite[Theorem 1.1]{Zhang.2020} is jointly concave, since, $0\leq \frac{1}{r}\leq 1$ and $\frac{p}{2}\leq \frac{r}{2}$, as $\frac{1}{r}=\frac{1}{p}-\frac{1}{q}\leq \frac{1}{p}$.
If the initial optimization is only over positive input states, then this simplifies the infimum and supremum here to $A=B$, which by joint concavity or convexity still yields a concave, or convex functional in $A$ over which, respectively, the infimum or supremum is taken. 
\end{proof}

Now \cref{lem:cb.norm.Choi.norm} and \cref{lem:computing.two.indexed.Schatten.norns}, together with a quantitative analysis of the ellipsoid method, yield \cref{thm:efficient.cb.1top}. Details can be found in \cref{app:ellipsoid.method.details}. 

\begin{theorem}\label{thm:cb.ellipsoid}
Given a linear map $\Psi:\C^{d\times d}\to \C^{d\times d}$, then $\|\Psi\|_{cb,1\to p}$ and $\|\Psi\|_{cb,1\to p}^+$ can be computed to relative error $(1+\epsilon)$ efficiently in time $\mathcal{O}\big(d^7\log\frac{\sqrt{d}}{\epsilon}\big)$.
\end{theorem}

\subsection{Efficiency of non-hypercontractive cb norms of CP maps}

We end this work with two simple observations: first, in the non-hypercontractive regime, the completely bounded norm coincides with the non-completely bounded one, at least for CP maps, which directly implies via \cref{thm:ellipsoid.result} or \cref{thm:Quantum.Boyds} that all non-hypercontractive mixed cb norms of CP maps are efficiently computable.

\begin{claim}[Theorem 13 from \cite{Devetak.2006}]\label{lem:cb.equals.non.cb}
Let $\Phi$ be a CP map and $q\geq p$, then
\begin{align}
    \|\Phi\|_{cb,q\to p} = \|\Phi\|_{q\to p}
\end{align}
\end{claim}

The second simple observation that we want to emphasize is that, like the matrix and non-completely bounded Schatten setting, the $cb,2\to 2$ of any (not necessarily CP) linear map is efficiently computable, as it reduces to the classical case:
\begin{claim}
Let $\Phi:\cS_2(\cH)\to \cS_2(\cK)$ be a linear map, then it holds that
\begin{align}
    \|\Phi\|_{cb,2\to 2}= \|\Phi\|_{2\to 2}= \|J_\Phi^{\Gamma}\|_\infty.
\end{align}
\end{claim}
\begin{proof}
This is a simple consequence of the fact that for $p=q=2$ the completely bounded norm coincides with the non-cb due to \cite[Theorem 4]{Watrous.2004} and the fact that $\|\Phi\|_{2\to 2}=\|J_\Phi^\Gamma\|_\infty$, where $J_\Phi^\Gamma$ is the partial transpose of the Choi operator of $\Phi$, is efficiently computable as a largest singular value.
\end{proof}

\section{Outlook}\label{sec:outlook}

In this work, we initiated the systematic study of the computational complexities of the norms of linear maps between Schatten spaces. For a summary of our results, see \cref{sec:summary.of.results}.  Some questions remain open and we mention a few of them here. The most natural one for this work concerns the hypercontractive $q \to p$ norms, i.e. for $q \leq p$, for positivity preserving maps. Its hardness or efficiency remains open even for $\ell_q \to \ell_p$ and is relevant for applications such as Gibbs state preparation via modified logarithmic Sobolev constants. It seems plausible to believe that most of these norms should also be hard to compute outside the regions where we proved efficiency. However, there might be some special points where non-trivial algorithms exist, for example $\infty \to 1$. In fact, the $\ell_\infty\to \ell_1$-norm can be efficiently approximated via the Groethendieck inequality, and in fact the Groethendieck relaxation (which can be written as a semidefinite program) is closely related to the completely bounded $\ell_{\infty} \to \ell_1$-norm. This begs the question of the complexity of approximating the (completely bounded) $\cS_{\infty} \to \cS_1$-norm.

Which begs two more questions, one, are they efficiently computable for more values of $q,p$ than we prove in this work
and secondly, how good of an approximation are the mixed cb-norms to their non cb-counterparts?

Concerning quantum Boyd's algorithm it is a natural question to ask if, as in the classical case, it can be extended to the whole region $q\geq p$. This would be tantamount to proving a suitable upper bound on $\kappa(A\mapsto A^r)$, see \cref{equ:def.contraction.coeff} and \cref{rem:kappa.greater.1}. Even though we have given an efficient algorithm to compute these norms of CP maps such an algorithm would be more amenable to changes in $q,p$ and possibly allow bounds on hypercontractive mixed norms of suitably contractive linear CP maps. 

Lastly one might wonder if our techniques also allow to prove a constant factor hardness concerning the $(1\to p)$ and $(1^+\to 1)$ Schatten norms. While we cannot disprove this to be true our techniques also don't allow to prove it. The issue arises when proving that an $\mathcal{O}(\varepsilon)$-approximate optimal $1\to p$ norm must have resulted from approximate optimal proper states $\tilde{\psi}$. In \eqref{equ:approx0} a $\poly(d)$-factor is introduced, which will need to be compensated for in $\varepsilon$ and hence obfuscates a constant, i.e. $d$ independent, scaling. We do however, believe that one can improve the $\poly(d)$-factor to an optimal $\mathcal{O}(d)$ one.


\hfill \newline\textbf{Acknowledgments:} We thank Salman Beigi for discussions regarding the hardness of the classical and quantum positivity preserving hypercontractive case, Hamza Fawzi for discussions on optimization and the ellipsoid method, and Akshay Ramachandran for discussions on related hardness proofs. We thank anonymous reviewers for helpful feedback. JK acknowledges support from the Program QuanTEdu-France n° ANR-22-CMAS-0001 France 2030. OF acknowledges funding from the European Research Council (ERC Grant AlgoQIP, Agreement No. 851716). CR acknowledges the support of the Deutsche Forschungsgemeinschaft (DFG, German Research Foundation) - Project-ID 470903074 - TRR352.

\bibliographystyle{abbrv}
\bibliography{lit}

@book{book.Grotschel.1993,
  title = {Geometric {{Algorithms}} and {{Combinatorial Optimization}}},
  author = {Gr\"otschel, Martin and Lov\'asz, L\'aszl\'o and Schrijver, Alexander},
  date = {1993},
  series = {Algorithms and {{Combinatorics}}},
  volume = {2},
  publisher = {Springer Berlin Heidelberg},
  location = {Berlin, Heidelberg},
  doi = {10.1007/978-3-642-78240-4},
  url = {http://link.springer.com/10.1007/978-3-642-78240-4},
  urldate = {2025-05-07},
  isbn = {978-3-642-78242-8 978-3-642-78240-4},  
}

@book{Book.HornJohnson.1991, place={Cambridge}, title={Topics in Matrix Analysis}, publisher={Cambridge University Press}, author={Horn, Roger A. and Johnson, Charles R.}, year={1991}}

@article{book.Bubeck.2015,
author = {Bubeck, S\'ebastien},
title = {Convex Optimization: Algorithms and Complexity},
year = {2015},
issue_date = {Nov 2015},
publisher = {Now Publishers Inc.},
address = {Hanover, MA, USA},
volume = {8},
number = {3–4},
issn = {1935-8237},
url = {https://doi.org/10.1561/2200000050},
doi = {10.1561/2200000050},
journal = {Found. Trends Mach. Learn.},
month = nov,
pages = {231–357},
numpages = {130}
}

@book{book.Watrous.2018, place={Cambridge}, title={The Theory of Quantum Information}, publisher={Cambridge University Press}, author={Watrous, John}, year={2018}}

@book{Book.Pisier.1998,
     author = {Pisier, Gilles},
     title = {Non-commutative vector valued $L_p$-spaces and completely $p$-summing map},
     series = {Ast\'erisque},
     publisher = {Soci\'et\'e math\'ematique de France},
     number = {247},
     year = {1998},
     mrnumber = {1648908},
     zbl = {0937.46056},
     language = {en},
     url ={http://www.numdam.org/item/AST_1998__247__R1_0/}
}

@book{book:Bhatia.1997,
  title = {Matrix {{Analysis}}},
  author = {Bhatia, Rajendra},
  year = {1997},
  series = {Graduate {{Texts}} in {{Mathematics}}},
  volume = {169},
  publisher = {Springer New York},
  address = {New York, NY},
  doi = {10.1007/978-1-4612-0653-8},
  url = {http://link.springer.com/10.1007/978-1-4612-0653-8},
  urldate = {2024-06-28},
  copyright = {http://www.springer.com/tdm},
  isbn = {978-1-4612-6857-4, 978-1-4612-0653-8},
  langid = {english}
}

@inbook{book:Pisier.2003,
  title={Introduction to Operator Space Theory},
  author={Gilles Pisier},
  year={2003},
  url={https://api.semanticscholar.org/CorpusID:53732225},
publisher={Cambride University Press}
}

@book{book:boyd.2004,
  title={Convex optimization},
  author={Boyd, Stephen and Vandenberghe, Lieven},
  year={2004},
  publisher={Cambridge university press}
}

@misc{Dan.2019,
      title={Optimal Analysis of Subset-Selection Based $L_p$ Low Rank Approximation}, 
      author={Chen Dan and Hong Wang and Hongyang Zhang and Yuchen Zhou and Pradeep Ravikumar},
      year={2019},
      eprint={1910.13618},
      archivePrefix={arXiv},
      primaryClass={cs.LG},
}

@misc{Flamary.2014,
      title={Mixed-norm Regularization for Brain Decoding}, 
      author={R\'emi Flamary and Nisrine Jrad and Ronald Phlypo and Marco Congedo and Alain Rakotomamonjy},
      year={2014},
      eprint={1403.3628},
      archivePrefix={arXiv},
      primaryClass={cs.LG}, 
}

@inproceedings{Saketha.2009,
 author = {Jagarlapudi, Saketha and G, Dinesh and S, Raman and Bhattacharyya, Chiranjib and Ben-tal, Aharon and K.r., Ramakrishnan},
 booktitle = {Advances in Neural Information Processing Systems},
 editor = {Y. Bengio and D. Schuurmans and J. Lafferty and C. Williams and A. Culotta},
 pages = {},
 publisher = {Curran Associates, Inc.},
 title = {On the Algorithmics and Applications of a Mixed-norm based Kernel Learning Formulation},
 volume = {22},
 year = {2009}
}

@article{Reeb.2011,
   title={Hilbert’s projective metric in quantum information theory},
   volume={52},
   ISSN={1089-7658},
   url={http://dx.doi.org/10.1063/1.3615729},
   DOI={10.1063/1.3615729},
   number={8},
   journal={Journal of Mathematical Physics},
   publisher={AIP Publishing},
   author={Reeb, David and Kastoryano, Michael J. and Wolf, Michael M.},
   year={2011},
   month=aug }

@article{shor2004equivalence,
  title={Equivalence of additivity questions in quantum information theory},
  author={Shor, Peter W},
  journal={Communications in Mathematical Physics},
  volume={246},
  number={3},
  pages={453--472},
  year={2004},
  publisher={Springer}
}

@misc{Capel.2021,
      title={The modified logarithmic Sobolev inequality for quantum spin systems: classical and commuting nearest neighbour interactions}, 
      author={\'Angela Capel and Cambyse Rouz\'e and Daniel Stilck França},
      year={2021},
      eprint={2009.11817},
      archivePrefix={arXiv},
      primaryClass={quant-ph},
      url={https://arxiv.org/abs/2009.11817}, 
}

@article{Beigi.2015,
   title={Hypercontractivity and the logarithmic Sobolev inequality for the completely bounded norm},
   volume={57},
   ISSN={1089-7658},
   url={http://dx.doi.org/10.1063/1.4934729},
   DOI={10.1063/1.4934729},
   number={1},
   journal={Journal of Mathematical Physics},
   publisher={AIP Publishing},
   author={Beigi, Salman and King, Christopher},
   year={2015},
   month=nov }

@article{Eveson.Nussbaum.1995, title={An elementary proof of the Birkhoff-Hopf theorem}, volume={117}, DOI={10.1017/S0305004100072911}, number={1}, journal={Mathematical Proceedings of the Cambridge Philosophical Society}, author={Eveson, Simon P. and Nussbaum, Roger D.}, year={1995}, pages={31–55}}

@article{Hendrickx.2010,
author = {Hendrickx, Julien M. and Olshevsky, Alex},
title = {Matrix $p$-Norms Are NP-Hard to Approximate If {$p \neq 1,2,\infty$}},
journal = {SIAM Journal on Matrix Analysis and Applications},
volume = {31},
number = {5},
pages = {2802-2812},
year = {2010},
doi = {10.1137/09076773X},
URL = {https://doi.org/10.1137/09076773X},
eprint = {https://doi.org/10.1137/09076773X}
}

@misc{Biswal.2011,
      title={Hypercontractivity and its applications}, 
      author={Punyashloka Biswal},
      year={2011},
      eprint={1101.2913},
      archivePrefix={arXiv},
      primaryClass={cs.DM},
      url={https://arxiv.org/abs/1101.2913} 
}

@misc{Fawzi.2025,
      title={Additivity and chain rules for quantum entropies via multi-index Schatten norms}, 
      author={Omar Fawzi and Jan Kochanowski and Cambyse Rouz\'e and Thomas Van Himbeeck},
      year={2025},
      eprint={2502.01611},
      archivePrefix={arXiv},
      primaryClass={quant-ph},
      url={https://arxiv.org/abs/2502.01611} 
}

@misc{Shahverdikondori.2022,
      title={Computing mixed Schatten norm of completely positive maps}, 
      author={Mohammad ShahverdiKondori and Sio On Chan},
      year={2022},
      eprint={2209.07504},
      archivePrefix={arXiv},
      primaryClass={math.NA},
      url={https://arxiv.org/abs/2209.07504} 
}

@misc{Watrous.2004,
      title={Notes on super-operator norms induced by Schatten norms}, 
      author={John Watrous},
      year={2004},
      eprint={quant-ph/0411077},
      archivePrefix={arXiv},
      primaryClass={quant-ph},
 url={https://arxiv.org/abs/quant-ph/0411077}
}

@inproceedings{Bhaskara.2011,
author = {Bhaskara, Aditya and Vijayaraghavan, Aravindan},
title = {Approximating matrix p-norms},
year = {2011},
publisher = {Society for Industrial and Applied Mathematics},
address = {USA},
booktitle = {Proceedings of the Twenty-Second Annual ACM-SIAM Symposium on Discrete Algorithms},
pages = {497–511},
numpages = {15},
location = {San Francisco, California},
series = {SODA '11}
}

@misc{Beigi.2008,
      title={On the Complexity of Computing Zero-Error and Holevo Capacity of Quantum Channels}, 
      author={Salman Beigi and Peter W. Shor},
      year={2008},
      eprint={0709.2090},
      archivePrefix={arXiv},
      primaryClass={quant-ph}
}

@article{Khanna.2001,
author = {Khanna, Sanjeev and Sudan, Madhu and Trevisan, Luca and Williamson, David P.},
title = {The Approximability of Constraint Satisfaction Problems},
journal = {SIAM Journal on Computing},
volume = {30},
number = {6},
pages = {1863-1920},
year = {2001},
doi = {10.1137/S0097539799349948},
URL = {https://doi.org/10.1137/S0097539799349948},
eprint = {https://doi.org/10.1137/S0097539799349948}
}

@article{Harrow.2013,
author = {Harrow, Aram W. and Montanaro, Ashley},
title = {Testing Product States, Quantum Merlin-Arthur Games and Tensor Optimization},
year = {2013},
issue_date = {February 2013},
publisher = {Association for Computing Machinery},
address = {New York, NY, USA},
volume = {60},
number = {1},
issn = {0004-5411},
url = {https://doi.org/10.1145/2432622.2432625},
doi = {10.1145/2432622.2432625},
journal = {J. ACM},
month = feb,
articleno = {3},
numpages = {43}
}

@article{Hiai.2013,
title = {Concavity of certain matrix trace and norm functions},
journal = {Linear Algebra and its Applications},
volume = {439},
number = {5},
pages = {1568-1589},
year = {2013},
issn = {0024-3795},
doi = {https://doi.org/10.1016/j.laa.2013.04.020},
url = {https://www.sciencedirect.com/science/article/pii/S0024379513002826},
author = {Fumio Hiai}
}

@article{Bhattiprolu.2023,
author = {Bhattiprolu, Vijay and Ghosh, Mrinal Kanti and Guruswami, Venkatesan and Lee, Euiwoong and Tulsiani, Madhur},
title = {Inapproximability of Matrix \(\boldsymbol{p \rightarrow q}\) Norms},
journal = {SIAM Journal on Computing},
volume = {52},
number = {1},
pages = {132-155},
year = {2023},
doi = {10.1137/18M1233418},
URL = {https://doi.org/10.1137/18M1233418
},
eprint = {https://doi.org/10.1137/18M1233418
}
}

@article{Beigi.2023,
  title = {Operator-Valued {{Schatten}} Spaces and Quantum Entropies},
  author = {Beigi, Salman and Goodarzi, Milad M.},
  date = {2023-08-29},
  journaltitle = {Letters in Mathematical Physics},
  shortjournal = {Lett Math Phys},
  volume = {113},
  number = {5},
  eprint = {2207.06693},
  eprinttype = {arXiv},
  eprintclass = {quant-ph},
  pages = {91},
  issn = {1573-0530},
  doi = {10.1007/s11005-023-01712-9},
  url = {http://arxiv.org/abs/2207.06693},
  urldate = {2024-06-28},
  langid = {english},
  year={2023}
}

@article{Diaconis.1996,
  title = {Logarithmic Sobolev inequalities for finite Markov chains},
  volume = {6},
  ISSN = {1050-5164},
  url = {http://dx.doi.org/10.1214/aoap/1034968224},
  DOI = {10.1214/aoap/1034968224},
  number = {3},
  journal = {The Annals of Applied Probability},
  publisher = {Institute of Mathematical Statistics},
  author = {Diaconis,  P. and Saloff-Coste,  L.},
  year = {1996},
  month = aug 
}

@article{Zhang.2020,
   title={From Wigner-Yanase-Dyson conjecture to Carlen-Frank-Lieb conjecture},
   volume={365},
   ISSN={0001-8708},
   url={http://dx.doi.org/10.1016/j.aim.2020.107053},
   DOI={10.1016/j.aim.2020.107053},
   journal={Advances in Mathematics},
   publisher={Elsevier BV},
   author={Zhang, Haonan},
   year={2020},
   month=may, pages={107053} }

@article{Boyd.1974,
  title = {The Power Method for $L_p$ Norms},
  author = {Boyd, David W.},
  date = {1974-01-01},
  journal = {Linear Algebra and its Applications},
  volume = {9},
  pages = {95--101},
  issn = {0024-3795},
  doi = {10.1016/0024-3795(74)90029-9},
  url = {https://www.sciencedirect.com/science/article/pii/0024379574900299},
  urldate = {2024-07-04}
}

@article{Devetak.2006,
   title={Multiplicativity of Completely Bounded p-Norms Implies a New Additivity Result},
   volume={266},
   ISSN={1432-0916},
   url={http://dx.doi.org/10.1007/s00220-006-0034-0},
   DOI={10.1007/s00220-006-0034-0},
   number={1},
   journal={Communications in Mathematical Physics},
   publisher={Springer Science and Business Media LLC},
   author={Devetak, Igor and Junge, Marius and King, Christopher and Ruskai, Mary Beth},
   year={2006},
   month=may, pages={37–63} }

@article{Bardet.2024,
  title = {Entropy {{Decay}} for {{Davies Semigroups}} of a {{One Dimensional Quantum Lattice}}},
  author = {Bardet, Ivan and Capel, \'Angela and Gao, Li and Lucia, Angelo and P\'erez-García, David and Rouz\'e, Cambyse},
  date = {2024-02-10},
  journaltitle = {Communications in Mathematical Physics},
  shortjournal = {Communications in Mathematical Physics},
  journal={Communications in Mathematical Physics},
  volume = {405},
  number = {2},
  pages = {42},
  issn = {1432-0916},
  doi = {10.1007/s00220-023-04869-5},
  url = {https://doi.org/10.1007/s00220-023-04869-5},
  year={2024},
}

@article{Bardet.2022,
  author    = {Ivan Bardet and Cambyse Rouz\'e},
  title     = {Hypercontractivity and Logarithmic Sobolev Inequality for Non-Primitive Quantum Markov Semigroups and Estimation of Decoherence Rates},
  journal   = {Annales Henri Poincar\'e},
  volume    = {23},
  number    = {11},
  pages     = {3839--3903},
  year      = {2022},
  month     = {nov},
  doi       = {10.1007/s00023-022-01196-8},
  url       = {https://doi.org/10.1007/s00023-022-01196-8}
}

@misc{Brandao.2015,
      title={Estimating operator norms using covering nets}, 
      author={Fernando G. S. L. Brandao and Aram W. Harrow},
      year={2015},
      eprint={1509.05065},
      archivePrefix={arXiv},
      primaryClass={quant-ph},
      url={https://arxiv.org/abs/1509.05065}, 
}

@misc{Gautier.2020,
      title={Computing the norm of nonnegative matrices and the log-Sobolev constant of Markov chains}, 
      author={Antoine Gautier and Matthias Hein and Francesco Tudisco},
      year={2020},
      eprint={2002.02447},
      archivePrefix={arXiv},
      primaryClass={math.NA},
      url={https://arxiv.org/abs/2002.02447}, 
}

@inproceedings{Bhattiprolu.2019,
author = {Vijay Bhattiprolu and Mrinalkanti Ghosh and Venkatesan Guruswami and Euiwoong Lee and Madhur Tulsiani},
title = {Approximability of p → q Matrix Norms: Generalized Krivine Rounding and Hypercontractive Hardness},
booktitle = {Proceedings of the 2019 Annual ACM-SIAM Symposium on Discrete Algorithms (SODA)},
chapter = {},
pages = {1358-1368},
doi = {10.1137/1.9781611975482.83},
URL = {https://epubs.siam.org/doi/abs/10.1137/1.9781611975482.83},
eprint = {https://epubs.siam.org/doi/pdf/10.1137/1.9781611975482.83},
year={2019},
}

@article{Daureen.2007,
  title={COMPUTATION OF MATRIX NORMS WITH APPLICATIONS TO ROBUST OPTIMIZATION},
  author={Daureen Steinberg},
  year={2007},
  url={https://api.semanticscholar.org/CorpusID:51772380}
}

@inproceedings{Barak.2012,
author = {Barak, Boaz and Brandao, Fernando G.S.L. and Harrow, Aram W. and Kelner, Jonathan and Steurer, David and Zhou, Yuan},
title = {Hypercontractivity, sum-of-squares proofs, and their applications},
year = {2012},
isbn = {9781450312455},
publisher = {Association for Computing Machinery},
address = {New York, NY, USA},
url = {https://doi.org/10.1145/2213977.2214006},
doi = {10.1145/2213977.2214006},
booktitle = {Proceedings of the Forty-Fourth Annual ACM Symposium on Theory of Computing},
pages = {307–326},
numpages = {20},
location = {New York, New York, USA},
series = {STOC '12}
}

@misc{Watrous.2012,
      title={Simpler semidefinite programs for completely bounded norms}, 
      author={John Watrous},
      year={2012},
      eprint={1207.5726},
      archivePrefix={arXiv},
      primaryClass={quant-ph}
}

\appendix

\section{Classical embedding proof} \label{app:classicalembedding}

\begin{proof}[Proof of \cref{prop:classical.quantum.embedding} and \cref{cor:generic.Schatten.hardness}]
Given a matrix $A=(A_{ij})_{ij}\in\C^{n\times m}$ define 
\begin{align}
    \Phi_A(\rho):=\sum_{ij}A_{ij}|i\rangle\langle j|\rho |j\rangle\langle i| : \cB(\C^n)\to \cB(\C^m)
\end{align} for some fixed ONBs $\{|i\rangle\}_{i=1}^n\subset\C^n, \{|j\rangle\}_{j=1}^m\subset\C^m$.
Notice that $\Phi\circ\Pi=\Phi$, when $\Pi(\rho):=\sum_j|j\rangle\langle j|\rho |j\rangle\langle j|$ is the pinching map w.r.t the fixed basis. Now we will prove that $\|A\|_{q\to p}=\|\Phi_A\|_{q\to p}$. First notice that
\begin{align}
    \frac{\|\Phi_A(\omega)\|_p}{\|\omega\|_q}= \frac{\|\Phi_A\circ\Pi(\omega)\|_p}{\|\omega\|_q}=\frac{\|\Phi_A\circ\Pi(\omega)\|_p}{\|\Pi(\omega)\|_q}\underbrace{\frac{\|\Pi(\omega)\|_q}{\|\omega\|_q}}_{\leq 1} \leq \frac{\|\Phi_A\circ\Pi(\omega)\|_p}{\|\Pi(\omega)\|_q} \leq \|A\|_{q\to p}.
\end{align}  
The last inequality follows from the fact that $\Pi(\omega)$ is diagonal in the $\{|j\rangle\}_j$ basis. 
The converse inequality follows by
\begin{align}
    \|\Phi_A\|_{q\to p} \geq \sup_{ \omega=\sum_jw_j|j\rangle\langle j|}\frac{\|\Phi(\omega)\|_p}{\|\omega\|_q} = \|A\|_{q\to p}\,.
\end{align}
To finish the proof we notice that clearly this works for any $A\in\C^{n\times m}$ and further if $A=(A_{ij})_{ij}$ is positivity preserving, i.e.,
  $A_{ij}\geq 0 \quad \forall i,j$, then $\Phi_A$ is CP, since it can be written with Kraus operators $K_{ij}:=\sqrt{A_{ij}}|i\rangle\langle j|$. 

The hardness statement in the Corollary now follows from
\cite[Chapter 2]{Daureen.2007}, \cite[Theorems 1.1, 1.2]{Bhattiprolu.2019}, \cite[Theorems 2.4, 2.5]{Brandao.2015}, see also \cite[Theorem 1.3 and Section 4.2]{Bhattiprolu.2023}, \cite{Barak.2012}, and \cite[Theorems 6.6, 6.8 and Proposition 6.1]{Bhaskara.2011}.
\end{proof}

\section{Ellipsoid method proof details}\label{app:ellipsoid.method.details}
Convex optimization is usually done in real vector spaces, however, the following association 
\begin{align}
    \C^{d\times d}\ni X\mapsto \hat{X}:=\frac{1}{\sqrt{2}}\begin{pmatrix}
    \Re(X) & -\Im(X) \\ \Im(X) & \Re(X)
\end{pmatrix} \in \R^{2d\times 2d},
\end{align} where if $X=X_+-X_-+iX_i-iX_{-1}$ is its Jordan decomposition we have $\Re(X):=X_+-X_-$ and $\Im(X)=X_i-X_{-i}$,
we can circumvent this problem. Now one can easily check that for any $X,Y\in\C^{d\times d}$
\begin{align}
    \Tr[\hat{X}^T\hat{Y}] = \Re(\Tr[X^*Y]),
\end{align} hence by working with this inner product we are effectively turning our vector space real, without having to explicitly carry around that association \cite{book:boyd.2004}. Thus we can use all literature results on real vector spaces.

For details on the black box approach to the ellipsoid method for convex optimization problems we refer to \cite[Sections 1 and 2.2]{book.Bubeck.2015}.
Essentially we are using \cite[Theorem 2.4]{book.Bubeck.2015} which gives the following.
\begin{theorem}[Ellipsoid method, Theorem 2.4 \cite{book.Bubeck.2015} adapted]
Given a convex function $f:\mathcal{X}\to [-M,0]$ on a convex set $\cX\subset \mathbb{C}^{d\times d}$ then the ellipsoid method gives an algorithm with $t\geq 2d^4\log(\frac{R}{r})$ calls to a separation and a sub-gradient oracle that outputs a value $\tilde{f}_t$ s.t.
\begin{align}
    \tilde{f}_t-m \leq \frac{2mR}{r}\exp\left(-\frac{t}{2d^4}\right),
\end{align}
where $m:=\min_{x\in\cX}f(x)$ and $r\cX-y_1\subset \{X\in\mathbb{C}^{d\times d}| \|X\|_2\leq 1\}\subset R\cX-y_2$ for some $y_1,y_2\in \mathbb{C}^{d\times d}$.
\end{theorem}
For our $\cX=\cD(\mathbb{C}^d)$ we have by $\|X\|_1\geq \|X\|_2\geq d^{-\frac{1}{2}}\|X\|_1$ that $\frac{R}{r}\leq2\sqrt{d}$ and since we care about $\epsilon$ relative approximation this implies after $t=2d^4\log(\frac{4\sqrt{d}}{\epsilon})$ calls to the separation and sub-gradient oracles we are done. Hence a polynomial complexity of implementing these will finish the respective arguments.

Here we give the explicit constructions of the separation and the sub-gradient oracle \cite[Section 1.4]{book.Bubeck.2015} required to finish the proof of \cref{thm:ellipsoid.result} and \cref{thm:cb.ellipsoid}.

\subsection{The separation oracle for quantum states}
Since both optimization problems are over $\cD(\C^d):=\{X\in\C^{d\times d}|X\geq0, \|X\|_1\leq 1\}$ we can use the same separation oracle.

The separation oracle is a function, which given a matrix $X\in \C^{d\times d}$ decides whether it is in $\cD(\C^d)$ and outputs \textsc{YES} if so, else outputs a separating hyperplane between $X$ and $\cD(\C^d)$, parametrized by some matrix $W=W^*\in \C^{d\times d}$ s.t. $\cD(\C^d)\subset\{Y|\Re(\Tr[(Y-X)W])\leq 0\}$. 
We construct it the following way:

Given $X\in\C^{d\times d}$ compute its spectral decomposition and with it 
$\|X\|_1:=\Tr[|X|]$. If $X\geq 0$ and $\|X\|_1\leq 1$ output \textsc{YES}. \\ \noindent Else if just $X\geq0$, then output $\1$, else output $P_+-\1$, where $P_+$ is defined as the  projection on to the support of $X_+$ the positive part of $X$. This can easily be constructed form the spectral decomposition, which takes $\mathcal{O}(d^3)$ steps to compute. 

This works because if $X\geq 0$, but $\|X\|_1>1$, then the constructed algorithm outputs $W=\1$, since and it holds that
\begin{align}
\cD(\C^d)=\{Y| \|Y\|_1\leq 1, Y\geq 0\}\subset \{Y|\Re(\Tr[(Y-X)\1])=\Re(\Tr[Y])-\|X\|_1\leq 0\}
\end{align} because for $Y\geq 0$ we have that $\Re(\Tr[Y-X])=\|Y\|_1-\|X\|_1\leq 0$ is equivalent to $\|Y\|_1\leq \|X\|_1$.
Now if $X\ngeq0$, then for $Y\geq 0$ we have similarly
\begin{align}
  \cD(\C^d) \subset \{Y|\Re(\Tr[(X-Y)(\1-P_+)])\leq 0\},
\end{align} since for $Y\geq 0$ it holds that both 
\begin{align}
    \Re(\Tr[(X-Y)(\1-P_+)]) &=\Re(\Tr[-X_-+iX_i-iX_{-1}-(\1-P_+))Y(\1-P_+)]) \\&= -\Tr[X_--(\1-P_+))Y(\1-P_+))]\leq 0.
\end{align} 

\subsection{The subgradient oracles}
The sub-gradient oracle, given a matrix $X\in\cD(\C^d)$, outputs a sub-gradient of the convex function $F$ at $X$. Since $F$ is convex $\nabla F(X)$, its gradient at $X$, exists and is an admissible sub-gradient at $X$, i.e. $\nabla F(X)$ given by $DF(X)(B)=\Re\Tr[\nabla F(X)B]$ satisfies $F(A)-F(B)\leq DF(A)(A-B)=\Re\Tr[\nabla F(A)(A-B)]$. Here $DF(X)(B)$ is the Fréchet derivative of $F$ at $X$ in direction $B$.

The core argument below will be a $\mathcal{O}(d^3)$-time computable\footnote{In principle one could be worried that the efficient algorithm is not a rational function of its inputs, however, this would require yet another level of rigorosity of which we don't think it provides an actual benefit to this work.} derivative of the function $\cB^+(\C^d)\to \cB^+(\C^d): A\mapsto A^\frac{1}{p}$ from \cite{Book.HornJohnson.1991}.

Concerning \cref{thm:ellipsoid.result} we have the convex function $F(X):=-\Tr[\Phi(X^\frac{1}{q})^p]$. We will show in the following, that its gradient is given as 
\begin{align}
  \nabla F(X) = pU\left((U^*\Phi^*(\Phi(X^\frac{1}{q})^{p-1}) U)\circ \left[\frac{\lambda_i^{\frac{1}{q}}-\lambda_j^{\frac{1}{q}}}{\lambda_i-\lambda_j}\right]^d_{i,j=1}\right) U^*,
\end{align} where $X=U\text{diag}(\{\lambda_i\}_i)U^*$ is the spectral decomposition of $X$ computable in $\mathcal{O}(d^3)$ and $\circ$ the Schur product of two matrices.
The gradient can hence be computed in $\mathcal{O}(d^3)$-time.

Denote with $j_p$ the power function $\lambda\mapsto\lambda^p$, and likewise for $j_{\frac{1}{q}}$.
By a chain-rule and definition of $F$ one can show that
\begin{align}
    DF(X)(B)=-\Tr[p\Phi(X^\frac{1}{q})^{p-1}\Phi(Dj_{\frac{1}{q}}(X)(B))] = -\Tr[p\Phi^*(\Phi(X^\frac{1}{q})^{p-1})(Dj_{\frac{1}{q}}(X)(B))],
\end{align} since $G(X):=\Tr[X^p]$ has derivative $DG(X)(B)=\Tr[pX^{p-1}B]$ due to the trace functional being present. See \cref{app:Boyds.analysis} for details on this. Hence to compute the gradient it remains to consider $Dj_{\frac{1}{q}}(X)(B)$. Recall that
\begin{align}
     Dj_{\frac{1}{q}}(X)(B)=\frac{\dif}{\dif t}F[X+tB]\bigg|_{t=0},  
\end{align}
which by \cite[Theorem 6.6.30]{Book.HornJohnson.1991} is given as
\begin{align}
   Dj_{\frac{1}{q}}(X)(B) = U\left(\left[\frac{\lambda_i^{\frac{1}{q}}-\lambda_j^{\frac{1}{q}}}{\lambda_i-\lambda_j}\right]^d_{i,j=1}\circ(U^*BU)\right)U^* \equiv U(J\circ(U^*BU))U^*,
\end{align}
where $\{\lambda_i\}_i$ are the eigenvalues of $X$ which can be computed in $\mathcal{O}(d^3)$ time and $U$ is the unitary in the spectral decomposition of $X$, which can also be computed in $\mathcal{O}(d^3)$-time. $\circ$ denotes the Schur product of two matrices, i.e. $[a_{ij}]\circ[b_{ij}]=[a_{ij}b_{ij}]$.
Putting everything together we get that
\begin{align}
    DF(X)(B) &= -\Tr[p\underbrace{\Phi^*(\Phi(X^\frac{1}{q})^{p-1})}_{\Xi}(Dj_{\frac{1}{q}}(X)(B))] \equiv -\Tr[p\Xi U(J\circ(U^*BU))U^*] \\ &= -\Tr[pU((U^*\Xi U)\circ J) U^*B] = -\Tr[\nabla F(X)B]
\end{align} which holds since $\Tr[A(B\circ C)]=\Tr[(A\circ B^T)C]$ and $J=J^T$ is symmetric.

Concerning \cref{thm:ellipsoid.result} we have the convex function
\begin{align}
    F(A,B):=-\Tr[((B^\frac{1}{2p}\otimes\1)J_\Phi^*(A^\frac{1}{p}\otimes
    \1)J_\Phi (B^\frac{1}{2p}\otimes\1))^\frac{p}{2}] = -\Tr[((A^\frac{1}{2p}\otimes\1)J_\Phi(B^\frac{1}{p}\otimes
    \1)J_\Phi^+ (A^\frac{1}{2p}\otimes\1))^\frac{p}{2}].
\end{align}
Due to the joint and thus also individual convexities it suffices to optimize first over say $B$ and then $A$. Hence the sub-gradients w.r.t. $A$ and $B$ separately suffice. Omitting the $\otimes\1$ for simplicity we get, respectively
\begin{align}
    D_AF(A,B)(Y) &=-\frac{p}{2}\Tr[(B^\frac{1}{2p}J_\Phi^*A^\frac{1}{p}J_\Phi B^\frac{1}{2p})^{\frac{p}{2}-1}(B^\frac{1}{2p}J_\Phi^* Dj_\frac{1}{p}(A)(Y)J_\Phi B^\frac{1}{2p})], \\
    D_BF(A,B)(Y) &=-\frac{p}{2}\Tr[(A^\frac{1}{2p}J_\Phi B^\frac{1}{p}J_\Phi^* A^\frac{1}{2p})^{\frac{p}{2}-1}(A^\frac{1}{2p}J_\Phi Dj_\frac{1}{p}(B)(Y)J_\Phi^* A^\frac{1}{2p})].
\end{align}
Using the same formula for $Dj_\frac{1}{p}(A)(Y)$, which is computable in $\mathcal{O}(d^3)$-time, as above we get the $\mathcal{O}(d^3)$-computable formulas for the respective subgradients.
\begin{align}
    \nabla_AF(A^\prime,B) &= -\frac{p}{2}U\Bigg\{(U^*J_\Phi B^\frac{1}{2p}(B^{\frac{1}{2p}}J_\Phi^*A^{\prime\frac{1}{p}}J_\Phi B^\frac{1}{2p})^{\frac{p}{2}-1}B^\frac{1}{2p}J_\Phi^*U)\circ \bigg[\frac{\lambda_i^\frac{1}{p}-\lambda_j^\frac{1}{p}}{\lambda_i-\lambda_j}\bigg]_{i,j=1}^d\Bigg\}U^*, \\
    \nabla_BF(A,B^\prime) &= -\frac{p}{2}U\Bigg\{(U^*J_\Phi^* A^\frac{1}{2p}(A^{\frac{1}{2p}}J_\Phi B^{\prime\frac{1}{p}}J_\Phi^* A^\frac{1}{2p})^{\frac{p}{2}-1}A^\frac{1}{2p}J_\Phi U)\circ \bigg[\frac{\lambda_i^\frac{1}{p}-\lambda_j^\frac{1}{p}}{\lambda_i-\lambda_j}\bigg]_{i,j=1}^d\Bigg\}U^*, \\
\end{align} where $A^\prime=U\text{diag}(\{\lambda_i\}_{i=1}^d)U^*$ in the first expression and $B^\prime=U\text{diag}(\{\lambda_i\}_{i=1}^d)U^*$ in the second are the spectral decompositions of, respectively, $A,B$.

We mention here, that if $\Phi$ in \cref{thm:cb.ellipsoid} is CP, then it suffices to optimize over $A=B$ and the optimization is over the function $F(A,A)$, which in practice may be simpler to do. Its derivative is computed analogously as above.

\section{Derivation and analysis of quantum Boyd's algorithm}\label{app:Boyds.analysis}
In this section we gather the technical part of the proof of the quantum Boyd algorithm. Recall the definition of the non-linear map whose iteration (and normalization) is the core part of the quantum Boyd's algorithm.
Recall the definition \cref{def:S}
\begin{align}
    S(\omega) :=\Lambda^*\left(\Lambda(\omega)^{p-1}\right)^{\frac{1}{q-1}} : \cB^+(\mathcal{H})\to \cB^+(\mathcal{H}) 
\end{align}    

Unless otherwise specified we assume in this section $\Lambda$ positivity improving s.t. $\Lambda(\omega)\geq \frac{\1}{Nn}\Tr[\omega]$ $\forall \omega\geq 0$, $\|\Lambda\|_{\infty\to\infty}\leq 1$, and $1\leq p\leq 2\leq q\leq \infty.$  
In this section we will denote with $\cS^{+,1}_q:=\{\rho\in \cB^+(\mathcal{H})|\|\rho\|_q=1\}$ the positive part of the $q-$Schatten unit sphere.

In \cref{lem:1} we had claimed that the stationary points of $f(\omega):=\frac{\|\Lambda(\omega)\|_p}{\|\omega\|_q}$ on $\cB^+(\mathcal{H})$ are fixed points of S and vice verca.
\begin{proof}[Proof of Lemma \ref{lem:1}]
For derivatives of matrix valued functions we will use some notation and elementary properties from \cite{book:Bhatia.1997}.
We calculate $(Df)(\omega)(B)$ and require it to vanish $\forall B\in\mathcal{B}(\mathcal{H})$. This will impose a condition which is equivalent to the fixed point condition under the map $S$ defined in \cref{def:S}. 
In the following we will use 
\begin{align}
\left.\frac{\dif}{\dif t}\right\vert_{t=0}\Tr[f(\omega+tH)]=\Tr[\left.\frac{\dif}{\dif t}\right\vert_{t=0}f(\omega+tH)]=\Tr[(Df)(\omega)(H)]
\end{align} holds for any differentiable function f, an integral representation of the $p-$th power of an operator, and the product rule. For all see e.g. \cite{book:Bhatia.1997}. Namely we have for $1<p<2$ and $j_p(\omega)_:=\omega^p=\int_0^\infty\dif\mu_{p-1}(\lambda)\omega^2(\omega+\lambda)^{-1}$, where $\dif\mu_{p-1}(\lambda)=\frac{\sin((p-1)\pi)}{\pi}\lambda^{p-2}\dif\lambda$, that
\begin{align}
 (Dj_p)(\omega)(H)= \int_0^\infty\dif\mu_{r}(\lambda)(\lambda+\omega)^{-1}\lambda\{\omega,H\}_+(\lambda+\omega)^++\int_0^\infty\dif\mu_r(\lambda)(\lambda+\omega)^{-1}\omega H\omega(\lambda+\omega)^{-1},   
\end{align}where $\{A,B\}_+=AB+BA$ is the anticommutator. \\
 \underline{\textsc{Claim:}} From this one can show that $\Tr[(Dj_p)(\omega)(H)]=p\Tr[H\omega^{p-1}]$ indeed holds. \\
 \underline{\textsc{Proof:}} Setting $r:=p-1$, we have
 \begin{align}
     \Tr[(Dj_p)(\omega)(H)] &=\int_0^\infty\dif\mu_r(\lambda)\left(\lambda\Tr[(AB+BA)(\lambda+A)^{-2}]+\Tr[BA(\lambda+A)^{-2}A]\right) \\
     &=\int_0^\infty\dif\mu_r(\lambda)\left(\lambda\Tr[B(\lambda+A)^{-2}A+BA(\lambda+A)^{-2}]+\Tr[BA(\lambda+A)^{-2}A]\right) \\
     &=\int_0^\infty\dif\mu_r(\lambda)\Tr[B(2\lambda A(\lambda+A)^{-2}+A^2(\lambda+A)^{-2})] \\
     &=\Tr\big[B\underbrace{\left(\int_0^{\infty}\dif\mu_r(\lambda)\lambda A(\lambda+A)^{-2}+A(\lambda+A)^{-1}\right)}_{=:F_\lambda(A)}\big] = \Tr[BpA^{p-1}]
 \end{align}
 The last equality follos after noticing that $F_\lambda(t)=\frac{\lambda t}{(\lambda+t)^2}+\frac{t}{\lambda+t}=1-\frac{\lambda^2}{(\lambda+t)^2}$ satisfies, for $t>0$
 \begin{align}
     \int_0^\infty\dif\mu_rF_\lambda(t)&=\int_0^\infty\dif\lambda\lambda^{r-1}\frac{\sin(r\pi)}{\pi}\left(\frac{\lambda t}{(\lambda+t)^2}+\frac{t}{\lambda+t}\right) = \int_0^\infty \dif\mu_r(\lambda)\frac{t}{\lambda+t}+\int_0^\infty\dif\lambda\lambda^r\frac{\sin(r\pi)}{\pi}\frac{t}{(t+\lambda)^2}\\&\overset{P.I.}{=} t^r+\int_0^\infty\dif\lambda r\lambda{r-1}\frac{\sin(r\pi)}{\pi}\frac{t}{\lambda+t}=t^r+rt^r=(r+1)t^r
 \end{align}
 By the chain rule we this get that $\Tr[(Dj_p\circ\Lambda)(\omega)(H)]=p\Tr[\Lambda(H)\Lambda(\omega)^{p-1}]=p\Tr[H\Lambda^*(\Lambda(\omega)^{p-1})]$. Since $\omega\geq 0$ and by assumption $\Lambda(\omega)>0$ it holds that $\|\omega\|_q=\Tr[\omega^q]^{\frac{1}{q}}$ and $\|\Lambda(\omega)\|_p=\Tr[\Lambda(\omega)^p]^{\frac{1}{p}}$ and hence $f(\omega)=\|\Lambda(\omega)\|_p\|\omega\|^{-1}_q=\Tr[\Lambda(\omega)^p]^{\frac{1}{p}}\Tr[\omega^{q}]^{-\frac{1}{q}}$.
 Having established these, we can put everything together with the product rule. For all $h$ we require
 \begin{align}
     0\overset{!}{=}(Df)(\omega)(H)&= \|\omega\|_q^{1-q}\|\Lambda(\omega)\|_p\Tr[\omega^{q-1}H]-\|\Lambda(\omega)\|_p^{1-p}\|\omega\|_q\Tr[\Lambda^*(\Lambda(\omega)^{p-1})H] \\&= \Tr\Big[H\underbrace{\left(\|\omega\|_q^{1-q}\|\Lambda(\omega)\|_p\omega^{q-1}-\|\Lambda(\omega)\|_p^{1-p}\|\omega\|_q\Lambda^*(\Lambda(\omega)^{p-1})\right)}_{ =0}\Big].
 \end{align}
 Hence for a critical $\tilde{\omega}\in \cB^+(\mathcal{H})$ we have
 \begin{align}
     0\overset{!}{=}(Df)(\tilde\omega)(H) &\overset{\forall H}{\Longleftrightarrow} \left(\|\omega\|_q^{1-q}\|\Lambda(\omega)\|_p\omega^{q-1}=\|\Lambda(\omega)\|_p^{1-p}\|\omega\|_q\Lambda^*(\Lambda(\omega)^{p-1})\right) \\ &\Longleftarrow \tilde{\omega}^{q-1} = \|\tilde{\omega}\|_q^q\|\Lambda(\tilde{\omega})\|_p^{-p}\Lambda^*(\Lambda(\tilde{\omega})^{p-1}),
 \end{align} which is nothing than the stationarity condition under the map $S(\omega):=\Lambda^*(\Lambda(\omega)^{p-1})^{\frac{1}{q-1}}\propto\omega$.
\end{proof}
The rest of this section is devoted to the analysis of the map $S$, this algorithm, and the proof of its convergence time as claimed in Theorem \ref{thm:strict.positive.q.Boyds}. 
\begin{lemma}
    The non-linear map $S,S^{q-1}:\cB^+(\mathcal{H})\to \cB^+(\mathcal{H})$ are operator monotone.
\end{lemma}
\begin{proof}
    The maps are operator monotone as a composition of operator monotone functions. $\Lambda^*$ is positive since $\Lambda$ is and positivity of a linear map is equivalent to operator monotonicity. It is well known that the function $t\mapsto t^r$ is operator monotone for $0\leq r\leq 1$ on $(0,\infty)$, see e.g. \cite{book:Bhatia.1997}. Hence since $1\leq p\leq 2\leq q\leq \infty$ both $t\mapsto t^{p-1}$ and $t\mapsto t^{\frac{1}{q-1}}$ are.
\end{proof}

\begin{claim}\label{rmk:positivity improving}
    Note also, that $\Lambda$ and hence $\Lambda^*$ are positivity improving by assumption and thus strictly operator monotone in the following sense.
    If $A\geq B\neq A$, then $\Lambda(A)>\Lambda(B)$ and $\Lambda^*(A)>\Lambda^*(B)$.
\end{claim}
\begin{proof}
    Assume $A\geq B\neq A$, i.e. $0\neq C:=A-B\geq 0$, then $\Lambda(C)=\Lambda(A)-\Lambda(B)\geq \frac{\1}{Nd}\Tr[C]>0$. This holds equally for $\Lambda^*$, since 
    \begin{align}
        \Lambda(\omega)\geq \frac{\1}{Nd}\Tr[\omega] &\Leftrightarrow \Lambda(\omega) -\frac{\1}{Nd}\Tr[\omega] \geq 0 \forall \omega 
        \\&\Leftrightarrow 0\leq \Tr[H\Lambda(\omega)-\frac{H}{Nd}\Tr[\omega]]= \Tr[\Lambda^*(H)\omega-\frac{\omega}{Nd}\Tr[H]] \forall \omega,H \\ &\Leftrightarrow \Lambda^*(H)-\frac{\1}{Nd}\Tr[H] \geq 0 \forall H \\ &\Leftrightarrow \Lambda^*(H)\geq \frac{\1}{Nd}\Tr[H].
    \end{align}
\end{proof}
Next we will show that iterating applying the map $S$ and renormalizing in $q-$norm converges with respect to projective Hilbert-metric. To do this we define 
\begin{align}
    m(\omega):=\sup\{\lambda\in\mathbb{R}|\omega^{q-1}\lambda^{q-1}\leq S^{q-1}(\omega)\} =\|\omega^{\frac{q-1}{2}}S(\omega)^{1-q}\omega^{\frac{q-1}{2}}\|^\frac{-1}{q-1}_\infty =  2^{-\frac{1}{q-1}D_{\text{max}}(\omega^{q-1}\|S(\omega)^{q-1})}, \\
    M(\omega):=\inf\{\lambda\in\mathbb{R}|\leq S^{q-1}(\omega)\leq \lambda^{q-1}\omega^{q-1}\} = \|\omega^{\frac{1-q}{2}}S(\omega)^{q-1}\omega^{\frac{1-q}{2}}\|^\frac{1}{q-1}_\infty = 2^{\frac{1}{q-1}D_{\text{max}}(S(\omega)^{q-1}\|\omega^{q-1})} .
\end{align}
Note that by operator monotonicity of the map $t\mapsto t^{\frac{1}{q-1}}$, if $\omega^{q-1}\lambda^{q-1}\leq S(\omega)^{q-1} \Rightarrow \omega\lambda\leq S(\omega)$ and hence we have both $\omega^{q-1}\lambda^{q-1}\leq S(\omega)^{q-1}\leq M(\omega)^{q-1}\omega^{q-1}$ and $\omega m(\omega)\leq S(\omega) \leq M(\omega)\omega $.
Also, obviously by this, $m(\omega)\leq M(\omega)$ and for any fixed point of $S$, say $\tilde{\omega}$, i.e. $S(\tilde{\omega})\propto\tilde{\omega}$, it holds that
\begin{align}
    m(\tilde{\omega})=M(\tilde{\omega}) = \left(\frac{\|\tilde{\omega}\|_q^q}{\|\Lambda(\tilde{\omega})\|_p^p}\right)^{\frac{1}{q-1}}=:\mu.
\end{align}
We will consider the quotient of those two, which is up to a logarithm an upper bound on the projective Hilbert metric between $\omega$ and $S(\omega)$ if $2\leq q$, since
\begin{align}
    \frac{M(\omega)}{m(\omega)}=\left(\|\omega^{\frac{1-q}{2}}S(\omega)^{q-1}\omega^{\frac{1-q}{2}}\|_\infty\|\omega^{\frac{q-1}{2}}S(\omega)^{1-q}\omega^{\frac{q-1}{2}}\|_\infty\right)^\frac{1}{q-1} \geq \exp(d_H(\omega,S(\omega))).
\end{align}
In the following we will explicitly show that the operation $\omega \mapsto \frac{S(\omega)}{\|S(\omega)\|_q}$ is contractive in (exponential) projective Hilbert metric.
\begin{remark}
One way of doing that \cite{Shahverdikondori.2022, Gautier.2020} is to observe that the contraction coefficient, say $\kappa(\frac{S(\cdot)}{\|S(\cdot)\|_q})=\kappa(S(\cdot))\leq \kappa(\Lambda^*)\kappa(\Lambda)\kappa(\cdot^{p-1})\kappa(\cdot^{\frac{1}{q-1}})<1$, is upper bounded by 1 for suitable $\Phi, q,p$, where the contraction coefficient is defined as
\begin{align}
    \kappa(\Phi):=\inf\{C| d_H(\Phi(A),\Phi(B)) \leq C d_H(A,B)\,\forall A,B\ge 0\}.
\end{align}
For the latter two contraction coefficients one may show that $\kappa(\cdot^{p-1})\kappa(\cdot^{\frac{1}{q-1}})=\frac{p-1}{q-1}$ iff $1\leq p\leq 2\leq q\leq\infty$, see \cref{lem:contraction.counterexample}. 
This, however, requires $\kappa(\Lambda)$. 
To give explicit worst-case bounds for arbitrary $\Lambda$ satisfying just the positivity improving assumption and without requiring computing $\kappa(\Lambda)$ we will, however, follow a slightly different route, inspired by \cite{Bhaskara.2011}.    
\end{remark}

Fist we show that the $q\to p$ norm we want to compute is sandwiched by the above defined $M,m$. Its proof has the advantage of also implying uniqueness of the optimal operator.

\begin{lemma}
    For any $\omega\in \cB^+(\mathcal{H})$ with $\|\omega\|_q=1$ it holds that $m(\omega)^{q-1}\leq \|\Lambda\|^p_{q\to p}<M(\omega)^{q-1}$.
    \label{lem:3}
\end{lemma}
\begin{proof}
    Set $\cS^{+,1}_q:=\{\rho\in \cB^+(\mathcal{H})|\|\rho\|_q=1\}$ be the strictly positive part of the $\|\cdot\|_q$-unit sphere. Let $\omega\in \cS^{+,1}_q$ and let $\tilde{\omega}\in \cS^{+,1}_q$ be the operator which maximizes $f$. We can assume $\omega\neq\tilde{\omega}$ since for that case we already know that the statement of the lemma is true by the above. By definition of $m(\omega)$ it follows
    \begin{align}
        m(\omega)^{q-1}&=m(\omega)^{q-1}\Tr[\omega^q]=\Tr[\omega m(\omega^{q-1})\omega^{q-1}]\leq \Tr[\omega S(\omega)^{q-1}]=\Tr[\omega \Lambda^*(\Lambda(\omega)^{p-1})] \notag \\ &=\Tr[\Lambda(\omega)\Lambda(\omega)^{p-1}]=\Tr[\Lambda(\omega)^p]\leq \|\Lambda\|^p_{q\to p}=m(\tilde{\omega})^{q-1}
    \end{align}
    The other inequality needs a bit more work. Write $\mu:=m(\tilde{\omega})=M(\tilde{\omega})=\left(\frac{\|\Lambda(\tilde{\omega})\|_p^p}{\|\tilde{\omega}\|_q^q}\right)^{\frac{1}{q-1}}=\|\Lambda\|_{q^\to p}^{\frac{p}{q-1}}$. By definition $S(\tilde{\omega})^{q-1}=\mu^{q-1}\tilde{\omega}^{q-1} \Leftrightarrow S(\tilde{\omega})=\mu\tilde{\omega}$.
    Now we define $\theta:=\sup\{\lambda\in\R|\omega^{q-1}\geq \lambda^{q-1}\tilde{\omega}^{q-1}\}=2^{-\frac{1}{q-1}D_{\text{max}}(\tilde{\omega}^{q-1}\|\omega^{q-1})}$. Since $\omega\neq\tilde{\omega}$ this is strictly positive and since both are $q$-normalized $0<\theta<1$. Now by the operator monotonicity of $t\mapsto t^{p-1}, t\mapsto t^{\frac{1}{q-1}}$ and the strict op-monotonicity of $\Lambda,\Lambda^*$ it follows
    \begin{align}
        \omega^{q-1}\geq \theta^{q-1}\omega^{q-1} &\Rightarrow \omega \geq \theta\tilde{\omega} \Rightarrow \Lambda(\omega)>\theta\Lambda(\tilde{\omega}) \Rightarrow \Lambda(\omega)^{p-1}>\theta^{p-1}\Lambda(\tilde{\omega})^{p-1} \\ &\Rightarrow \Lambda^*(\Lambda(\omega)^{p-1})=S(\omega)^{q-1}>\theta^{p-1}S(\tilde{\omega})^{q-1}=\theta^{p-1}\mu^{q-1}\tilde{\omega}^{q-1}.
    \end{align}
    Now by definition of $\theta$ there exists a vector $\psi\in\mathcal{H}\setminus\{0\}$ s.t. $(\omega^{q-1}-\theta^{q-1}\tilde{\omega}^{q-1})\psi=0$ and thus
    \begin{align}
        0< \langle \psi,(S(\omega)^{q-1}-\theta^{p-1}\mu^{q-1}\tilde{\omega}^{q-1})\psi\rangle = \langle \psi,(S(\omega)^{q-1}-\theta^{p-1}\mu^{q-1}\theta^{1-q}\omega^{q-1})\psi\rangle.
    \end{align}
    This now implies $M(\omega)^{q-1}>\mu^{q-1}\theta^{p-q}>\mu^{q-1}=\|\Lambda\|_{q\to p}^p$ since $q\geq p$ and $0<\theta<1$, which is what was to prove.
\end{proof}

This directly also yields the uniqueness of invariant operators under $S$, hence the algorithm will necessarily converge to the maximizer of $f$. 

\begin{corollary}\label{cor:unique.optimizer}
    There is a unique positive state $\tilde{\omega}\in \cB^+(\mathcal{H})$ which is (invariant) stationary w.r.t $S$. Hence this unique invariant state under $S$ maximizes $f$.
\end{corollary}
\begin{proof}
    Let $\tilde{\omega}\in \cS^{+,1}_q$ be the operator which maximizes $f$. Now in the proof of Lemma \ref{lem:3} we showed that for all other $\omega\in \cS^{+,1}_q\setminus\{\tilde{\omega}\}$ satisfy
    \begin{align}
        m(\omega)\leq m(\tilde{\omega})=M(\tilde{\omega}) < M(\omega).
    \end{align} Since every under $S$ invariant operator $\omega\in \cS^{+,1}_q$ satisfies $m(\omega)=M(\omega)$ this directly implies that the only stationary point of $S$ is the maximizer of the function $f$.
\end{proof}

Next we will show that the projective Hilbert metric is non-increasing under the application of $S$ and renormalizing.

\begin{lemma}\label{lem:BoydsMonotonicity}
    Let $\omega\in \cB^+(\mathcal{H})$, then $m\left(\frac{\omega}{\|\omega\|_q}\right)\leq m\left(\frac{S(\omega)}{\|S(\omega)\|_q}\right)$ and $M\left(\frac{\omega}{\|\omega\|_q}\right)\geq M\left(\frac{S(\omega)}{\|S(\omega)\|_q}\right)$ with equality iff $\omega$ is the fixed point of $S$, or equally the maximizer of $f$.
\end{lemma}

\begin{proof}
    Denote the maximizer of $f$ on $\cS^{+,1}_q$ as $\tilde{\omega}\in \cS^{+,1}_q$. If $\omega=\tilde{\omega}$, we have equality in the statement of the Lemma by stationarity under the map $S$. So it remains to be shown for $\omega\in \cS^{+,1}_q\setminus\{\tilde{\omega}\}$. We will prove the statement for $m$ since for $M$ it works equivalently. We want to show
    \begin{align}
        S\left(\frac{S(\omega)}{\|S(\omega)\|_q}\right)^{q-1} > m\left(\frac{\omega}{\|\omega\|_q}\right)^{q-1}\frac{S(\omega)^{q-1}}{\|S(\omega)\|_q^{q-1}},
    \end{align} which implies the statement in the Lemma. To show this we need the following facts. For $\lambda\geq 0, \sigma\in \cB^+(\mathcal{H})$ it is easy to show that $m(\lambda \sigma)^{q-1}\lambda^{q-p}=m(\sigma)^{q-1}$. From the operator monotonicity of $S^{q-1}$ and $\omega m(\omega)\leq S(\omega)$ it follows that $S(S(\omega))^{q-1}\geq S(\omega m(\omega))^{q-1}$. With these we can rewrite
    \begin{align}
        &S(S(\omega))^{q-1}\|S(\omega)\|_q^{1-p}= S\left(\frac{S(\omega)}{\|S(\omega)\|_q}\right)^{q-1} > m\left(\frac{\omega}{\|\omega\|_q}\right)^{q-1}\frac{S(\omega)^{q-1}}{\|S(\omega)\|_q^{q-1}} \\ &\hspace{7.05cm}= m(\omega)^{q-1}S(\omega)^{q-1}\|\omega\|_p^{q-p}\|S(\omega)\|_q^{1-q} \\ \Leftrightarrow & S(S(\omega))^{q-1}>m(\omega)^{p-1}m(\omega)^{q-p}S(\omega)^{q-1}\|\omega\|_p^{q-p}\|S(\omega)\|_q^{p-q}= m(\omega)^{q-1}S(\omega)^{q-1}\left(\frac{\|\omega\|_qm(\omega)}{\|S(\omega)\|_q}\right)^{q-p}.
    \end{align}
    Thus this reduces to
    \begin{align}
        \left(\frac{\|\omega\|_qm(\omega)}{\|S(\omega)\|_q}\right)^{q-p} \leq 1 \Leftrightarrow \|\omega\|_qm(\omega) \leq \|S(\omega)\|_q,
    \end{align}
    which is implied directly by $\omega m(\omega)\leq S(\omega)$.
    The proof for $M$ instead of $m$ goes analogous. Note that in both cases due to strict monotonicity, if $\omega\notin\textup{span}\tilde{\omega}$, then the inequalities are strict.
\end{proof}
In order to get a concrete bound on the runtime of this algorithm, we need to further quantify how contractive it is. We do this with the following powerful and immportant Lemma.

\begin{lemma}
Let $\omega\in \cS^{+,1}_q$ satisfy $\omega\geq \epsilon$ and assume WLOG that $\|\Lambda\|_{\infty\to\infty}\leq 1$. Suppose $M(\omega)\geq (1+\alpha)m(\omega)$, then $m\left(\frac{S(\omega)}{\|S(\omega)\|_q}\right) \geq (1+\frac{\alpha\epsilon}{Nd})m(\omega)$.  
\label{lem:5}
\end{lemma}
\begin{proof}
    What we want to show is $M(\omega)\geq(1+\alpha)m(\omega)$ implies
    \begin{align}
        S\left(\frac{S(\omega)}{\|S(\omega)\|_q}\right)^{q-1}\geq \left(1+\frac{\alpha\epsilon}{Nd}\right)^{q-1}m(\omega)^{q-1}\frac{S(\omega)^{q-1}}{\|S(\omega)\|_q^{q-1}}. \label{equ:lem5proof}
    \end{align} With the same manipulations as in the proof of Lemma \ref{lem:BoydsMonotonicity} we see that \eqref{equ:lem5proof} is equivalent to
    \begin{align}
        S(S(\omega))^{q-1} \geq \left(1+\frac{\alpha\epsilon}{Nn}\right)^{q-1}m(\omega)^{p-1}S(\omega)^{q-1}\left(\frac{\|\omega\|_qm(\omega)}{\|S(\omega)\|_q}\right)^{q-p} \geq \left(1+\frac{\alpha\epsilon}{Nd}\right)^{q-1}m(\omega)^{p-1}S(\omega)^{q-1}.
    \end{align} Here we used, in the inequality, that $p\leq q$ and $\omega m(\omega)\leq S(\omega)$. We will now prove that this inequality follows from the assumptions of the Lemma. For simplicity we write $\lambda:=m(\omega)$. We have $\lambda\omega\leq S(\omega)\leq M(\omega)\omega=(1+\alpha)\lambda\omega$. This is equivalent to $\lambda \leq \omega^{-\frac{1}{2}}S(\omega)\omega^{-\frac{1}{2}}\leq (1+\alpha)\lambda$, where each inequality is tight. Now let $\{|\psi\rangle_i\}_i$ be the normalized eigenbasis of $\omega^{-\frac{1}{2}}S(\omega)\omega^{-\frac{1}{2}}$ s.t. we have
    \begin{align}
        \omega^{-\frac{1}{2}}S(\omega)\omega^{-\frac{1}{2}} = \sum_{i=1}^d\lambda(1+\epsilon_i)|\psi_i\rangle\langle\psi_i|, \text{ where } 0=\epsilon_1\leq \epsilon_i\leq \epsilon_n=\alpha \quad \forall i.
    \end{align} 
    Setting $|\tilde{\psi}\rangle:=\omega^{\frac{1}{2}}|\psi_d\rangle$ we get $S(\omega)\geq \lambda\omega+\alpha\lambda|\tilde{\psi}\rangle\langle\tilde{\psi}|$.
    Now by the strict positivity $\Lambda(\rho)\geq \frac{\1}{Nn}\Tr[\rho]$ for any input $\rho$ it follows that
    \begin{align}
        \Lambda(S(\omega))\geq \lambda\Lambda(\omega)+\alpha\lambda\Lambda(|\tilde{\psi}\rangle\langle\tilde{\psi}|)\geq \lambda\Lambda(\omega)+\alpha\lambda\frac{\1}{Nd}\langle\psi_d|\omega|\psi_d\rangle\geq \lambda\left(1+\frac{\alpha\epsilon}{Nd}\right)\Lambda(\omega),
    \end{align} where we used that $\langle\psi_n|\omega|\psi_n\rangle\geq \epsilon$ and $\Lambda(\omega)\leq \Lambda(\1)\leq \1$ since $\|\Lambda\|_{\infty\to\infty}\leq 1$.
    Thus we have 
    \begin{align}
        S(S(\omega))^{q-1}\geq \lambda^{p-1}\left(1+\frac{\alpha\epsilon}{Nd}\right)^{p-1}S(\omega)^{q-1}=m(\omega)^{p-1}\left(1+\frac{\alpha\epsilon}{Nd}\right)^{p-1}S(\omega)^{q-1},
    \end{align} which is what was to show.
\end{proof}
\begin{remark}
We note that classically this is true without the requirement on the smallest eigenvalue $\epsilon$ of $\omega$. If one could also show this in the quantum setting, the runtime of the algorithm would improve to $\tilde{\mathcal{O}}(Nd)$ up to logarithmic corrections.    
\end{remark}

Before we can get a concrete runtime out of this we need a bound on the $\epsilon$ in the above Lemma, which we do in the following Lemma.

\begin{lemma}\label{lem:eigenvalue.lower.bound}
Given any $\omega$, s.t. $\|\omega\|_q=1$, it holds that
\begin{align}
    \frac{S(\omega)}{\|S(\omega)\|_q}\geq \frac{1}{Nd\sqrt{d}},
\end{align} if $\Lambda$ is TP and
\begin{align}
    \frac{S(\omega)}{\|S(\omega)\|_q}\geq \frac{1}{(Nd)^{\frac{p}{q-1}}} \geq \frac{1}{N^2d^2}
\end{align} else.
In particular, this applies to the unique fixed point, which is the optimal operator $\tilde{\omega}$.
\end{lemma}
\begin{proof}
Here we use the positivity improving property and the bound on $\|\Lambda\|_{\infty\to\infty}$ to get the desired result.
Note first that $\|\Lambda^*\|_{\infty\to\infty}=\|\Lambda\|_{1\to 1}\leq d\|\Lambda\|_{\infty\to\infty}\leq d$ and in the case where $\Lambda$ is TP we have that $\Lambda^*(\1)=\1$. In the TP case we have 
\begin{align}
    \1 \geq \Lambda(\omega)^{p-1} \geq \1 (Nd)^{1-p}\Tr[\omega]^{p-1} \Rightarrow \1 \geq S(\omega) \geq (Nd)^\frac{1-p}{q-1}\Tr[\omega]^{\frac{p-1}{q-1}} \geq (Nd)^{\frac{1-p}{q-1}},
\end{align} which implies that
\begin{align}
    \frac{S(\omega)}{\|S(\omega)\|_q} \geq (Nd)^{\frac{1-p}{q-1}}d^{-\frac{1}{q}} \geq \frac{1}{Nd\sqrt{d}},
\end{align} where we used that $p\leq 2\leq q$.
In the cases where $\Lambda$ is not TP we have, similarly
\begin{align}
 \|\omega\|_\infty^{p-1}\1 \geq \Lambda(\omega)^{p-1} \geq \1 (Nd)^{1-p}\Tr[\omega]^{p-1} \Rightarrow d^\frac{1}{q-1}\1 \geq S(\omega) \geq d^{\frac{1-p}{q-1}}N^{-\frac{p}{q-1}}\Tr[\omega]^{\frac{p-1}{q-1}},
\end{align} which implies that
\begin{align}
    \frac{S(\omega)}{\|S(\omega)\|_q} \geq d^{\frac{1-p}{q-1}}N^{\frac{-p}{q-1}}d^\frac{-1}{q-1} = \frac{1}{(Nd)^\frac{p}{q-1}}\geq \frac{1}{N^2d^2},
\end{align} where we again used that $p\leq 2\leq q$.
\end{proof}

To bound the initial distance we observe, similarly, the following.

\begin{corollary}[\cref{lem:initial.distance.boyds}]\label{cor:initial.distance} It holds that
\begin{align}
\exp(d_H(\1,S(\1)))=\frac{M(\1)}{m(\1)} = \|S(\1)\|_\infty\|S(\1)^{-1}\|_\infty \leq 
\begin{cases}
    & N^\frac{p-1}{q-1} \leq N  \quad \text{ if } \Lambda \text{ is TP}, \\ &dN^{\frac{p}{q-1}}\leq N^2d \quad \text{ else}.
\end{cases}   
\end{align}
\end{corollary}
\noindent The proof follows as the proof above.

Given this and the epsilon we now directly get the explicit runtime quoted in the theorem. We do this in the following way. Along the way we also prove \cref{lemcontractivitykappa}.
\begin{proof}[Proof of \cref{thm:strict.positive.q.Boyds} and \cref{lemcontractivitykappa}]\label{proof.quantum.Boyds}
Denote the output of the $j$-th iteration of the algorithm as $\omega_j$ on initial input $\omega_0=\frac{\1}{\|\1\|_q}$ for $j\in\N$ and denote the optimal operator as $\tilde{\omega}$ as before.
First notice that \cref{lem:5} implies that
\begin{align}
  \exp(d_H(\omega_j,\omega_{j+1})) \leq \frac{M(\omega_j)}{m(\omega_j)}=: 1+\alpha_{j}\leq 1+\alpha\eta^j,
\end{align}
where $\eta= {1-\frac{\epsilon}{Nd}}\equiv 1-\mu$ and $\alpha+1=\frac{M(\1)}{m(\1)}=\frac{M(\omega_0)}{m(\omega_0)}$. 
Combining this with the lower bounds on $\epsilon$ in \cref{lem:eigenvalue.lower.bound} directly yields \cref{lemcontractivitykappa}.
This follows by induction, where \cref{lem:5} is the induction start and the induction step is
\begin{align}
    \frac{M(\omega_j)}{m(\omega_j)} &\overset{\cref{lem:5}}{\leq} \frac{M(\omega_{j-1})}{m(\omega_{j-1})}\frac{1}{1+\alpha_{j-1}\frac{\epsilon}{Nd}} = \frac{1+\alpha_{j-1}}{1+\alpha_{j-1}\frac{\epsilon}{Nd}} \\ &\hspace{0.7cm}= 1+\alpha_{j-1}\frac{1-\frac{\epsilon}{Nd}}{1+\alpha_{j-1}\frac{\epsilon}{Nd}} < 1+\alpha_{j-1}\left({1-\frac{\epsilon}{Nd}}\right),
\end{align} where we denoted $1+\alpha_j:=\frac{M(\omega_j)}{m(\omega_j)}$.
We note that this is effectively proving that $\kappa(S)\leq \eta$.
Now it is easily seen that $\tilde{\omega}\leq \lambda \omega_K$ for
\begin{align}
    \lambda = \|\omega_K^{-1/2}\tilde{\omega}\omega_K^{-1/2}\| &\leq \exp(d_H(\omega_K,\tilde{\omega})) \leq \exp\left(\sum_{j=K}^\infty d_H(\omega_j,\omega_{j+1})\right) \\
    &\leq \prod_{j=K}^\infty \frac{M(\omega_j)}{m(\omega_j)} \leq \prod_{j=K}^\infty 1+\alpha\eta^j =\exp\left(\sum_{j=K}^\infty\log(1+\alpha\eta^j) \right) \\
    &\leq \exp\left(\alpha\sum_{j=K}^\infty\eta^j\right) \leq \exp\left(\alpha\frac{\eta^K}{1-\eta}\right)= 1+\alpha{\eta^K} + o(\eta^{K}).
\end{align}

Now we observe that for $\omega_K$, with the above $\lambda$ we have that
\begin{align}
   \|\Lambda\|_{q\to p} = \|\Lambda(\tilde{\omega})\|_{p} \leq \lambda\|\Lambda(\omega_K)\|_p.
\end{align}
Thus for $K\geq\frac{\log\frac{(1-\eta)\varepsilon}{\alpha}}{\log(\eta)} = \frac{1}{\log(1-\mu)}\log\left(\frac{\mu\varepsilon}{\alpha}\right) = \mathcal{O}\left(\frac{1}{\mu}\log\left(\frac{\alpha}{\mu\epsilon}\right)\right)= \mathcal{O}(\poly(N,d)\log(N,d,\frac{1}{\varepsilon}))$
we have
\begin{align}
    (1-\varepsilon)\|\Lambda\|_{q\to p} \leq \|\Lambda(\omega_K)\|_p\leq \|\Lambda\|_{q\to p}.
\end{align}
\end{proof}

Combining \cref{lem:initial.distance.boyds} and the above we have showed the following.
\begin{corollary}
Let $\Lambda,p,q$ be as assumed in this section. If $\Lambda$ is TP, then after at most $K=\mathcal{O}(N^2d^2\sqrt{d}\log\frac{N^3d^2\sqrt{d}}{\epsilon})$ iterations does the algorithm output a state that approximates $\|\Lambda\|_{q\to p}$ to multiplicative precision $\epsilon$ and for general $\Lambda$ we have the same after at most $K=\mathcal{O}\left(N^3d^3\log(\frac{N^5d^4}{\epsilon})\right)$.    
\end{corollary}

\subsection{Proof of \cref{lem:Schatten.mixed.continuity.bound}}\label{lemcontinuity.bound}

If $\Lambda$ is TP we have
\begin{align}
\|\Lambda\|_{q\to p} =\sup_{\rho}\frac{\|\Lambda(\rho)\|_p}{\|\rho\|_q} \geq \sup_{\rho}\frac{d^{\frac{1}{p}-1}\|\Lambda(\rho)\|_1}{\|\rho\|_1} \geq d^{\frac{1}{p}-1},
\end{align} where we used that $\|\rho\|_1\geq \|\rho\|_p\geq d^{\frac{1}{p}-1}\|\rho\|_1$. Else we have that
\begin{align}
    \|\Lambda\|_{q\to p} =\sup_{\rho}\frac{\|\Lambda(\rho)\|_p}{\|\rho\|_q} \geq \sup_{\rho}\frac{\|\Lambda(\rho)\|_\infty}{d^\frac{1}{q}\|\rho\|_\infty} \geq d^{-\frac{1}{q}}c.
\end{align}
On the one hand we clearly have
\begin{align}
    \|\Lambda_\delta\|_{q\to p} &\leq (1-\delta)\|\Lambda\|_{q\to p} + \delta \left\|\frac{\1}{d}\right\|_p\sup_\rho\frac{\|\rho\|_1}{\|\rho\|_q} \\ &\leq (1-\delta)\|\Lambda\|_{q\to p} + \delta d^{\frac{1}{p}-\frac{1}{q}} 
\end{align} where we used that $\|\rho\|_1\leq d^{1-\frac{1}{q}}\|\rho\|_q$ and $\|\1\|_p=d^\frac{1}{p}$. On the other hand, denoting with $\tilde{\omega}$ the optimal operator that achieves $\|\Lambda\|_{q\to p}$ we have, similarly
\begin{align}
    \|\Lambda_\delta\|_{q\to p} &\geq \left|(1-\delta)\|\Lambda\|_{q\to p}-\delta\|\frac{\1}{d}\|_p\|\tilde{\omega}\|_1 \right| \\ &\geq (1-\delta)\|\Lambda\|_{q\to p}-\delta d^{\frac{1}{p}-\frac{1}{q}} 
\end{align}
Combining these inequalities we get in the TP case that 
\begin{align}
    (1-\delta+\delta d^{1-\frac{1}{q}})^{-1}\|\Lambda_\delta\|_{q\to p} \leq \|\Lambda\|_{q\to p} \leq \|\Lambda_\delta\|_{q\to p}(1-\delta-\delta d^{1-\frac{1}{q}})^{-1}.
\end{align} So having $\|\Lambda_\delta\|_{q\to p}$ to multiplicative error $(1+\frac{\epsilon}{2})$ means having $\|\Lambda\|_{q\to p}$ to multiplicative error
\begin{align}
    \left(1+\frac{\epsilon}{2}\right)\frac{1-\delta-\delta d^{1-\frac{1}{q}}}{1-\delta+\delta d^{1-\frac{1}{q}}}=1+\frac{\epsilon}{2}+2\delta d^{1-\frac{1}{q}} +\mathcal{O}(\delta^2,\epsilon \delta). 
\end{align} Picking $\delta=d^{\frac{1}{q}-1}\frac{\epsilon}{4}\geq\frac{\epsilon}{4d}$ yields the claim.

In the non TP case we analogously get
\begin{align}
    (1-\delta+\delta d^{\frac{1}{p}}c)^{-1}\|\Lambda_\delta\|_{q\to p} \leq \|\Lambda\|_{q\to p} \leq \|\Lambda_\delta\|_{q\to p}(1-\delta-\delta d^{\frac{1}{p}}c)^{-1}.
\end{align} So having $\|\Lambda_\delta\|_{q\to p}$ to multiplicative error $(1+\frac{\epsilon}{2})$ means having $\|\Lambda\|_{q\to p}$ to multiplicative error
\begin{align}
    \left(1+\frac{\epsilon}{2}\right)\frac{1-\delta-\delta d^{\frac{1}{p}}c}{1-\delta+\delta d^{\frac{1}{p}}c}=1+\frac{\epsilon}{2}+2\delta d^{\frac{1}{p}}c +\mathcal{O}(\delta^2,\epsilon \delta). 
\end{align} Picking $\delta=d^{-\frac{1}{p}}\frac{\epsilon}{4c}\geq\frac{\epsilon}{4dc}$ yields the claim.

\end{document}